\documentclass[12pt,a4paper]{article}

  \usepackage{a4wide}
  \usepackage{latexsym}
  \usepackage{epsf}
  \usepackage{amssymb}
  \usepackage{graphicx}
  \usepackage{amsmath, cite}
  \usepackage{slashed,epsfig}
  \usepackage{bbm}
  \usepackage{bbm}
  \usepackage{amsmath,amssymb,amsthm}
  \usepackage{pst-all}
  \usepackage{here}
  \usepackage{overpic}
\usepackage{slashbox}
\usepackage{tabularx}
\usepackage{multirow}
\usepackage{rotating}
\usepackage{scalefnt}

  \definecolor{AV}{rgb}{0.65,0.0,0}

\usepackage[
      colorlinks=true,
      linkcolor=blue,
      urlcolor=blue,
      filecolor=blue,
      citecolor=red,
      pdfstartview=FitV,
      pdftitle={},
      pdfauthor={},
      pdfsubject={},
      pdfkeywords={},
      pdfpagemode=None,
      bookmarksopen=true
]{hyperref}

\renewcommand{\d}{\textrm{d}}

\newcommand{\e}{\textrm{e}}
\renewcommand{\d}{\textrm{d}}

\newcommand{\Sp}{\mathop{\rm {}Sp}}

\newcommand{\SL}{\mathop{\rm SL}}

\newcommand{\beq}{\begin{equation}}
\newcommand{\eeq}{\end{equation}}
\newcommand{\be}{\begin{equation}}
\newcommand{\ee}{\end{equation}}
\newcommand{\beqa}{\begin{eqnarray}}
\newcommand{\eeqa}{\end{eqnarray}}
\newcommand{\beqar}{\begin{eqnarray*}}
\newcommand{\eeqar}{\end{eqnarray*}}
\newcommand{\bea}{\begin{eqnarray}}
\newcommand{\eea}{\end{eqnarray}}

\newcommand{\nn}{\nonumber}

\begin{document}
%\begin{flushright}
%\small
%number\\
%\date \\
%\normalsize
%\end{flushright}

\begin{center}

\vspace{.5cm}
{\LARGE {\bf Black hole solutions to the $F_4$-model and their orbits (I)}} \\

\vspace{1cm}

{\large W. Chemissany$^{a,b}$, P. Giaccone$^{c}$, D. Ruggeri$^c$,  M. Trigiante$^{c}${}~,\\
\vspace{1cm}
 $^a$ {\small\slshape Instituut voor Theoretische Fysica, Katholieke Universiteit Leuven,\\ Celestijnenlaan 200D, B-3001 Leuven, Belgium }\\\vspace{0.2cm}
  $^b$ {\small\slshape Department of Physics and Astronomy, University of Waterloo, Waterloo, Ontario, Canada, N2L 3G1 }\\\vspace{0.2cm}

$^c${\small\slshape Laboratory of Theoretical Physics, Dipartimento di Scienze Applicate e Tecnologia, Politecnico di Torino,
C.so Duca degli Abruzzi, 24, I-10129 Torino, Italy}}\\
\vspace{0.4cm}
{\bf Abstract} \end{center} {\small In this paper we continue the
program of the classification of nilpotent orbits using the approach
developed in arXiv:1107.5986, within the study of black hole
solutions in $D=4$ supergravities. Our goal in this work is to
classify static, single center black hole solutions to a specific $N=2$
four dimensional ``magic'' model, with special K\"ahler scalar
manifold ${\rm Sp}(6,\mathbb{R})/{\rm U}(3)$, as orbits of geodesics
on the pseudo-quaternionic manifold ${\rm F}_{4(4)}/[{\rm
SL}(2,\mathbb{R})\times {\rm Sp}'(6,\mathbb{R})]$ with respect to the
action of the isometry group ${\rm F}_{4(4)}$. Our analysis amounts
to the classification of the orbits of the geodesic ``velocity''
vector with respect to the isotropy group $H^*={\rm
SL}(2,\mathbb{R})\times {\rm Sp}'(6,\mathbb{R})$, which include a thorough classification of the \emph{nilpotent orbits} associated with extremal solutions  and reveals a richer
structure than the one predicted by the $\beta-\gamma$ labels alone,
based on the Kostant Sekiguchi approach. We provide a general proof
of the conjecture made in ${\rm hep-th}/0908.1742$ which states that
regular single center solutions belong to orbits with coinciding
$\beta-\gamma$ labels. We also prove that the reverse is not true by
finding distinct orbits with the same $\beta-\gamma$ labels, which
are distinguished by suitably devised tensor classifiers. Only one
of these is generated by regular solutions.  Since regular static
solutions only occur with nilpotent degree not exceeding 3, we only
discuss representatives of these orbits in terms of black hole
solutions. We prove that these representatives can be found in the
form of a purely dilatonic four-charge solution (the generating
solution in $D=3$) and this allows us to identify the orbit
corresponding to the regular four-dimensional metrics. $H^*$-orbits
with degree of nilpotency greater than 3 are analyzed solely from a
group theoretical point of view, leaving a systematic analysis of
their possible interpretation in terms of static multicenter or
stationary non-static solutions to a future work. We just limit
ourselves to give (singular) single-center  representatives of these
orbits, to be possibly interpreted as singular limits of regular
multicenter solutions.
%We also comment on the relations between
%these orbits illustrating how their representatives can be can be
%decomposed into the sum of generators belonging to orbits with lower
%degree of nilpotency.
%This ``sum-rule'' is relevant in view of a
%future orbit analysis of multicenter solutions.
We provide the explicit transformations mapping the various
$H^*$-orbits and in particular BPS into non-BPS regular solutions
showing that they in general belong to the complexification of the
global symmetry group in $D=3$.}

\newpage

\pagestyle{plain} \tableofcontents

\section{Introduction}
Dimensional reduction along time offers a powerful way to study
stationary solutions of 4D symmetric supergravity models via
group-theoretical methods
\cite{Breitenlohner:1987dg,Chemissany:2010zp,Chemissany:2010ay,Maison:2000fj,
Bergshoeff:2008be,Chemissany:2009hq,
Chemissany:2009af,Bossard:2009bw,
Bossard:2009mz,Bossard:2009my,Breitenlohner:1998cv,Bossard:2009we,Bossard:2009at,
Ferrara:1997ci,Gimon:2007mh}. In this way the black hole solutions
are identified with the geodesics on a pseudo-Riemannian coset
manifold $G/H^*$ and the corresponding geodesic equations are best
approached when they are cast into the Lax form
\cite{Chemissany:2010zp,Chemissany:2010ay}. More precisely, it is
known that  regular extremal black holes associated with Lax
operators $L(\tau)$ ($\tau=-1/r$) that are nilpotent all along their
radial evolution
\cite{Gunaydin:2007bg,Gaiotto:2007ag,Bergshoeff:2008be,Chemissany:2009af}.
Since geodesics are totally defined by their ``initial point''
$P(0)$ and ``initial velocity'' $L(0)$, and since the action of
$G/H^*$ on $P(0)$ is transitive on the manifold, we can fix $P(0)$
to coincide with the origin $O$ and classify the geodesics by the
orbits of the corresponding initial velocity $L(0)$ with respect to
the isotropy group $H^*$. Hence the classification of extremal black
holes requires a classification of the orbits of nilpotent elements
of the coset space $\mathfrak{K}^*$ (isomorphic to the tangent space to the manifold in $O$) with respect to the adjoint action
of the stability subgroup ${H}^*$ of $G.$ Using the simplificative
analogy of $L(\tau)$ with a velocity vector in Minkowsky space,
where the stability group is ${\rm SO}(1,3)$, we can say that
``time-like'' $L(0)$ correspond to non-extremal four dimensional
solutions, while extremal ones correspond to ``light-like''
geodesics on the manifold $G/H^*$, generated by nilpotent matrices.
As opposed to the Minkowski case, in the problem at hand, however, the
``light-like'' vectors may actually fall in a variety of
$H^*$-orbits. Classifying these is the main goal of the present
work. Just as the velocity of a photon can be made simplest by going
to a suitable frame of reference in which one of the axes coincide
with the direction of propagation, we construct, for the class of
orbits which are relevant for static black holes, a frame in which
the velocity of geodesic is simplest. This frame is defined by the
``generating solution'' and considerably simplifies the analysis of
the correspondence between static black holes ad orbits in $D=3$. In
the present work we shall restrict to a specific $\mathcal{N}=2$
symmetric supergravity coupled to six vector multiplets, whose
scalar fields span the special K\"ahler manifold $G_4/H_4={\rm
Sp}(6,\mathbb{R})/{\rm U}(3)$. Upon timelike-reduction to $D=3$ and
dualization of vectors into scalar fields, the target space of the
resulting Euclidean sigma-model is the symmetric manifold
$G/H^*=\mathrm{F}_{4(4)}/[\mathrm{SL}(2,\mathbb{R})\times {\rm
Sp}'(6,\mathbb{R})]$.\par Our approach to the problem is a synthesis
of the ones followed in \cite{Fre:2011uy,Fre:2011ns}.
 In \cite{Fre:2011uy}, in order to achieve
a classification of nilpotent orbits, the authors of
  thoroughly discussed the static spherical
symmetric black-hole solutions of the simplest $\mathcal{N}=2$
supergravity model with one vector multiplet coupling, often dubbed
the $S^3$-model\footnote{The obtained results generalized previous
ones in \cite{Kim:2010bf}.}. In  this paper it has been shown that a
complete classification of the nilpotent $H^*$-orbits in
$\mathfrak{K}^*$ can be effected using the signatures of
symmetric-covariant ${H}^*$-tensors, named \emph{tensor classifiers}
(TC). The tensor structures used for the orbit analysis in  the
$G_{2}$-model are not enough however to provide a complete
classification of the orbits in more general case: New tensor
classifiers have to be devised. The standard approach to the study
of the relevant $H^*$-nilpotent orbits in the tangent space to the
manifold (coset space) was based on the description of a nilpotent
generator $E$ of the coset as part of a triplet of
$\mathrm{SL}(2,\mathbb{R})$-generators $\{E,F,h\}$, named
\emph{standard triple}, and on the classification of such triples
with respect to the so called $\gamma-\beta$-labels, which are
$H^*$- invariant quantities \cite{Kim:2010bf}.  As we shall prove in
the present work, this orbit analysis is by no means exhaustive:
Distinct orbits are found with the same $\gamma-\beta$-labels.

We shall apply a new constructive algorithm, devised in
\cite{Fre:2011ns}, which combines the method of standard triples
with new techniques based on the Weyl group. After a general group
theoretical analysis of the model this novel approach allows for a
systematic construction of the various nilpotent orbits by solving
suitable matrix equations in nilpotent generators $E$. Solutions to
these equations comprise representatives $E$ of the various orbits
and the final part of the analysis is to group them under the action
of suitable compact subgroups of $H^*$. Solutions which are
 not  connected by the action of such subgroups are then
found to be distinguished by certain $H^*$-invariants, which
comprise, besides the $\gamma-\beta$-labels, also the signatures of suitable \emph{tensor
classifiers}. This guarantees the completeness of the classification.
The tensor classifiers introduced here play an essential role in our
analysis. Although they still do not exhaust all possible
$H^*$-tensor structures which can be devised, they provide by
themselves, without the use of the $\gamma-\beta$-labels, an almost
complete classification of the orbits. Their use allows to find the
orbit of a nilpotent generator $E$ without the need of constructing
the correspondent standard triple for the computation of the
relative $\gamma-\beta$-labels. Most importantly, they allow to
distinguish orbits with the same $\gamma-\beta$-labels!\par This
approach  was applied in \cite{Fre:2011ns} to the analysis of the
$\frac{\textrm{SO}(4,n)}{\textrm{SO}(2,n-2)\times \textrm{SO}(2,2)}$
model with $n>4$. In this work it was shown that the pattern of the
nilpotent orbits is a universal property depending on the
\emph{Tits-Satake (TS) universality class} \cite {Fre:2006eu} of the
model, which is defined by the coset
$\frac{\textrm{SO}(4,5)}{\textrm{SO}(2,3)\times \textrm{SO}(2,2)}$.

 The number of such classes was found for the
$\mathcal{N}=2$ symmetric models to be \emph{five} (see Table 2 in
\cite{Fre:2011ns}) and are defined by the Tits-Satake  algebra
associated with the $D=3$ isometry algebra $\mathfrak{g}$. It is
tempting to conjecture that the pattern of $H^*$-orbits found here
for the $\mathrm{F}_{4(4)}$-model captures the orbit structure of
all the models within the same universality class. We postpone an
answer to this question to a future work.
\par
The paper is organized as follows. In Sect.\ref{sksc} we discuss the geometry of the special K\"ahler manifold of the  $D=4$ $\mathcal{N}=2$ model, we review the $r$ and $c^*$-maps  and the general description of static $D=4$ black holes
as geodesics on the pseudo-Riemannian scalar manifold of the Euclidean $D=3$ theory obtained through time-reduction of the $D=4$ one. In Sect. \ref{npo}, we review our approach to the classification of the $H^*$-nilpotent orbits in $\mathfrak{K}^*$. The results are listed in Tables \ref{alpha1}-\ref{alpha15} and in Table \ref{orbitstc} of Appendix \ref{AA}. In Sect. \ref{generatingsol}, we review the construction of the generating solutions and show how it provides representatives of all the $H^*$-orbits with degree of nilpotency less or equal to $3$. We also identify those $H^*$-orbits containing the geodesics corresponding to regular and small $D=4$ solutions, in light of the known classifications.  We end  Sect. \ref{generatingsol} with a discussion of the orbits of non-extremal solutions.
In Sect. \ref{owhdn} we also provide examples of solutions corresponding to orbits with higher degree of nilpotency and show that they correspond to singular $D=4$ solutions. We end with some concluding remarks.
\par
During the final writing stage of the present work, we became aware of the interesting paper \cite{Bossard:2012ge} whose analysis has, in some points, an overlap with ours.

% Furthermore they developed a new algorithm for
%the classification and construction of the nilpotent orbits for each
%universality classes which heavily relies on an essential use of the
%Weyl group $\mathcal{W}$  of the TS subalgebra
%$\mathbbm{G}_{\textrm{TS}}$ and on a certain subgroup thereof
%\cite{Fre:2009zz}.  The orbits of the full group $G$ split into
%sub-orbits with respect to the stability subgroup $H^*.$ Such a
%splitting is dictated by the structure of the discrete coset
%$\frac{\mathcal{W}}{\mathcal{W}_{H}}.$
%
%For the case of the universality class
%$\frac{\textrm{SO}(4,5)}{\textrm{SO}(2,3)\times \textrm{SO}(2,2)}$
%the complete list of the nilpotent orbits has been derived in
%\cite{Fre:2011ns}, where also  the algorithm has been applied to the
%TS class $G_{2(2)}$ reproducing the previously obtained result which
%comprise seven nilpotent orbit. A very non-trivial and important
%case is the subject of this paper, namely the universal manifold
%$\frac{\textrm{F}_{4(4)}}{\textrm{Sp}(6)\times \textrm{SU}(1,1)}.$
%This very interesting Lie algebra problem is provided by the fifth
%and last TS universality class of table 2 in \cite{Fre:2011ns}.
\section{Static Black Holes in $D=4,\,\mathcal{N}=2$ Supergravity}\label{sksc}
We consider a four dimensional supergravity theory whose bosonic sector consists
of the graviton
$g_{\mu\nu}(x)$, $n_v$ vector fields $A^\Lambda_\mu(x)$,
$\Lambda=0,\dots, n_v-1$,  and $n_s$ scalar fields $\phi^r(x)$,
$r=1,\dots, n_s$.
The general form of the bosonic action reads\footnote{Here we are using the
``mostly plus'' signature for the metric $g_{\mu\nu}$ and the
 convention $\epsilon_{0123}=1$.}: \begin{align}
S_4&= \int d^4x\,\sqrt{|g|} \,\mathcal{L}_4=\int
d^4x\,\sqrt{|g|}\left[\frac{R[g]}{2}-\frac{1}{2}\,G_{rs}(\phi)\,\partial_\mu
\phi^r\partial^\mu \phi^s+\right.\nonumber\\
&+\left.\frac{1}{4} F^\Lambda_{\mu\nu}
I_{\Lambda\Sigma}(\phi)\,F^{\Sigma\, \mu\nu}+\frac{1}{8
\sqrt{|g|}}\,\epsilon_{\mu\nu\rho\sigma} F^{\Lambda\,\mu\nu}
R_{\Lambda\Sigma}(\phi)\,F^{\Sigma\, \rho\sigma}\right]\,.
\end{align}
 The scalar-dependent
matrix $G_{rs}(\phi)$ represents the positive definite metric on the Riemannian (simply connected) scalar manifold
$\mathcal{ M}^{(D=4)}_{scal} $, and  we have collectively denoted the
scalar fields by the short-hand notation $\phi\equiv (\phi^r)$. The
vector field strengths are defined as usual:
$F^\Lambda_{\mu\nu}\equiv
\partial_\mu A^\Lambda_\nu-\partial_\nu A^\Lambda_\mu$. The $n_v\times n_v$ matrices
$R_{\Lambda\Sigma}(\phi^r),\,I_{\Lambda\Sigma}(\phi^r)$ are the real
and imaginary parts of the  complex \emph{kinetic matrix}
$\mathcal{N}_{\Lambda\Sigma}(\phi^r)$ of the vector fields: $
R_{\Lambda\Sigma}\equiv {\rm Re}(\mathcal{N}_{\Lambda\Sigma})$, $
I_{\Lambda\Sigma}\equiv {\rm Im}(\mathcal{N}_{\Lambda\Sigma})$.\par
 In
$\mathcal{N}=2$ the scalar manifold is the product of a special
K\"ahler manifold $\mathcal{M}_{SK}$, parametrized by the  scalars
sitting in the vector multiplets,
 and a quaternionic K\"ahler manifold spanned by the hypermultiplet  scalars.
The latter do not contribute to the black hole solutions since they
do not enter the vector kinetic matrix
$\mathcal{N}_{\Lambda\Sigma}$. We shall therefore restrict ourselves
to an $\mathcal{N}=2$ supergravity coupled just to vector multiplets
and no hypermultiplets.
 We shall moreover  restrict ourselves to models
exhibiting a homogeneous symmetric  (special K\"ahler) scalar
manifold of the form $\mathcal{
M}^{(D=4)}_{scal}=\mathcal{M}_{SK}=G_4/H_4$ (symmetric models). The
action of an isometry transformation $g\in G_4$ on the scalar fields
$\phi^r$ parametrizing $\mathcal{ M}^{(D=4)}_{scal} $is defined by
means of a \emph{coset representative} $\mathbb{L}_4(\phi)\in
G_4/H_4$ as follows:
\begin{equation}
g\cdot \mathbb{L}_4(\phi^r)=\mathbb{L}_4(g\star\phi^r)\cdot
h(\phi^r,g)\,,\label{coset}
\end{equation}
where $g\star\phi^r$ denote the transformed scalar fields,
non-linear functions of the original ones $\phi^r$, and
$h(\phi^r,g)$ is a \emph{compensator} in $H_4$. The coset
representative is defined modulo right action of $H_4$ and  is fixed
by the chosen parametrization of the manifold.
\subsection{The Special K\"ahler Geometry of the $D=4$ Model}\label{skg4m}
In the present section  we shall compute  the main geometric
quantities related to the special K\"ahler geometry of the model
under consideration. Recall that a special K\"ahler manifold
$\mathcal{M}_{SK}$ \cite{Castellani:1990zd,D'Auria:1990fj,Andrianopoli:1996cm, deWit:1984pk,de Wit:1983rz}, of
complex dimension $n$, is a Hodge-K\"ahler manifold on which a flat,
holomorphic, symplectic vector structure is defined, with structure
group ${\rm Sp}(2n+2,\mathbb{R})$. If $\Omega(z^a)$ is a holomorphic
section of this bundle:
\begin{align}
\Omega(z^a)&=(\Omega^M(z^a))=\left(\begin{matrix}X^\Lambda (z^a)\cr
F_\Lambda(z^a)\end{matrix}\right)\,,\\
&\Lambda=0,\dots, n\,;\,\,\, a=1,\dots, n\,;\,\,\,M=1,\dots, 2n+2\,,
\end{align}
the K\"ahler potential is expressed as follows:
\begin{equation}
\mathcal{K}(z^a,\bar{z}^a)=-\log\left(-i\,\Omega\mathbb{C}\bar{\Omega}\right)=
-\log\left[-i\,(X^\Lambda \bar{F}_\Lambda-F_\Lambda
\bar{X}^\Lambda)\right]\,.\label{Kom}
\end{equation}
$\mathbb{C}=(\mathbb{C}_{MN})$ being the ${\rm
Sp}(2n+2,\mathbb{R})$-invariant matrix:
\begin{equation}
\mathbb{C}=\left(\begin{matrix}{\bf 0} & {\bf 1} \cr -{\bf 1}& {\bf
0}\end{matrix}\right)\,.
\end{equation}
The complex vector field $\Omega(z^a)$ also belongs to a
\emph{holomorphic line bundle}, namely it transforms by
multiplication times a holomorphic function $\Omega(z^a)\rightarrow
e^{-f(z)}\,\Omega(z^a)$. This implies, according to eq. (\ref{Kom}),
a K\"ahler transformation on the potential $\mathcal{K}$:
$\mathcal{K}\rightarrow \mathcal{K}+ f(z)+\bar{f}(\bar{z})$. It is
useful to introduce a section of a ${\rm U}(1)$-bundle over the
scalar manifold, $V(z^a,\bar{z}^a)\equiv
e^{\frac{\mathcal{K}}{2}}\,\Omega(z^a)$, which, as
$\Omega(z^a)\rightarrow e^{-f(z)}\,\Omega(z^a)$, transforms under a
${\rm U}(1)$- transformation: $V(z^a,\bar{z}^a)\rightarrow
e^{-i\theta}\,V(z^a,\bar{z}^a)$, where $\theta(z,\bar{z})={\rm
Im}(f)$. This vector satisfies the property of being
\emph{covariantly holomorphic} with respect to the ${\rm
U}(1)$-connection:
\begin{equation}
\nabla_{\bar{a}}V\equiv
(\partial_{\bar{a}}-\frac{1}{2}\,\partial_{\bar{a}}\mathcal{K})V=0\,,
\end{equation}
where $\partial_a\equiv \frac{\partial}{\partial {z}^a}$ and
$\partial_{\bar{a}}\equiv \frac{\partial}{\partial \bar{{z}}^a}$. If
we define $$U_a=(U_a{}^M)\equiv
\nabla_{{a}}V=(\partial_a+\frac{1}{2}\,\partial_a\mathcal{K})V\,,$$
the following properties hold:
\begin{align}
V\mathbb{C}\bar{V}&=i\,\,;\,\,\,U_a\mathbb{C}\bar{V}=\bar{U}_{\bar{a}}\mathbb{C}\bar{V}=0\,\,;\,\,\,
U_a\mathbb{C}\bar{U}_{\bar{b}}=-i\,g_{a\bar{b}}\,.\label{props}
\end{align} If
$E_a{}^I$, $I=1,\dots, n$,  is the complex vielbein matrix of the manifold,
$g_{a\bar{b}}=\sum_{I}E_a{}^I\bar{E}_{\bar{b}}{}^I$, and ${E}_I{}^a$
its inverse, we introduce the quantities $U_I\equiv E_I{}^a\,U_a$,
in terms of which the following $(2n+2)\times(2n+2)$ matrix
$\hat{\mathbb{L}}_4=(\hat{\mathbb{L}}_4{}^M{}_N)$ is defined:
\begin{equation}
\hat{\mathbb{L}}_4(z,\bar{z})=\sqrt{2}\,\left({\rm Re}(V),\,{\rm
Re}(U_I),\,-{\rm Im}(V),{\rm Im}(U_I)\right)\,,
\end{equation}
which, by virtue of eq.s (\ref{props}), is symplectic:
$\hat{\mathbb{L}}^T_4\mathbb{C}\hat{\mathbb{L}}_4=\mathbb{C}$. In
terms of this matrix one can construct the symmetric, symplectic,
negative definite matrix $\mathcal{M}_4=(\mathcal{M}_{4\,MN})$
\begin{equation}
\mathcal{M}_4=\mathbb{C}\hat{\mathbb{L}}_4\hat{\mathbb{L}}_4^T\mathbb{C}\,.\label{M4}
\end{equation}
This matrix is related to  $ R_{\Lambda\Sigma}$ and $
I_{\Lambda\Sigma}$ as follows:
\begin{equation}
\mathcal{M}_4=\left(\begin{matrix}I+R I^{-1}R & -R I^{-1}\cr -
I^{-1}R & I^{-1}\end{matrix}\right)\,.
\end{equation}
For symmetric homogeneous special K\"ahler manifolds, the symplectic
bundle defines an embedding of the isometry group $G_4$ into
$Sp(2n+2,\mathbb{R})$, realized by the symplectic representation
${\bf R}$ by which $G_4$ acts on the symplectic section $V$ as part
of the structure group. The global symmetries of the $D=4$ model
(duality symmetries) consist in the simultaneous action of $G_4$ on
the scalar fields and on the symplectic vector of the electric field
strengths and their magnetic duals in the representation ${\bf
R}$.\par
\paragraph{Special coordinates.}
 One can always, by suitably fixing the
symplectic gauge, choose a section $\Omega(z^a)$ in which
$X^\Lambda(z^a)$ can be regarded as projective coordinates for the
manifold. In particular, in a local patch in which $X^0\neq 0$,
$X^a/X^0$ are independent functions of $z^a$ and can be thus used as
coordinates, known as \emph{special coordinates}. In the special
coordinate patch we can then choose $z^a\equiv X^a/X^0$ in the first
place. Moreover the lower components can be expressed in terms of a
\emph{prepotential} $F(X)$: $F_\Lambda=\frac{\partial F}{\partial
X^\Lambda}$, $F(X)$ being a homogeneous function of degree 2 in the
$X^\Lambda$. Of particular relevance are the \emph{cubic} models in
which: $F(X)=\frac{1}{6}\,d_{abc} X^a X^b X^c/X^0$. For these models
one defines
$$\mathcal{F}(z^a)=F(X)/(X^0)^2=\frac{1}{6}\,d_{abc} z^a z^b
z^c\,,$$ in terms of which the holomorphic section has the simple
form:
\begin{equation}
\Omega(z^a)=\left(\begin{matrix}1\cr z^a\cr -\mathcal{F}(z^a) \cr
\frac{\partial }{\partial z^a}\mathcal{F}\end{matrix}\right)\,,
\end{equation}
Writing the complex scalars in terms of their real and imaginary
parts, $z^a=\alpha^a-i\,\lambda^a$,  the K\"ahler potential and the
hermitian metric $g_{a\bar{b}}$ read:
\begin{align}
e^{-\mathcal{K}}&=-\frac{4}{3}\,d_{abc} \lambda^a \lambda^b
\lambda^c>0\,\,\,,\,\,\,\,
g_{a\bar{b}}=\partial_a\partial_{\bar{b}}\mathcal{K}=-\frac{3}{2}\,\left(d_{ab}-\frac{3}{2}\frac{d_a
d_b}{d}\right)>0\,,
\end{align}
where $d\equiv d_{abc} \lambda^a \lambda^b \lambda^c$ should be a
negative number, and we have used the short hand notation:
$d_a\equiv d_{abc} \lambda^b \lambda^c$, $d_{ab}\equiv d_{abc}
\lambda^c$.\par
 Cubic models originate from dimensional reduction of five
dimensional supergravities. The real scalars $\alpha^a={\rm Re}(z^a)$ are
the internal components of the five-dimensional vectors, while
$\lambda^a=-{\rm Im}(z^a)$ are functions of the scalars in the
five-dimensional vector multiplets and the radial modulus of the
compact fifth dimension. This defines a relation of inclusion of the
scalar manifold in five-dimensions $\mathcal{M}_{D=5}$ spanned by
the vector-multiplet scalars and the special K\"ahler manifold
$\mathcal{M}_{SK}$ in $D=4$ known as \emph{r-map}
\cite{deWit:1991nm}. More specifically $\mathcal{M}_{D=5}$ is
geometrically characterized as a \emph{very special real manifold}
\cite{Gunaydin:1983bi,deWit:1992cr} of real dimensions $n-1$,  and
the r-map is a correspondence between this manifold and the special
K\"ahler one, of complex dimension $n$, originating from reduction
over a circle. The tensor $d_{abc}$ characterizes the very special
geometry of $\mathcal{M}_{D=5}$ and, for symmetric manifolds
$\mathcal{M}_{D=5}=G_5/H_5$, it is invariant with respect to the
isometry group $G_5$. In this case the special K\"ahler manifold
$\mathcal{M}_{SK}$, image to $\mathcal{M}_{D=5}$ through the r-map,
is symmetric as well, namely it as the form
$\mathcal{M}_{SK}=G_4/H_4$, with $G_5\subset G_4$. If we further
compactify the four dimensional theory with only vector multiplets
to three-dimensions, and we dualize vectors into scalars,  we end up
with a sigma model in which the target space is a \emph{quaternionic
K\"ahler} manifold $\mathcal{M}_{QK}$ or a \emph{para-quaternionic
K\"ahler} manifold $\mathcal{M}_{QK}^*$, depending on whether the
internal circle is space-like or time-like, respectively. The
inclusion relation between  $\mathcal{M}_{SK}$ and
$\mathcal{M}_{QK}$ \cite{Bagger:1983tt} ($\mathcal{M}_{QK}^*$) is
called \emph{c-map} \cite{Cecotti:1988qn,Ferrara:1989ik}
(\emph{c*-map}). The property of the  manifold in $D=4$ of being
homogeneous or homogeneous-symmetric is preserved by both the c- and
the c*-maps. Thus if we consider a symmetric special K\"ahler
manifold of the form $\mathcal{M}_{SK}=G_4/H_4$, its image through
the c-map is a manifold of the form $\mathcal{M}_{QK}=G/H$, $H$
being the maximal compact subgroup of the isometry group $G$, and
through the c*-map will have the form  $\mathcal{M}_{QK}^*=G/H^*$,
$H^*$ still being maximal in $G$, though no longer compact (in fact
it is a different real form of the complexification of $H$). A
common feature of $\mathcal{N}=2$ four dimensional symmetric
supergravities is that, upon time-like dimensional reduction to
$D=3$, the isotropy group $H^*$ has the general form: $H^*={\rm
SL}(2,\mathbb{R})\times G'_4$, where the prime in $G'_4$ is used to
distinguish it from the four-dimensional duality group $G_4$, the
two, though being the same Lie group, are distinct inside $G$.
Stationary solutions in four-dimensions can be described as
solutions of the $D=3$ sigma-model obtained through time-reduction
\cite{Breitenlohner:1987dg}, see also Sect. \ref{sbhg}.
\par We can always represent \emph{locally} the manifold
$\mathcal{M}_{QK}^*$ as follows:
\begin{equation}
\mathcal{M}_{QK}^*= \left[{\rm O}(1,1)\times
\mathcal{M}_{SK}\right]\ltimes  \exp(Heis)\,,\label{heisdec}
\end{equation}
where ${\rm O}(1,1)$ is parametrized by the radial modulus $e^U$ of
the internal timelike circle, the corresponding generator being
denoted by $T_0$. $Heis$ denotes a $(2n+3)$-dimensional Heisenberg
algebra \cite{Ferrara:1989ik} parametrized by the $2n+2$ scalar
fields $\mathcal{Z}^M$ originating from the four dimensional vectors
(their time components and the scalars dual to their
three-dimensional descendants) and the scalar $a$ dual to the Kaluza
Klein vector. If $T_M,\,T_\bullet$ denote the corresponding
generators, the following characteristic commutation relations hold:
\begin{equation}
[T_0,\,T_M]=\frac{1}{2}\,T_M\,\,;\,\,\,[T_0,\,T_\bullet]=T_\bullet\,\,;\,\,\,[T_M\,T_N]=\mathbb{C}_{MN}\,T_\bullet\,,\label{relc1}
\end{equation}
all other commutators being zero. If $\mathcal{M}_{SK}$ normal
homogeneous, see below,  denoting by $T_r$ the generators of
solvable Lie group of isometries acting transitively on the manifold
we have:
\begin{equation}
[T_0,T_r]=[T_\bullet,T_r]=0\,\,;\,\,\,[T_r,T_M]=T_r{}^N{}_M\,T_N\,\,;\,\,\,[T_r,T_s]=-
T_{rs}{}^{s'} T_{s'}\,,\label{relc2}
\end{equation}
$T_r{}^N{}_M$ representing the symplectic representation of $T_r$ on
contravariant symplectic vectors.\par
 While the corresponding representation
(\ref{heisdec}) is global for the manifold $\mathcal{M}_{QK}$, image
through the c-map, it is only local for $\mathcal{M}_{QK}^*$ and
defines the \emph{physical patch} of the manifold, spanned by the
physical scalar fields $U,a,\mathcal{Z}^M,z^a,\bar{z}^a$. At the
boundary of this patch $e^{-U}$ vanishes, in general signalling a singularity
in the four-dimensional stationary space-time metric. We can
therefore safely restrict ourselves to this parch when considering
non-singular four-dimensional solutions.
\par
 The special coordinates $z^a$, in light of their
five-dimensional origin, can be characterized as transforming in a
linear representation of the subgroup $G_5$ of $G_4$. This feature
is useful in order to identify the $z^a$ within a parametrization of
the manifold $\mathcal{M}_{QK}$ or $\mathcal{M}_{QK}^*$.\par
  In
the problem under consideration, the four dimensional $N=2$ model
contains $n=6$ vector multiplets and no hypermultiplets. We have the
following inclusion relations:
\begin{equation}
\mathcal{M}_{D=5}=\frac{{\rm SL}(3,\mathbb{R})}{{\rm
SO}(3)}\stackrel{r-map}{\longrightarrow}\mathcal{M}_{SK}=\frac{{\rm
Sp}(6,\mathbb{R})}{{\rm
U}(3)}\stackrel{c*-map}{\longrightarrow}\frac{{\rm F}_{4(4)}}{{\rm
SL}(2,\mathbb{R})\times {\rm Sp}'(6,\mathbb{R})}\,.
\end{equation}
In this case the global symmetry group of the $D=3$ sigma-model, namely the isometry group of the corresponding target space, is $G={\rm F}_{4(4)}$,  $H^*={\rm SL}(2,\mathbb{R})\times G_4'={\rm SL}(2,\mathbb{R})\times {\rm Sp}'(6,\mathbb{R})$ and the  maximal compact subgroup of $G$ is $H={\rm SU}(2)\times {\rm USp}(6)$.
The complex dimension of the  scalar manifold spanned by the $D=4$ vector
multiplets' scalars is $n=6$, $G_4={\rm Sp}(6,\mathbb{R})$ and
$H_4={\rm U}(3)$. The representation ${\bf R}$ by which $G_4$ is
embedded in the structure group ${\rm Sp}(14,\mathbb{R})$ is the ${\bf
14}'$ of ${\rm Sp}(6,\mathbb{R})$.
 The special coordinates $z^a$, $a=1,\dots, 6$, transform in the
${\bf 6}$ of $G_5={\rm SL}(3,\mathbb{R})$ and thus can be identified
with the six independent entries of a complex symmetric matrix
$z^a\equiv z^{i,j}=z^{j,i}$, $i,j=1,2,3$. The cubic prepotential
$\mathcal{F}(z^{ij})$, being ${\rm SL}(3,\mathbb{R})$-invariant, can
only have the following form:
\begin{equation}
\mathcal{F}(z^{i,j})=\epsilon_{i_1 j_1 k_1}\epsilon_{i_2 j_2
k_2}\,z^{i_1,i_2}z^{j_1,j_2}z^{k_1,k_2}\,.
\end{equation}
If we make the identification
$(z^a)=(z^{1,1},z^{1,2},z^{1,3},z^{2,2},z^{2,3},z^{3,3})$, then the
scalars corresponding to the diagonal entries parametrize a
characteristic submanifold $\mathcal{M}^{(STU)}$ of the special K\"ahler manifold:
\begin{equation}
z^1=z^{1,1}=s=a_1-i\,e^{\varphi_1}\,\,;\,\,\,z^4=z^{2,2}=t=a_2-i\,e^{\varphi_2}\,\,;\,\,\,z^6=z^{3,3}=u
=a_3-i\,e^{\varphi_3}\,,\label{stupara}
\end{equation}
where $\varphi_i$ are the three dilatonic scalars parametrizing the
three dimensional Cartan subalgebra in the coset and we have set
$a_1=\alpha^1,\,a_2=\alpha^4,\,a_3=\alpha^6$. This submanifold has the form:
\begin{equation}
\mathcal{M}^{(STU)}=\left(\frac{{\rm SL}(2,\mathbb{R})}{{\rm SO}(2)}\right)^3\,,
\end{equation}
and defines the STU truncation of the model, describing an
$\mathcal{N}=2$ supergravity coupled to $3$ vector multiplets.
\par
 We can then
describe the special coordinates in terms of the $2n$ real scalar
fields $(\phi^r)=(\varphi_i, \alpha^a, \lambda^\ell)$, where
$\ell=2,3,5$ and the corresponding scalars $z^\ell$ are the off
diagonal entries $z^{i,j}$, $i\neq j$. In terms of $z^a$ and
$\phi^r$, the prepotential and the K\"ahler potential, respectively,
read:
\begin{align}
\mathcal{F}(z^a)&=\frac{1}{6}\,d_{abc}\,z^a z^b z^c={z^1}{z^4}{z^6}+
2{z^2}{z^3}{z^5}- ({z^3})^2\,{z^4}   - ({z^2})^2\,{z^6} -
({z^5})^2{z^1}\,,\nonumber\\
\frac{e^{-\mathcal{K}}}{8}&=-\frac{1}{6}\,d_{abc}\,\lambda^a
\lambda^b \lambda^c=e^{\varphi_1+\varphi_2+\varphi_3}+2\,
\lambda^2\lambda^3\lambda^5- e^{\varphi_1}\,(\lambda^5)^2-
e^{\varphi_2}\,(\lambda^3)^2- e^{\varphi_3}\,(\lambda^2)^2\,.
\end{align}
Positive definiteness of $g_{a\bar{b}}$ implies, besides
$\lambda^1,\lambda^4,\lambda^6>0$, which is  consistent with our
position (\ref{stupara}), also
$e^{\varphi_2+\varphi_3}-(\lambda^5)^2>0$,
$e^{\varphi_1+\varphi_3}-(\lambda^3)^2>0$,$e^{\varphi_1+\varphi_1}-(\lambda^2)^2>0$.
 We identify the origin $O$ of
the manifold with the point
$\varphi_i=a_i=\alpha^\ell=\lambda^\ell=0$, and construct the coset
representative $\mathbb{L}_4(\phi^r)=(\mathbb{L}_4(\phi^r)^M{}_N)$
as follows:
\begin{equation}
\mathbb{L}_4(\phi^r)=\hat{\mathbb{L}}_4(\phi^r)\,\hat{\mathbb{L}}_4(O)^{-1}\,\,;\,\,\,\,\mathbb{L}_4(O)={\bf
1}\,.
\end{equation}
The construction of this matrix applies to the most general
symmetric homogeneous special K\"ahler manifold. The symplectic
matrix $\mathbb{L}_4(\phi^r)$ is continuously connected to the
identity matrix. In fact it can be verified that $\mathbb{L}_4$ is
an element of the  \emph{solvable} subgroup $\mathcal{S}_4$ of the
isometry group $G_4$ which acts transitively on the
manifold\footnote{Such solvable group of isometries with a simple
and transitive action on the manifold exists in all the homogeneous
special K\"ahler manifolds which are relevant to supergravity. The
existence of this group defines the so called \emph{normal
homogeneous} manifolds, which were classified in
\cite{alek,deWit:1991nm,cortes}. Scalar fields arising from the
dimensional reduction of higher dimensional string excitations are
parameters of this solvable group and define the (global)
\emph{solvable parametrization}
\cite{Andrianopoli:1996bq,Andrianopoli:1996zg} of the manifold. }.
One can also verify that $\mathbb{L}_4(z,\bar{z})$ represents the
symplectic transformation which maps the symplectic section $V$
computed in $O$ into the one evaluated at a generic point:
$$
\mathbb{L}_4(z,\bar{z}) V(O)=V(z,\bar{z})\,.
$$
The solvable group of isometries, which $\mathbb{L}_4$ belongs to,
for symmetric manifolds $\mathcal{M}_{SK}=G_4/H_4$ is defined by the
Iwasawa decomposition of the semisimple group $G_4$ with respect to
its maximal compact subgroup $H_4$. If we write
$\mathcal{S}_4=\exp(Solv_4)$, where $Solv_4$ is a solvable Lie
algebra, its parameters must be in relation with the scalars
$\phi^r$. This relation is readily computed for the model under
consideration. First define the generators $T_r=(T_r)^M{}_N$ as
follows:
\begin{equation}
T_r=\left.\frac{\partial \mathbb{L}_4}{\partial
\phi^r}\right\vert_{\phi^r\equiv 0}\,.
\end{equation}
One can verify that they close a solvable algebra $Solv_4$ defined
by the Iwasawa decomposition of $\mathfrak{g}_4=\mathfrak{sp}(6)$
with respect to its maximal compact subalgebra
$\mathfrak{H}_4=\mathfrak{u}(3)$. This construction is general and
applies to any symmetric homogeneous special K\"ahler manifold.
$Solv_4$ is the Borel subalgebra of $\mathfrak{sp}(6)$ and is
spanned by the three diagonal  Cartan generators ${\bf h}_i$,
$i=1,2,3$, and by the $9$ shift generators $E_\beta$ corresponding
to the positive roots $\beta$. The latter can be split into ${\bf
a}_k$,  $k=1,\dots, 6$, ${\bf b}_\ell$, $\ell=2,3,5$. The generators
$T_r=\{{\bf h}_i,{\bf a}_a,\,{\bf b}_\ell\}$ are defined as follows:
\begin{equation}
{\bf h}_i=\left.\frac{\partial \mathbb{L}_4}{\partial
\varphi_i}\right\vert_{\phi^r\equiv 0}\,\,;\,\,\,{\bf
a}_a=\left.\frac{\partial \mathbb{L}_4}{\partial
\alpha^a}\right\vert_{\phi^r\equiv 0}\,\,;\,\,\,{\bf
b}_\ell=\left.\frac{\partial \mathbb{L}_4}{\partial
\lambda^\ell}\right\vert_{\phi^r\equiv 0}\,.
\end{equation}
The coset representative $\mathbb{L}_4$ can be constructed as an
element of the solvable group $\exp(Solv_4)$ through the following
exponential map:
\begin{align}
\mathbb{L}_4&=\exp(\sum_{a=1}^6 \alpha^a\ {\bf
a}_a)\exp(\sum_{\ell=2,3,5} f^\ell(\lambda,\varphi)\, {\bf
b}_\ell)\exp(\sum_{i=1}^3
\Phi^i(\lambda,\varphi)\,{\bf h}_i)\,,\nonumber\\
f^2(\lambda,\varphi)&=\frac{e^{\varphi_3} \lambda^2-\lambda^3
\lambda^5}{D}\,\,;\,\,\,f^3(\lambda,\varphi)=\frac{e^{\varphi_2}
\lambda^3-\lambda^2
\lambda^5}{2\,D}+\frac{e^{-\varphi_3}\,\lambda^3}{2}\,\,;\,\,\,f^5(\lambda,\varphi)=
e^{-\varphi_3}\,\lambda^5\,,\nonumber\\
\Phi^1(\lambda,\varphi)&=\log\left(\frac{e^{-\mathcal{K}}}{8\,D}\right)\,\,;
\,\,\,\Phi^2(\lambda,\varphi)=\varphi_2+
\log(e^{-\varphi_2-\varphi_3}\,D)\,\,;\,\,\,\Phi^3(\lambda,\varphi)=\varphi_3\,,\label{scsolv}
\end{align}
where $D\equiv e^{\varphi_2+\varphi_3}-(\lambda^5)^2>0$. Eq.s
(\ref{scsolv}) define, for our specific model, the precise relation
between special coordinates and the solvable parametrization.\par
Once we have the solvable
generators $T_r=\{{\bf h}_i,\,E_\beta\}$ in the symplectic representation ${\bf R}$ , the full Lie algebra $\mathfrak{g}_4$ generating the
 group $G_4$ in the same representation, is simply obtained as
follows: \begin{equation} \mathfrak{sp}(6)={\rm Span}({\bf
h}_i,\,E_\beta,\,E_{-\beta})\,,
\end{equation}
where  $E_{-\beta}=\eta_4\, E_\beta^T\,\eta_4^{-1}$, and
\begin{equation}
\eta_4\equiv\hat{\mathbb{L}}_4(O)\hat{\mathbb{L}}_4(O)^T={\rm
diag}(1,1,\frac{1}{2},\frac{1}{2},1,\frac{1}{2},1,1,1,2,2,1,2,1)\,.
\end{equation}
Let us give the precise correspondence between the generators $h_i,{\bf a}_a,\,{\bf b}_\ell $ in terms of $\mathfrak{sp}(6)$-roots $\pm\beta$. If $\epsilon_i$, $i=1,2,3$, is an orthonormal basis of the root space, so that $\beta=\beta^i \,\epsilon_i$, and $H_i$ the corresponding generators in the Cartan subalgebra, so that $\beta(H_i)=\beta^i$, the basis $\{H_i,\,E_\beta,E_{-\beta}\}$ of the algebra is defined by the usual commutation relations:
\begin{equation}
 [H_i,E_{\pm\beta}]=\pm \beta^i\,E_{\pm\beta}\,\,;\,\,\,[E_{\beta},\,E_{-\beta}]=\beta^i\,H_i\,.\label{normchev}
\end{equation}
The Cartan generators $H_i$ are related to $h_i$ as follows:
\begin{equation}
h_i=\frac{1}{2}\,H_i\,.
\end{equation}
The relation between $E_\beta$ and ${\bf a}_a,\,{\bf b}_\ell$ is summarized in the Table below.
\begin{center}
\begin{tabular}{|l|c|c||c|}
  \hline
   $E_\beta$ &$\beta$ & $\beta^i$& solvable generator \\
 \hline
$E_{\beta_1}$ & $\beta_1$ &$(1,-1,0)$ & ${\bf b}_2$\\
$E_{\beta_2}$ & $\beta_2$ &$(0,1,-1)$& ${\bf b}_5$\\
$E_{\beta_3}$ & $\beta_3$ &$(0,0,2)$& $\sqrt{2}\,{\bf a}_6$\\
$E_{\beta_1+\beta_2}$ & $\beta_1+\beta_2$ &$(1,0,-1)$& ${\bf b}_3$\\
$E_{\beta_2+\beta_3}$ & $\beta_2+\beta_3$ &$(0,1,1)$& ${\bf a}_5$\\
$E_{\beta_1+\beta_2+\beta_3}$ & $\beta_1+\beta_2+\beta_3$ &$(1,0,1)$& ${\bf a}_3$\\
$E_{2\beta_2+\beta_3}$ & $2\beta_2+\beta_3$ &$(0,2,0)$& $\sqrt{2}\,{\bf a}_4$\\
$E_{\beta_1+2\beta_2+\beta_3}$ & $\beta_1+2\beta_2+\beta_3$ &$(1,1,0)$& ${\bf a}_2$\\
$E_{2\beta_1+2\beta_2+\beta_3}$ & $2\beta_1+2\beta_2+\beta_3$ &$(2,0,0)$& $\sqrt{2}\,{\bf a}_1$\\
\hline
\end{tabular}
\end{center}
\vskip 5mm
The solvable generators of the STU truncation are then ${\bf h}_i,\,{\bf a}_1,\,{\bf a}_4,\,{\bf a}_6$.\par
In terms of $\mathbb{L}_4(\phi^r)$ matrix $\mathcal{M}_4$ reads:
$\mathcal{M}_4=\mathbb{C}\mathbb{L}_4(\phi^r)\,\eta_4\,\mathbb{L}_4(\phi^r)^T\mathbb{C}$,
as it can easily be derived from Eq. (\ref{M4}).\par
 Since, with
respect to ${\rm O}(1,1)\times {\rm SL}(3,\mathbb{R})$ the ${\bf
14}'$ of ${\rm Sp}(6,\mathbb{R})$ branches as:
\begin{equation}
{\bf 14}'\rightarrow {\bf 1}_{-3}+{\bf 6}_{-1}+{\bf
1}_{+3}+\bar{{\bf 6}}_{+1}\,,
\end{equation}
we can split the index $\Lambda$ labeling the vector fields
$A^\Lambda_\mu$ as well as the upper  component of $V$,
consequently:
\begin{equation}
A^\Lambda_\mu=\{A^0_\mu,\,A^a_\mu\}=\{A^0_\mu,\,A^{i,j}_\mu\}\,,
\end{equation}
$A^0_\mu$ being the graviphoton in the ${\bf 1}_{-3}$ and
$A^a_\mu\equiv A^{i,j}_\mu$ the remaining six vectors in the ${\bf
6}_{-1}$. Just as for the scalar fields, the truncation to the STU
model is effected by setting all $A^{i,j}_\mu$, with $i\neq j$, to
zero, or, equivalently, $A^\ell_\mu\rightarrow 0$, $\ell=2,3,5$. The
bosonic content of the STU model then consists, besides of the
metric, of $z^{i,i}$, $A^0_\mu$ and $A^{i,i}_\mu$, $i=1,2,3$, corresponding to $A^a_\mu$, with $a=1,4,6$.
\par The above
analysis is useful for defining a one to one correspondence between
(solvable) coordinates of $\mathcal{M}_{QK}^*$ in the three
dimensional theory, and four dimensional fields, which we shall need
to \emph{oxidize} geodesic solutions on $\mathcal{M}_{QK}^*$ to
$D=4$ static black holes. Indeed we now know how to intrinsically
define the special coordinates $z^a$ as a subset of the $D=3$ fields
$\phi^I$ in a suitable parametriation. To this end we locally
represent $\mathcal{M}_{QK}^*$ in the physical patch as a solvable
metric Lie group $\exp(Solv)$, with:
\begin{equation}
Solv=[\mathfrak{o}(1,1)\oplus Solv_4]\oplus_s Heis\,,
\end{equation}
where $\oplus_s$ denotes a semidirect sum. As usual for symmetric
homogeneous manifolds, $Solv$ is defined by the Iwasawa
decomposition of $\mathfrak{f}_{4(4)}$ with respect to its maximal
compact subalgebra. Then we choose as generators of $Solv$ the
matrices $T_I=\{T_0,\,T_r,\,T_M,\,T_\bullet\}$ satisfying the
general relations (\ref{relc1}),(\ref{relc2}), $T_r$ being the
generators of $Solv_4$,\footnote{We shall describe the generators of
the Lie algebra $\mathfrak{g}$ of $G$ in the fundamental
representation of this group, which is the ${\bf 26}$ of ${\rm
F}_{4(4)}$. With an abuse of notation we use for the $T_r$
generators in $\mathfrak{g}$, the same symbol used for the abstract
generators of $\mathfrak{g}_4$.} and $T_M$ are chosen so that the
adjoint action of $T_r$ on them, described by the $2n$ matrices
$T_{r}{}^M{}_N$, realizes the symplectic representation ${\bf R}$ of
$T_r$ computed above and pertaining to the special coordinate frame.
Let us give the weights $\gamma_M$ associated with the
representation ${\bf R}$ in this basis, defined by:
\begin{equation}
(H_i)^M{}_N=\gamma_M(H_i)\delta^M_N\,\,\,\,\mbox{no summation over M}\,.
\end{equation}
In the table below we list the weights $\gamma_M$ in the orthonormal
basis $(\epsilon_i)$ and give the correspondence of the
corresponding charge entry with $D0,D2,D4,D6$-charges in Type IIA
theory.
\begin{center}
\begin{tabular}{|l|c|c||c|}
  \hline
   $\gamma_M$ &$\gamma^i_M$ & $(p^\Lambda,q_\Lambda)$ & $Dp$-charge\\
 \hline
$\gamma_{1}$ &$(-1,-1,-1)$ &$p^0$ & $D6$\\
$\gamma_{2}$ &$(1,-1,-1)$ &$p^1$ & $D4$\\
$\gamma_{3}$ &$(0,0,-1)$ &$p^2$ & $D4$\\
$\gamma_{4}$ &$(0,-1,0)$ &$p^3$ & $D4$\\
$\gamma_{5}$ &$(-1,1,-1)$ &$p^4$ & $D4$\\
$\gamma_{6}$ &$(-1,0,0)$ &$p^5$ & $D4$\\
$\gamma_{7}$ &$(-1,-1,1)$ &$p^6$ & $D4$\\
$\gamma_{8}$ &$(1,1,1)$ &$q_0$ & $D0$\\
$\gamma_{9}$ &$(-1,1,1)$ &$q_1$ & $D2$\\
$\gamma_{10}$ &$(0,0,1)$ &$q_2$ & $D2$\\
$\gamma_{11}$ &$(0,1,0)$ &$q_3$ & $D2$\\
$\gamma_{12}$ &$(1,-1,1)$ &$q_4$ & $D2$\\
$\gamma_{13}$ &$(1,0,0)$ &$q_5$ & $D2$\\
$\gamma_{14}$ &$(1,1,-1)$ &$q_6$ & $D2$\\
\hline
\end{tabular}
\end{center}
The truncation to the STU model is effected by restricting to the
weights $\gamma_1,\,\gamma_2,\,\gamma_5,\,\gamma_7,\,\gamma_8$,
$\gamma_9,\,\gamma_{12},\,\gamma_{14}$, consistently with our
previous discussion about the vector fields. Upon time-reduction to $D=3$ and dualizations of vectors into scalars, the $STU$ truncation yields the following quaternionic K\"ahler submanifold:
\begin{equation}
\mathcal{M}^{*\,(STU)}_{QK}=\frac{{\rm SO}(4,4)}{{\rm SO}(2,2)\times {\rm SO}(2,2)}\,.\label{STUQK}
\end{equation}
Having characterized the special coordinates in an intrinsic
algebraic way, and  knowing how to embed $Solv_4$ inside $Solv$, we
can construct the corresponding coset representative
$\mathbb{L}_4(\phi^r)$ as an element of $\exp(Solv)$ in the
fundamental representation ${\bf 26}$ of $G={\rm F}_{4(4)}$. The
coset representative $\mathbb{L}(\phi^I)$ of ${\rm F}_{4(4)}/[{\rm
SL}(2,\mathbb{R})\times {\rm Sp}(6,\mathbb{R})]$ in the solvable
parametrization can be defined by the following exponential map:
\begin{equation}
\mathbb{L}(\phi^I)=\exp(-a T_\bullet)\,\exp(\sqrt{2}
\mathcal{Z}^M\,T_M)\,\mathbb{L}_4(\phi^r)\,\exp(2U
T_0)\,.\label{cosetr3}
\end{equation}
We can define the  involutive automorphism $\sigma$ on the algebra
$\mathfrak{g}$ of $G$ which leaves the algebra $\mathfrak{H}^*$
generating $H^*$ invariant. This involution in the fundamental
representation of $G$ has the form $\sigma(M)=-\eta M^T\eta$, $\eta$
being an $H^*$-invariant metric, and induces the (pseudo)-Cartan
decomposition of $\mathfrak{g}$ of the form:
\begin{equation}
\mathfrak{g}=\mathfrak{H}^*\oplus \mathfrak{K}^*\,,\label{pseudoC}
\end{equation}
where $\sigma(\mathfrak{K}^*)=-\mathfrak{K}^*$, and the following
relations hold
\begin{equation}[\mathfrak{H}^*,\mathfrak{H}^*]\subset\mathfrak{H}^*,
\quad [\mathfrak{H}^*,\mathfrak{K}^*]\subset \mathfrak{K}^*,\quad
[\mathfrak{K}^*,\mathfrak{K}^*] \subset
\mathfrak{H}^*.\label{HKrels}\end{equation} We see that $H^*$ has a
linear adjoint action  in the space $\mathfrak{K}^*$ which is thus the
carrier of an $H^*$-representation. As previously pointed out,
$\mathcal{N}=2$ symmetric models, $H^*={\rm SL}(2,\mathbb{R})\times
G'_4$ and its adjoint action on $\mathfrak{K}^*$ realizes the
representation ${\bf (2,R)}$.\par
The decomposition (\ref{pseudoC}) has to be contrasted with the ordinary Cartan decomposition of $\mathfrak{g}$
\begin{equation}
\mathfrak{g}=\mathfrak{H}\oplus \mathfrak{K}\,,\label{Cartan}
\end{equation}
into its maximal compact subalgebra $\mathfrak{H}$ generating $H$ and its orthogonal non-compact complement $\mathfrak{K}$. This decomposition is effected through the Cartan involution $\tau$ of which $\mathfrak{H}$  and $\mathfrak{K}$ represent the eigenspaces with eigenvalues $+1$ and $-1$ respectively. In the real matrix representation in which we shall work, the action of $\tau$ can be implemented as: $\tau(X)=-X^T$.
\par Next we construct
the left invariant one-form and the vielbein
$P^{\mathcal{A}}=P_I{}^{\mathcal{A}} d\phi^I$:
\begin{equation}
\mathbb{L}^{-1}d\mathbb{L}=P^{\mathcal{A}}\,T_{\mathcal{A}}=P^{\mathcal{A}}
\mathbb{K}_{\mathcal{A}}+\Omega_{H^*}\,\,;\,\,\,{\mathcal{A}}=1,\dots,
4n+4\,.\label{li1f}
\end{equation}
where we have introduced the basis $\{\mathbb{K}_\mathcal{A}\}$ of
$\mathfrak{K}^*$ to be defined below in eq. (\ref{Kge}).
Following the prescription of \cite{Chemissany:2010zp}, the
normalization of the $H^*$-invariant metric on the tangent space of
$\mathcal{M}_{QK}^*$ is chosen as follows
\begin{equation}
g_{{\mathcal{A}}{\mathcal{B}}}=\frac{1}{2 {\rm Tr}[T_0^2]}\, {\rm
Tr}[\mathbb{K}_{\mathcal{A}} \mathbb{K}_{\mathcal{B}}]=\frac{1}{6}\, {\rm
Tr}[\mathbb{K}_{\mathcal{A}} \mathbb{K}_{\mathcal{B}}]\,,
\end{equation}
being ${\rm Tr}[T_0^2]=3$ in our model. The metric of the $D=3$
sigma-model has the familiar form:
\begin{align}
ds^2&=P^{\mathcal{A}} P^{\mathcal{B}}
g_{{\mathcal{A}}{\mathcal{B}}}=2 dU^2+2 g_{a\bar{b}} dz^a
d\bar{z}^b+\frac{e^{-4U}}{2} \omega^2+e^{-2U}
d\mathcal{Z}^T\mathcal{M}_4(\phi^r)d\mathcal{Z}\,,\\
\omega&= da+ \mathcal{Z}^T\mathbb{C}d\mathcal{Z}\,.
\end{align}
\subsection{Static Black Holes and Geodesics}\label{sbhg}
We shall now restrict our discussion to static, spherically
symmetric and asymptotically flat black hole solutions. The general
ansatz for the metric has the following form:
\begin{eqnarray}
ds^2&=&-e^{2U}\,dt^2+e^{-2U}\,\left(\frac{c^4}{\sinh^4(c\tau)}\,d\tau^2+\frac{c^2}{\sinh^2(c\tau)}\,d\Omega^2\right)\,,\label{dstau}
\end{eqnarray}
where $U=U(\tau)$ and the coordinate $\tau$ is related to the radial
coordinate $r$ by the following relation:
\begin{eqnarray}
\frac{c^2}{\sinh^2(c\tau)}&=&(r-r_0)^2-c^2=(r-r^-)\,(r-r^+)\,.\label{rtau}
\end{eqnarray}
Here $c^2\equiv 2ST$ is the extremality parameter of the solution,
with $S$ the entropy and $T$ the temperature of the black hole. When
$c$ is non vanishing the black hole has two horizons located at
$r^{\pm}=r_0\pm c$. The outer horizon is located at $r_H=r^+$
corresponding to  $\tau\rightarrow -\infty$. The extremality limit
 at which the two horizons coincide, $r_H=r^+=r^-=r_0$, is $c\rightarrow
 0$. For extremal solutions eq. (\ref{rtau}) reduces to
 $\tau=-1/(r-r_0)$.
 Spherical symmetry also requires the scalar fields in the solution
 to depend only on $\tau$: $\phi^r=\phi^r(\tau)$. The solution is also
characterized by a set of electric and magnetic charges defined as
follows:
\begin{equation}
p^\Lambda = \frac{1}{4\pi}\,\int _{S^2}  F^\Lambda \qquad \qquad
q_\Lambda = \frac{1}{4\pi}\,\int _ {S^2}  G _\Lambda\,,\label{em}
\end{equation}
where $S^2$ is a spatial two-sphere in the space-time geometry of
the dyonic solution (for instance, in Minkowski space-time the
two-sphere at radial infinity $S^2_\infty$). In terms of these
charges the general ansatz for the electric-magnetic field strength
vector ${F}^\Lambda,\,G_\Lambda$ reads:
\begin{align}
\mathbb{F}&=\left(\begin{matrix}F^\Lambda_{\mu\nu}\cr
G_{\Lambda\,\mu\nu}
\end{matrix}\right)\,\frac{dx^\mu\wedge
dx^\nu}{2}=e^{2\,U}\mathbb{C}\cdot\mathcal{M}_4(\phi^r)\cdot\Gamma\,
dt\wedge d\tau+\Gamma\,\sin(\theta)\,d\theta\wedge
d\varphi\,,\nonumber\\
\Gamma&= (\Gamma^M)=\left(\begin{matrix}p^\Lambda \cr
q_\Lambda\end{matrix}\right)=\frac{1}{4\pi}\,\int _{S^2}
\mathbb{F}\,.
\end{align}
In $D=4$ these solutions are described by the following effective action
\begin{eqnarray}
S^{(4)}_{eff}&=&\int\mathcal{L}^{(4)}_{eff}\,d\tau=\int\left(\dot{U}^2+\frac{1}{2}\,G_{rs}(\phi)\,\dot{
\phi}^r\,\dot{
\phi}^s+e^{2\,U}\,V(\phi;\,\Gamma)\right)\,d\tau\,,\label{lag}
\end{eqnarray}
where the upper dot stands for the derivative of the field with
respect to $\tau$ and the effective potential $V(\phi;\,\Gamma)$
reads:
\begin{equation}
V(\phi;\,\Gamma)=-\frac{1}{2}\,\Gamma^T\,\mathcal{M}_4(\phi)\,\Gamma>0\,.
\end{equation}
In the $D=3$ Euclidean theory the effective action reads
\begin{align}
S_{eff}&=\int\mathcal{L}_{eff}\,d\tau\,,\label{Seff}\\
\mathcal{L}_{eff}&=\frac{1}{2}\,g_{IJ}(\phi)\,\dot{\phi}^I\,\dot{\phi}^J=
\dot{U}^2+ g_{a\bar{b}} \dot{z}^a
d\dot{\bar{z}}^b+\frac{e^{-2U}}{2}\,
\dot{\mathcal{Z}}^T\mathcal{M}_4(\phi^r)\dot{\mathcal{Z}}\,.\nonumber
\end{align}
The two effective actions $\mathcal{L}_{eff}$ and
$\mathcal{L}^{(4)}_{eff}$ are related by a Legendre transformation
trading the cyclic variables $\mathcal{Z}^M$ with their conserved
conjugate momenta, which are the quantized charges $\Gamma^M$.\par
Solutions $\phi^I(\tau)$ to the $D=3$ theory are geodesics on the
symmetric homogeneous manifold $\mathcal{M}^*_{QK}$ with
pseudo-Riemannian metric $g_{IJ}(\phi)$. The ``velocity vector'' of
the geodesic can be described by the $\mathfrak{K}^*$-matrix
\begin{equation}
L( \tau)=
\dot{\phi}^I(\tau)\,P_I{}^{\mathcal{A}}(\phi(\tau))\,\mathbb{K}_\mathcal{A}=\Delta^{\mathcal{A}}(\tau)\,\mathbb{K}_\mathcal{A}\,,\label{laxdef}
\end{equation}
in terms of which the effective action reads:
\begin{equation}
\mathcal{L}_{eff}=\frac{C}{2}\,{\rm
Tr}(L^2)=\frac{1}{2}\,g_{{\mathcal{A}}{\mathcal{B}}}\,\Delta^{\mathcal{A}}\,\Delta^{\mathcal{B}}\,,
\end{equation}
where $C\equiv 1/(2 {\rm Tr}(T_0^2))$ depends on the chosen
representation for the $\mathfrak{g}$-generators. For our model,
having chosen to represent all matrices in the fundamental ${\bf
26}$ representation of ${\rm F}_{4(4)}$, $C=1/6$.\par The geodesic
equations derived from (\ref{Seff}) can be cast into the following
equivalent forms:
\begin{align}
\mathcal{M}^{-1}\dot{\mathcal{M}}&=2\,Q^T=\mbox{const.}\,,\label{geosi1}\\
\dot{L}-[W,L]&=0\,,\label{geosi2}
\end{align}
where $\mathcal{M}(\tau)\equiv
\mathbb{L}(\phi(\tau))\,\eta\,\mathbb{L}(\phi(\tau))^T$, $Q$
 is the $\mathfrak{g}$-matrix of the Noether charges of the solution
 and $W(\tau)$ is a compensator matrix, defined as
 $W=\dot{\phi}^I\,\Omega_{H^*,\,I}$, $\Omega_{H^*}$ being the $\mathfrak{H}^*$-valued 1-form introduced in (\ref{li1f}).
  Eq. (\ref{geosi2}) is a
 Lax-pair equation in the Lax matrix $L(\tau)$, whose relation to
 the Noether charge matrix $Q$ is
 \begin{equation}
Q=\mathbb{L}(\phi)\,L\,\mathbb{L}(\phi)^{-1}\,.\label{noe}
 \end{equation}
 Using the notation of \cite{Chemissany:2010zp} the ADM mass, the scalar charges, the quantized charges and the NUT charge are computed as traces of $Q$ with the solvable generators $T_0,\,T_r,\,T_M,\,T_\bullet$, respectively.
 In particular the electric-magnetic charges $(\Gamma^M)=(p^\Lambda,\,q_\Lambda)$ can be evaluated as follows:
 \begin{equation}
 \Gamma^M=\sqrt{2}\,C\,\mathbb{C}^{MN}{\rm Tr}(Q\,T_N)\,,\label{GQ}
 \end{equation}
 while the ADM mass reads:
 \begin{equation}
 M_{ADM}=C\,{\rm Tr}(Q\,T_0)\,.\label{MADM}
 \end{equation}
Let us consider the subspace $\mathfrak{K}^{*(R)}$ of
$\mathfrak{K}^*$ spanned by the compact generators $K_A=(T_A+\eta
T_A^T\eta)/2= (T_A- T_A^T)/2$, $A=\,\dots,\, 2n+2$. The components $\mathbb{Y}=(Y^A)$ of the Lax matrix within this
subspace are expressed as follows:
\begin{align}
\left.L\right\vert_{\mathfrak{K}^{*(R)}}=Y^A\,K_A\,\,;\,\,\,\,
\mathbb{Y}(\phi^I,\Gamma)=(Y^A)=\sqrt{2}\,e^U\,\hat{\mathbb{L}}(O)\,{\bf
Z}(\phi^I,\Gamma)\,,
\end{align}
where ${\bf Z}=({\bf Z}^A)$ is the symplectic vector  defined as:
\begin{equation}
{\bf Z}(\phi^I,\Gamma)=\hat{\mathbb{L}}(\phi^r)^T\mathbb{C}\,\tilde{\Gamma}\,\,\,\,;\,\,\,\,\tilde{\Gamma}=\Gamma-\mathfrak{n}\,\mathcal{Z}\,,
\end{equation}
where $\mathfrak{n}$ is the NUT charge, which we shall consider to be zero on our solutions. If $\mathfrak{n}=0$, ${\bf Z}$ is the symplectic vector consisting  of the real and imaginary parts
of the central charge $Z$ and the matter charges $Z_I$, defined as:
\begin{equation}
Z=V^T\,\mathbb{C}\,\Gamma\,\,;\,\,\,Z_I=U_I\,\mathbb{C}\,\Gamma=E_I{}^a\,\nabla_a\,Z\,,\label{ZZI}
\end{equation}
 and depends on $\phi^r$
and $\Gamma$, see \cite{Andrianopoli:2006ub,Andrianopoli:2010bj} for the notation. If a global symmetry transformation $g_4\in G_4$ of the $D=4$ theory is applied to the solution, it will act non-linearly (as an isometry) on the scalars $\phi^r$  and linearly the charge vector $\Gamma$ and $\mathcal{Z}$ through symplectic matrix ${\bf R}[g_4]$ representing $g_4$ in ${\bf R}$, while the other scalars $U,\, a$ will be left unaffected. Using (\ref{coset}) one finds that the central/matter-charge vector transforms only through the compensator $h\in H_4$: $\mathbb{Y}(g_4\star \phi^I,{\bf R}[g_4]\Gamma)={\bf R}[h]\,\mathbb{Y}(\phi^I,\,\Gamma)$.\par
In general the components $Y^A$ transform in a representation ${\bf R}'$ \cite{Bergshoeff:2008be} under the larger group $H_c={\rm U}(1)_E\times H_4={\rm U}(1)_E\times {\rm U}(3)$ which is the maximal compact subgroup of $H^*$, where ${\rm U}(1)_E$ is the maximal compact subgroup of the Ehlers group ${\rm SL}(2,\mathbb{R})_E$. The representation ${\bf R}'$ is
\begin{equation}
{\bf R}'={\bf 1}_{-1}+{\bf 6}_{-1}+{\bf 1}_{+1}+\bar{{\bf 6}}_{+1}\,,
\end{equation}
where the grading refers to ${\rm U}(1)_E$. The space $\mathfrak{K}^{*(R)}$ is in fact the carrier of the representation ${\bf R}'$  with respect to the adjoint action of  $H_c$.

 \paragraph{Global Symmetry and Geodesics}
 A geodesic, solution to eq. (\ref{geosi1}) or, equivalently, eq.
 (\ref{geosi2}), is uniquely determined by its initial conditions
 defined by the values $\phi_0^I=\phi^I(\tau=0)$ of the scalar fields and of the Lax matrix
 $L_0=L(\tau=0)$ at radial infinity $\tau=0$. Let us denote by $\phi^I[\tau;\,\phi_0,\,L_0]$ the unique geodesic
 with initial conditions $(\phi_0^I,\,L_0)$. The global symmetry
 group of the Euclidean $D=3$ theory is the isometry group $G$. For
 a generic isometry $g\in G$, let us denote by $g\star \phi^I$ the
 transformed scalars, non-linear functions of the original ones
 $\phi^I$, defined by
 \begin{equation}
g\cdot \mathbb{L}(\phi^I)=\mathbb{L}(g\star\phi^I)\cdot
h(\phi^I,g)\,,\label{coset3}
\end{equation}
where $h(\phi^I,g)$ is a \emph{compensator} in $H^*$. Under the
above transformation, the vielbein matrix
$P=P^{\mathcal{A}}\,\mathbb{K}_\mathcal{A}$ transforms under the
compensator only
\begin{equation}
P(\phi)\rightarrow P(g\star\phi)=h(\phi,g)\star P(\phi)\equiv
h(\phi^I,g)\,P(\phi)\,h(\phi^I,g)^{-1}\,.
\end{equation}
 Given a geodesic
$\phi^I[\tau;\,\phi_0,\,L_0]$ and an isometry $g\in G$,
$g\star\phi^I[\tau;\,\phi_0,\,L_0]$ is the unique geodesic with
boundary conditions $(g\star \phi_0^I,\,h(\phi,g)\star L_0)$:
\begin{equation}
g\star\phi^I[\tau;\,\phi_0,\,L_0]=\phi^I[\tau;\,g\star
\phi_0^I,\,h(\phi_0,g)\star L_0]\,.\label{Gactiongeo}
\end{equation}
Thus in order to classify geodesic solutions with respect to the
action of the  global symmetry group $G$, which is the main purpose
of the present work, we can restrict to the action of $G$ on the
initial conditions $(\phi_0,\,L_0)$. Notice that the action of
transformations in $G/H^*$ is transitive on the manifold. This means
that we can always map, by means of a suitable $G/H^*$
transformation, any geodesic into one originating in the origin $O$:
$\phi^I_0\equiv 0$. We are left with the action of the stability
group $H^*$ of $O$ on the solution which only affects the initial
velocity vector on the tangent space $T_O\mathcal{M}_{QK}^*$:
\begin{equation}
h\in H^*\,:\,\,\,\phi^I[\tau;\,O,\,L_0]\longrightarrow
\phi^I[\tau;\,O,\,h^{-1}\,L_0\,h]\,.
\end{equation}
Thus we have reduced the problem of classifying the geodesics with
respect to the action of $G$ to that of classifying the orbits of
the initial velocity vector $L_0$ with respect to the adjoint action
of $H^*$.\par In the $D=3$ theory there are $n_v$ fermion fields
$\lambda^A$ transforming under supersymmetry as follows:
\begin{equation}
\delta_\epsilon \lambda^A= \Delta^{a
A}(\tau)\,\epsilon_a\,,\label{susytrans}
\end{equation}
where $\epsilon_a$ is a doublet of supersymmetry parameters and  we
have written the tangent space index $\mathcal{A}$, labeling the Lax
components $\Delta^{\mathcal{A}}$, as a couple of indices
$\mathcal{A}=(a,A)$, in which $a=1,2$, and $A=1,\dots,\,2 n_v$
labels the representation ${\bf 2}$ and ${\bf R}={\bf 14}'$ of the subgroups
${\rm SL}(2,\mathbb{R})$ and $G'_4={\rm Sp}(6,\mathbb{R})$ of
$H^*={\rm SL}(2,\mathbb{R})\times G'_4$. BPS solutions are
characterized by the property of preserving a fraction of
supersymmetry, that is there exists a spinor $\epsilon_a$ satisfying
the Killing spinor equation: $\delta_\epsilon \lambda^A=0$, or,
equivalently, that the rectangular matrix $\Delta^{a,A}$ have a
null-eigenvector $\epsilon_a$: $\Delta^{a,A}\,\epsilon_a=0$. It is
straightforward to prove that this is the case if and only if
$\Delta^{a,A}$ factorizes as follows:
$\Delta^{a,A}=\epsilon^a\,\Delta^A$, \cite{Gunaydin:2007bg}, where
$\epsilon^a=\epsilon^{ab}\,\epsilon_b$. This property is not
affected by the action of $H^*$ and,  since  the action of $G$ on a
geodesic amounts to the action of an $H^*$-compensator on $L_0$
(i.e.on  $\Delta^{a\,A}(\tau=0)$), according to eq.
(\ref{Gactiongeo}), we conclude that the geodesics corresponding to
BPS black holes sit in a same $G$-orbit. Note that the existence of
a residual supersymmetry is clearly independent on $\tau$, since the
evolution in $\tau$ of the Lax matrix, solution  to (\ref{geosi2}),
is governed by a suitable $H^*$-transformation $\mathcal{O}(\tau)$:
$L(\tau)=\mathcal{O}(\tau)^{-1}\,L_0\,\mathcal{O}(\tau)$
\cite{Chemissany:2009hq,Chemissany:2009af}.\par
Having set the point at radial infinity to coincide with  the origin $O$ ($\phi^I\equiv 0$) of the manifold, the components $\mathbb{Y}_0=(Y_0^A)$ along $\mathfrak{K}^{*(R)}$ of the Lax matrix $L_0$ at $\tau=0$,  coincides, modulo basis redefinition, with (the real and imaginary parts of) the central and matter charges which, in turn, are expressed solely as combinations of the quantized ones $\Gamma$, being $\phi^r\equiv 0$. With an abuse of notation we shall sometimes use the same symbol ${\bf R}$ for the representation of the electric and magnetic charges under $G_4$ and for the representation ${\bf R}'$ of $H_c$.
\paragraph{Regularity.}
Not all Lax matrices $L$ generate  geodesics corresponding to
regular $D=4$ solutions, or their \emph{small} limits. A necessary
condition for regularity was given in \cite{Bossard:2009at} in terms
of the following matrix equation:
\begin{equation}
L(\tau)^3=c^2\,L(\tau)\,\,\Leftrightarrow\,\,\,\,L_0^3=c^2\,L_0\,,\label{regcond}
\end{equation}
for $L_0$ evaluated in the fundamental representation of the algebra
$\mathfrak{g}$ (for all models except the one with
$\mathfrak{g}=\mathfrak{e}_8$). The non-extremality parameter can
itself be expressed in terms of $L_0$:
\begin{equation}
c^2=\frac{C}{2}\,{\rm Tr}(L_0^2)\,.
\end{equation}
For extremal solutions ($c=0$), the regularity condition requires
$L_0$ (or, equivalently, $Q$) to be a nilpotent matrix, with degree
of nilpotency not exceeding 3:
\begin{equation}
L_0^3=0\,,\label{regcondnil}
\end{equation}
 This condition was first proven in
\cite{Gaiotto:2007ag}.

\subsection{Group Theoretic Structure}
In this subsection we review some algebraic and geometric properties of the $F_{4(4)}$-model.  $F_{4(4)}$ is an exceptional, maximally split group whose Lie algebra $\mathfrak{f}_{4(4)}$ is generated by \footnote{We use for this basis the same normalization used for the $\mathfrak{sp}(6)$ generators in (\ref{normchev}).}
\begin{equation} \{H_{i},E_{\alpha},E_{-\alpha}\},\qquad i=1,\cdots4;\,\, \alpha=1,\cdots, 24.\end{equation}
The complex Lie algebra $\mathfrak{f}_4^\mathbb{C}$ has rank four and it is defined by the $4\times 4$
Cartan matrix encoded in the following Dynkin diagram
\vspace{10mm}
\begin{center}
\begin{picture}(110,30)
\put (-60,20){$\mathfrak{g}_2$}
\put (10,25){\circle {10}}
\put (6,40){$\alpha_{1}$}
\put (15,25){\line (1,0){20}}
\put (40,25){\circle {10}}
\put (36,40){$\alpha_{2}$}
\put (45,27){\line (1,0){20}}
\put (50,22.3){{$>$}}
\put (45,24){\line (1,0){20}}
\put (71,25){\circle {10}}
\put (67,40){$\alpha_{3}$}
\put (76,25){\line (1,0){20}}
\put (101,25){\circle {10}}
\put (97,40){$\alpha_{4}$}
\put (120,21){$=\quad\quad\left (\begin{array}{cccc}
  2 & -1&0&0\\
  -1 & 2&-2&0\\
0&-1&2&-1\\
0&0&-1&2
\end{array} \right)$}
\end{picture}
\end{center}
The corresponding root space is spanned by
\begin{equation}\Delta_{\textrm{simple}}=\lbrace{\alpha_1=\epsilon_2-\epsilon_3,\quad \alpha_2=\epsilon_3-\epsilon_4,\quad
\alpha_3=\epsilon_4,\quad \alpha_4=\frac{1}{2}(\epsilon_1-\epsilon_2-\epsilon_3-\epsilon_4)
\rbrace}\end{equation}
where the set of roots reads

\begin{equation}\Delta_{F_{4(4)}} = \left\{ \begin{array}{l}
         \pm \epsilon_{i}\\
        \pm\epsilon_{i}\pm \epsilon_{j}\\
\frac{1}{2}(\pm \epsilon_{1}\pm \epsilon_2\pm \epsilon_3 \pm
\epsilon_4)\end{array} \right\},\qquad \textrm{with}\,\, i<
j,\,\,\textrm{and}\,\,i,j=1,2,3,4\,, \end{equation} where
$(\epsilon_i)$ is a basis of four ortho-normal Euclidean vectors.
For the reader's convenience in the Table below we tabulate the
roots of $F_{4(4)}$ in two different bases.
\begin{table}[htbp]
\scalefont{1}
\begin{center}
\begin{tabular}{|l|c||c|}
  \hline
   Root & \begin{picture}(110,30)

\put (10,10){\circle {10}}
\put (6,18){$\alpha_{1}$}
\put (15,10){\line (1,0){20}}
\put (40,10){\circle {10}}
\put (36,18){$\alpha_{2}$}
\put (45,12){\line (1,0){20}}
\put (50,7.3){{$>$}}
\put (45,9){\line (1,0){20}}
\put (71,10){\circle {10}}
\put (67,18){$\alpha_{3}$}
\put (76,10){\line (1,0){20}}
\put (101,10){\circle {10}}
\put (97,18){$\alpha_{4}$}\end{picture} & Orthornomal Basis \\
 \hline
1 & 0\qquad 1\qquad 2\qquad 2 &$\epsilon_1-\epsilon_2$\\
2 & 1\qquad 1\qquad 2\qquad 2 &$\epsilon_1-\epsilon_3$\\
3 & 1\qquad 0\qquad 0\qquad 0 &$\epsilon_2-\epsilon_3$\\
4 & 1\qquad 2\qquad 2\qquad 2 &$\epsilon_1-\epsilon_4$\\
5 & 1\qquad 1\qquad 0\qquad 0 &$\epsilon_2-\epsilon_4$\\
6 & 0\qquad 1\qquad 0\qquad 0 &$\epsilon_3-\epsilon_4$\\
7 & 2\qquad 2\qquad 3\qquad 1 &$\epsilon_1$\\
8 & 1\qquad 1\qquad 1\qquad 0 &$\epsilon_2$\\
9 & 0\qquad 1\qquad 1\qquad 0 &$\epsilon_3$\\
10 & 0\qquad 0\qquad 1\qquad 0 &$\epsilon_4$\\
11 & 1\qquad 2\qquad 4\qquad 2 &$\epsilon_1+\epsilon_4$\\
12 & 1\qquad 1\qquad 2\qquad 0 &$\epsilon_2+\epsilon_4$\\
13 & 0\qquad 1\qquad 2\qquad 0 &$\epsilon_3+\epsilon_4$\\
14 & 1\qquad 2\qquad 2\qquad 0 &$\epsilon_2+\epsilon_3$\\
15 & 1\qquad 3\qquad 4\qquad 2 &$\epsilon_1+\epsilon_3$\\
16 & 2\qquad 3\qquad 4\qquad 2 &$\epsilon_1+\epsilon_2$\\
17 & 0\qquad 1\qquad 2\qquad 1 &$\frac{1}{2}(\epsilon_1-\epsilon_2+\epsilon_3+\epsilon_4)$\\
18 & 0\qquad 1\qquad 1\qquad 1 &$\frac{1}{2}(\epsilon_1-\epsilon_2+\epsilon_3-\epsilon_4)$\\
19 & 0\qquad 0\qquad 1\qquad 1 &$\frac{1}{2}(\epsilon_1-\epsilon_2-\epsilon_3+\epsilon_4)$\\
20 & 0\qquad 0\qquad 0\qquad 1 &$\frac{1}{2}(\epsilon_1-\epsilon_2-\epsilon_3-\epsilon_4)$\\
21 & 1\qquad 1\qquad 2\qquad 1 &$\frac{1}{2}(\epsilon_1+\epsilon_2-\epsilon_3+\epsilon_4)$\\
22 & 1\qquad 1\qquad 1\qquad 1 &$\frac{1}{2}(\epsilon_1+\epsilon_2-\epsilon_3-\epsilon_4)$\\
23 & 1\qquad 2\qquad 3\qquad 1 &$\frac{1}{2}(\epsilon_1+\epsilon_2+\epsilon_3+\epsilon_4)$\\
24 & 1\qquad 2\qquad 2\qquad 1 &$\frac{1}{2}(\epsilon_1+\epsilon_2+\epsilon_3-\epsilon_4)$\\
\hline
\end{tabular}
\caption{$\mathfrak{f}_{4(4)}$-positive roots $\alpha$, each represented by a number running from 1 to 24.}\label{f44posr}
\end{center}
\end{table}
The matrix form of $H_i$ and of the shift generators $E_\alpha$, $\alpha=1,\dots 24$, corresponding to the roots listed in Table \ref{f44posr} is given in Appendix \ref{apptech}.
In four dimensions the electric and magnetic charges together span
an irreducible symplectic representation  $\bold{R}$ of $\Sp(6).$
Upon dimensional reduction on the time direction and dualization of
the vector fields into scalars, the isometry group $F_{4}$ of the
resulting moduli space now contains $\SL_{E}(2,\mathbb{R})\times \Sp(6)
$ with respect to which it is adjoint
representations branches as follows
\begin{eqnarray}\label{branchsp}
{\bf Adj}[F_{4}]&\rightarrow &({\bf Adj}[{\rm SL}(2,\mathbb{R})_E],{\bf 1})\oplus ({\bf 1},{\bf Adj}[\Sp(6)])\oplus {\bf (2,R)}\,,
\end{eqnarray}
\begin{equation}\label{branch}\boldmath{52 \rightarrow (3,1)\oplus(1,21)\oplus (2,14')}\end{equation}

A suitable combination $H_{0}$
\begin{equation} H_{0}=H_{1}+H_{2}=2\, T_0\,,\end{equation}
being parametrized by the radial modulus of the internal circle is
the Cartan generator of the $\SL_{E}(2,\mathbb{R})$ factor (it is
twice the generator $T_0$ introduced in the previous section). The
positive roots $\alpha$ of $F_{4}$ naturally split into
\begin{itemize}
 \item[i)] the $\Sp(6)$ positive roots $\beta=\{1,6,9,10,13,17,18,19,20\},$ such that $\beta(H_{0})=0.$
\item[ii)] the roots $\gamma_{M}=\{3,2,21,22,12,8,5,15,14,24,23,4,7,11\},$ such that $\gamma_{M}(H_{0})=1,$ with $M=1,\cdots,\, 2 n_{v}.$
\item[iii)] the roots $\beta_{0}=\{16\}$ such that $\beta_{0}(H_{0})=2.$
\end{itemize}
The $\SL_{E}(2,\mathbb{R})$ group is generated by $H_{0}, E_{\pm \beta_{0}},$ being $H_{0}=H_{\beta_{0}}.$
Accordingly, the branching
(\ref{branch}) reads
\begin{equation} \boldmath{52\rightarrow 1_{(0)}\oplus 1_{(2)}\oplus 1_{(-2)}\oplus 21_{(0)}\oplus 14'_{(+1)}\oplus 14'_{(-1)}}\end{equation}
which means
\begin{itemize}
 \item
The space $\bold{R_{(+1)}=14'}$ is generated by the nilpotent
generators $T_{M}=\{\epsilon_M\,E_{\gamma_{M}}\}$ ($\epsilon_M\equiv 1$ except $\epsilon_2=\epsilon_4=\epsilon_7=\epsilon_{14}=-1$) being parametrized by the
scalar fields $\mathcal{Z}^{M}$ originating from the $D=4$ vector fields and
the corresponding conserved charges are the electric and magnetic
charges.
\item
The generator $E_{\beta_{0}}=E_{16}=T_\bullet$ is associated with
the axion dual $a$ to the Kaluza-Klein vector and the corresponding
conserved charge is the Taub-NUT charge. Thus the double grading
structure implies that the ${\bf R}_{(+1)}={\bf 14}'$ is  no longer an abelian
subalgebra but, together with $E_{\beta_{0}}$, closes a Heisenberg
algebra
\begin{equation} [T_{M},T_{N}]=\mathbbm{C}_{MN} E_{\beta_{0}}\end{equation}
where $\mathbb{C}_{MN}$ is the symplectic invariant matrix.
\item The generators $T_{r}$ with grading zero with respect to $H_{0},$ i.e.,
\begin{equation} Solv_4=\textrm{span}\{H_{1}-H_{2},H_{3},H_{4}, E_{1},E_{6}, E_{9},E_{10}, E_{13},E_{17},E_{18},E_{19},E_{20}\}
 \end{equation}
are associated with the four-dimensional scalar fields $\phi_{r}$
and the corresponding scalar charges.
\item We see that in $D=3$  the maximal compact subgroup $H_{c}$ of $H^*=\Sp'(6)\times \SL(2,\mathbb{R})$
can be written as $H_{c}=U(1)_{E}
\times U(3)$ where $U(1)_{E}$  factor is generated by $E_{\beta_{0}}-E_{-\beta_{0}}.$
\end{itemize}
The  solvable parametrization is defined by the coset representative
$\mathbb{L}(\phi^I)$ in (\ref{cosetr3}).
% We have defined  $\phi_{r}=(\varphi_{i},
%\chi_{\beta})$ so that
%\begin{equation}\e^{\phi^{r}T_{r}  }=\e^{\chi^{\beta} E_{\beta}}\e^{\varphi^{1}(H_{1}-H_{2})+\varphi^{2}H_{2}
% +\varphi^{3}H_{4}}
%\end{equation}
The matrix $\eta$ defining the decomposition through the involution
$\sigma$ has the following intrinsic expression in terms of $H_{0},$
\begin{equation}\label{T}\eta=\e^{2 T},\qquad T= H_{0}\ln(i)\end{equation}
yielding
\begin{equation}\eta=\textrm{diag}(-1,-1,-1,-1,1,-1,1,-1,1,1,1,1,1,1,1,1,1,1,-1,1-1,1,-1,-1,-1,-1).\end{equation}
A geodesic on the manifold $G/H^*$ is parametrically described by
the functions $\phi^I(\tau)$, $\tau$ being the affine parameter
related to the radial variable in the four dimensional black hole
solution. The pull-back of the left-invariant Cartan-Maurer form
along the geodesic  takes the form
\begin{equation}\Omega=\mathbb{L}^{-1} \frac{\d}{\d \tau} \mathbb{L},\qquad L=
\frac{1}{2}(\Omega+\eta \Omega^{T}\eta), \end{equation}
 where $L$ is Lax operator defined in (\ref{laxdef}). We denote the generators of solvable algebra by $T_{A}$ which, for our model, are
\begin{equation}T_{\mathcal{A}}=\{H_{i},E_{\alpha}\}\,,\,\,\qquad i=1,\cdots, 4;\quad \alpha=1,\cdots, 24\,,\nonumber\\
\end{equation}
The following relations then hold
\begin{equation}\label{Kge}\{\mathbb{K}_{\mathcal{A}}\}=\{H_i,\,K_\alpha\}\,\,;\,\,\,K_\alpha=\frac{1}{2}(E_{\alpha}+\eta E_{\alpha}^{T}\eta)\,.
\end{equation}  having $\mathbb{K}_{\mathcal{A}}$ denoted the generators of $\mathfrak{K}^*.$
We also define the generators of $H^{*}$ by $J_{\alpha}$
\begin{equation}J_{\alpha}=\frac{1}{2}(E_{\alpha}-\eta E_{\alpha}^{T}\eta).\end{equation}
The branching (\ref{branchsp}) implies that the tangent space $\mathfrak{K}^*$ of $G/H^*$  defined by the
pseudo-Cartan decomposition of $\mathfrak{g},$ transforms in the $(\bf{2},\bf{R})$ of $H^{*}.$ A generic element $L\in\mathfrak{K}^*$ thus has the form
has the form
\begin{equation}L=(\Delta^{(a A)}),\qquad \textrm{where},\quad  A=1,\cdots \textrm{dim} (\bold{ R}); \quad
a=1,2.\end{equation}
From the general form of $H^{*}$ we infer that
\begin{equation}p=\textrm{rank}\left (\frac{H^{*}}{H_{c}}\right)=\textrm{rank}\left(\frac{G'_{4}}{H'_{4}}\right)+1=\textrm{rank}\left(\frac{\Sp(6)'}{\textrm{U}'(3)}\right)+1=4\end{equation}
where $p$ is the dimension of the minimal space (\emph{normal space})  defined by the normal form of $\bold{R}'$ with respect to $H_{c}$, see discussion below eq. (\ref{MADM}). This will be relevant in see Sect. \ref{generatingsol} when we will define a submanifold $\mathcal{M}_N$ of  $\mathcal{M}_{QK}^*$ within which the generating geodesic of regular/small single center black holes unfolds.
 \footnote{Thus $p$ is the minimal number of components of $\mathbb{Y}_0$ (i.e. central and matter charges at radial infinity)  into which the most general vector $\mathbb{Y}_0$,
 in $\bold{R}',$ can be reduced by means of an $H_{c}$-transformation. }.
There are four roots $\gamma_{k}$ out of $\gamma_M$, which define
the normal form and which are mutually orthogonal. Our choice of the
$E_{\gamma_{k}}$ will be
\begin{equation}\label{egamma}E_{\gamma_{k}}=\{E_{3},E_{14},E_{4},E_{11}\}\,,\end{equation}
which define a set of four conserved quantized charges in $D=4$ and
which correspond to the generators $T_M$, $M=1,9,12,14$.  Out of
these  generators we can construct two $p$-dimensional abelian
spaces $\mathfrak{K}_{N}^{*(R)}$ and $\mathfrak{H}_{N}^{*(R)}$ whose
generators will be denoted by  $\{\mathcal{K}_\ell\}$ and
$\{\mathcal{J}_\ell\}$, $\ell=0,1,4,6$, respectively, and  defined
as (see Sect. \ref{generatingsol}):
\begin{align}\label{KJrel}\mathcal{K}_{0}=\frac{1}{2}\,(T_1+\eta T_1^T\eta)\,\,,\,\,\,\mathcal{K}_{1}=\frac{1}{2}\,(T_9+\eta T_9^T\eta)\,\,,\,\,\,\mathcal{K}_{4}=\frac{1}{2}\,(T_{12}+\eta T_{12}^T\eta)\,\,,\,\,\,\mathcal{K}_{6}=\frac{1}{2}\,(T_{14}+\eta T_{14}^T\eta)\,,\nonumber\\
\mathcal{J}_{0}=\frac{1}{2}\,(T_1-\eta T_1^T\eta)\,\,,\,\,\,\mathcal{J}_{1}=\frac{1}{2}\,(T_9-\eta T_9^T\eta)\,\,,\,\,\,\mathcal{J}_{4}=\frac{1}{2}\,(T_{12}-\eta T_{12}^T\eta)\,\,,\,\,\,\mathcal{J}_{6}=\frac{1}{2}\,(T_{14}-\eta T_{14}^T\eta)\,.\end{align}
\section{Nilpotent Orbits in $\mathfrak{K}^*$}\label{npo}
We have learned in the previous sections that $\mathfrak{K}^*$ is the
carrier of an $H^*$ representation, the action of $H^*$ on the
matrices in $\mathfrak{K}^*$ being the adjoint one. Constructing and
classifying  $H^*$-adjoint orbits in  $\mathfrak{K}^*$, with
particular reference to the nilpotent ones, is still an open problem
in mathematics. It amounts to grouping the elements of
$\mathfrak{K}^*$ in orbits $\mathcal{O}$ (or conjugacy classes) with
respect to the adjoint action of $H^*$:
\begin{equation}
k_1,\,k_2\,\in\mathcal{O}\subset
\mathfrak{K}^*\,\Leftrightarrow\,\,\,\exists h\in
H^*\,\,:\,\,\,k_2=h^{-1}\,k_1\,h\,.
\end{equation}
A valuable approach to this task makes use of the theory of adjoint
orbits within a real Lie algebra $\mathfrak{g}$ with respect to the action of the
Lie group $G$ it generates \cite{Collingwood}. In this respect the
Konstant-Sekiguchi theorem \cite{Collingwood} is of invaluable help
since it allows for a complete classification of such orbits. This
is however not enough for our purposes, since we are interested in
the adjoint action of  $H^*$ on $\mathfrak{K}^*$ and a same $G$-orbit
may branch into several $H^*$-orbits. To understand this splitting
one may use $H^*$-invariant quantities which are not $G$-invariant,
such as $\gamma$-labels \cite{Kim:2010bf} or tensor classifiers
\cite{Fre:2011uy}. These, however, cannot guarantee by themselves a
complete classification. Here we shall use a different approach to
such a classification, which was originally devised in
\cite{Fre:2011ns}. \par 
We start from the notion of standard triple
associated with a nilpotent element $E$ of a real Lie algebra
$\mathfrak{g}$: According to the Jacobson-Morozov theorem
\cite{Collingwood}, such element can be though of as part of a
\emph{standard triple} of $\mathfrak{sl}(2,\mathbb{R})$-generators
$\{E,\,F,\,h\}$, satisfying the following commutation relations:
\begin{equation}
[h,E]=2\,E\,\,;\,\,\,[h,F]=-2\,F\,\,;\,\,\,[E,F]=h\,.\label{sttriple}
\end{equation}
We shall refer all the properties of the generators of
$\mathfrak{g}$ to the corresponding matrices in the real fundamental
representation ${\bf 26}$ of $\mathfrak{f}_{4(4)}$. In particular the action of the Cartan involution on
a generator $X$ amounts to taking the opposite of the transpose of
the corresponding matrix: $\tau(X)=-X^T$. If we were interested in
the orbits in the complexification $\mathfrak{g}^{\mathbb{C}}$ of
$\mathfrak{g}$ with respect to the adjoint action of the group
$G^{\mathbb{C}}$ it generates, different $G^{\mathbb{C}}$-nilpotent
orbits correspond to inequivalent embeddings of
$\mathfrak{sl}(2,\mathbb{R})={\rm Span}(E,\,F,\,h)$ inside
$\mathfrak{g}$, and these would correspond to different branchings
of a given representation of $G^{\mathbb{C}}$ with respect to the
${\rm SL}(2,\mathbb{R})$-subgroup. These different branchings are
uniquely characterized by the spectrum of the adjoint action of $h$
on $\mathfrak{g}^{\mathbb{C}}$. Such spectrum is conveniently
described by fixing a Cartan subalgebra $\mathcal{C}$ of
$\mathfrak{g}^{\mathbb{C}}$, in which $h$, being a semisimple
generator, can be rotated by means of a
$G^{\mathbb{C}}$-transformation, and evaluating the values of the
simple roots $\alpha_i$ of $\mathfrak{g}^{\mathbb{C}}$, associated
with $\mathcal{C}$, on $h$:
\begin{equation}
\mbox{$G^{\mathbb{C}}$-Orbit of
$E$}\,\leftrightarrow\,\mbox{$G^{\mathbb{C}}$-orbits of
$h$}\,\leftrightarrow\,\mbox{Spectrum}\,{\rm
Adj}_h\,\leftrightarrow\,\{\alpha_i(h)\}\,.
\end{equation}
The integers $\alpha_i(h)$, which are conventionally evaluated after
$h$ is rotated in the fundamental domain, can only have values
$0,1,2$ and are called $\alpha$-labels. They provide a complete
classification of the nilpotent $G^{\mathbb{C}}$-adjoint orbits in
$\mathfrak{g}^{\mathbb{C}}$ and can be found, for instance, in
\cite{Collingwood}.\par
 When we consider the problem of classifying nilpotent
 $G$-adjoint  orbits in the real Lie algebra $\mathfrak{g}$, a same
 $G^{\mathbb{C}}$-orbit will in general branch with respect to the
 action of $G$. In this case we can still reduce the problem of
 classifying the orbits of nilpotent elements $E$ of $\mathfrak{g}$
 to that of classifying orbits of some characteristic semisimple
 generators. This time however the relevant semisimple generator associated with the triple of $E$, is no longer $h$, but
 $i\,(E-F)$. More specifically $E-F$ is a compact matrix, i.e. it has only imaginary eigenvalues, and is thus an
 element of the maximal compact subalgebra $\mathfrak{H}$ of
 $\mathfrak{g}$. Having denoted  by $H$ the maximal compact subgroup of $G$,
 let  $H^{\mathbb{C}}$ be  its complexification, generated by the complexification $\mathfrak{H}^{\mathbb{C}}=\mathfrak{H}+i\,\mathfrak{H}$ of $\mathfrak{H}$.
The Kostant-Sekiguchi (KS) theorem defines a one-to-one
 correspondence between $G$-orbits of a nilpotent element $E$ of
 $\mathfrak{g}$, and the orbit under the adjoint action of
 $H^{\mathbb{C}}$ on $\mathfrak{K}^\mathbb{C}$, where the latter is  the complexification of the space of non-compact $\mathfrak{g}$-generators $\mathfrak{K}$ defined by the Cartan decomposition (\ref{Cartan}): $\mathfrak{K}^\mathbb{C}=\mathfrak{K}+i\,\mathfrak{K}$. These orbits are in turn in one-to-one correspondence with the $H^{\mathbb{C}}$-adjoint orbit of the element $(E-F)$ of $\mathfrak{H}$. Such orbits are completely defined by
 the (real) spectrum of the adjoint action of $i\,(E-F)$ over $\mathfrak{H}^{\mathbb{C}}$,
or, equivalently, by the embedding of the same semisimple element
within a suitable Cartan subalgebra $\mathcal{C}_H$ of
$i\,\mathfrak{H}$. If $\beta_k$ are the simple roots of
$\mathfrak{H}^{\mathbb{C}}$, such embedding is defined by the so
called $\beta$-labels, which are the values $\beta_k(i(E-F))$. In
summary the KS theorem states the following correspondence:
\begin{equation}
\left[\mbox{$G$-Orbit of
$E$}\right]\,\leftrightarrow\,\left[\mbox{$H^{\mathbb{C}}$-orbits of
$i(E-F)$}\right]\,\leftrightarrow\,\left[\mbox{Spectrum}\,{\rm
Adj}_{i(E-F)}\vert_{\mathfrak{H}^{\mathbb{C}}}\right]\,\leftrightarrow\,\{\beta_k(i\,(E-F))\}\,.
\end{equation}
The labels $\beta_k(i\,(E-F))$ are conventionally evaluated once
$i\,(E-F)$ is rotated into the fundamental domain and are
non-negative integers. The $\alpha,\beta$-labels are classified in
the mathematical literature, for all Lie groups
\cite{Collingwood}.\par Let us now come back to our original
problem: What are the possible $H^*$-orbits of nilpotent elements
$E$ in $\mathfrak{K}^*$? We know that $E$ is part of a standard triple.
Since $E$ is in $\mathfrak{K}^*$, compatibility of (\ref{sttriple})
with (\ref{HKrels}) requires that $h\in \mathfrak{H}^*$ and $F\in
\mathfrak{K}^*$. In particular $h$ is a semisimple, non-compact
element of $\mathfrak{H}^*$ ($\tau(h)=-h^T=-h$, $\sigma(h)=h$), and thus can be
chosen (modulo $H^*$-transformations of the triple) within a given
non-compact Cartan subalgebra $\mathcal{C}_{H^*}$ of
$\mathfrak{H}^*$. Clearly different $G^{\mathbb{C}}$ or $G$-orbits
(uniquely defined by $\alpha,\,\beta$-labels, respectively)
correspond to different $H^*$-orbits. However a same $G$-orbit may
branch with respect to the action of $H^*$. In \cite{Kim:2010bf},
the case $G={\rm G}_{2(2)}$, $H^*={\rm SL}(2,\mathbb{R})^2$ was
studied in detail, and the so called $\gamma$-labels were introduced
to distinguish between different $H^*$-orbits. The notion of
$\gamma$-labels is similar to that of $\beta$-labels. Let us denote by $\mathfrak{H}^{*\mathbb{C}}=\mathfrak{H}^{*}+i\,\mathfrak{H}^{*}$ the complexification of $\mathfrak{H}^{*}$, generating the subgroup $H^{*\,\mathbb{C}}$ of $G^{\mathbb{C}}$. The $\gamma$-labels identify
the $H^*$-orbits of $h$ within $\mathfrak{H}^*$ and can either be described in terms of the spectrum of the
adjoint action of $h$ on $\mathfrak{H}^*$, or in terms of the values
of the simple roots $\beta'_k$ of $\mathfrak{H}^{*\mathbb{C}}$ (referred now to the Cartan subalgebra
$\mathcal{C}_{H^*}$) on $h$, taken in the fundamental domain:
\begin{equation}
\mbox{$\gamma$-labels}\,\leftrightarrow \,
\left[\mbox{Spectrum}\,{\rm
Adj}_{h}\vert_{\mathfrak{H}^*}\right]\,\leftrightarrow\,\{\beta'_k(h)\}\,.
\end{equation}
These quantities are clearly invariant with respect to the adjoint
action of $H^*$ (and in general of its complexification
$H^{*\,\mathbb{C}}$) on the whole triple and in particular on $h$, and
thus different $\gamma$-labels correspond to different $H^*$-orbits
of $E$. Clearly the sets of all possible $\beta$- and
$\gamma$-labels coincide. In Table \ref{f44orbitar} we give a list of the $\alpha $ and $\beta$- (and thus also of the $\gamma$-) labels for the ${\rm F}_{4(4)}$-model \cite{Collingwood}. There is no mathematical property
guaranteeing  that $\gamma$-labels, together with the $\alpha$ and
$\beta$ ones, provide a complete classification of the
$H^*$-nilpotent orbits in $\mathfrak{K}^*$. And indeed here we provide
the first counterexample: different $H^*$- orbits sharing the same
$\alpha,\,\beta,\,\gamma$-labels.\par Let us now review the
constructive procedure introduced in \cite{Fre:2011ns}. Given a
nilpotent element $E$ of $\mathfrak{K}^*$ we shall adopt the working
assumption that there exists an element $E'$ in the same
$H^*$-orbit, whose triple $\{E',F',h'\}$ have the property that
$F'=E'^T$.\footnote{Although we do not have a proof for this for
generic $G/H^*$ spaces, it is proven for spaces of the form ${\rm
GL}(n,\mathbb{R})/{\rm SO}(p,q)$ \cite{Djokovic:1981bh,Bergshoeff:2008be},
using the $\eta$-symmetric normal forms. The most general $G/H^*$
manifold, can be thought of as a totally geodesic submanifold of a
${\rm GL}(n,\mathbb{R})/{\rm SO}(p,q)$ space, for some $n,p,q$.} We
shall then restrict to triples of this kind.\par The neutral element
$h$ of a  triple $\{E,F,h\}$, should fall in one of the $H^*$-orbits
uniquely defined by the $ \gamma$-labels. We then take a
representative $h$ of each such orbits and solve the matrix
equations in the unknown $E$:
\begin{eqnarray}
[h,E]\,&=& 2E\, \label{eq1},\\
\small[E,E^T\small] &=& h\,.\label{eq2}
\end{eqnarray}
Using a MATHEMATICA code, for each $h$ we find a set of solutions to
(\ref{eq1}), (\ref{eq2}). We group these solutions under the action
of the compact part $H^{little}_c[h]$ of the little group of $h$. In
all cases we could find that solutions which were not connected by
the adjoint action of $H^{little}_c[h]$, could be distinguished by
$H^*$-invariant quantities. Such quantities are the signatures of
certain symmetric covariant (or contravariant) $H^*$-tensors, called \emph{ tensor
classifiers}, to be discussed in detail in Subsect. \ref{tcs}.
In principle, if one is able to find tensor classifiers capable of distinguishing between
solutions $E$ to (\ref{eq1}), (\ref{eq2}) which share the same $\beta$-label (i.e. fall in the same $G$-orbit) but are not related by $H^{little}_c[h]$, the resulting classification of the $H^*$-orbits can be claimed to be complete. In our case a set of tensor classifiers fulfilling this task were constructed.
 They even allow for an \emph{almost} complete distinction among the various orbits without the use of $\alpha, \beta,\gamma$ labels. The main advantage of a complete classification effected by only using  tensor classifiers is that given a nilpotent element $E$ in $\mathfrak{K}^*$, the computation of the $\alpha, \beta,\gamma$ -labels would  require the determination of the whole standard triple $\{E,F,h\}$, which in general is a non-trivial task, since $F\neq E^T$. Tensor classifiers computed on $E$ would give the answer straight away.
 In our model, in order for a tensor-classifier-based classification to be complete, probably tensors of higher degree in the Lax components would have to be constructed. The analysis is however complete once the use of the  tensor classifiers is complemented with the $\alpha, \beta,\gamma$ -labels.
The different $H^*$-orbits are grouped into $G$-orbits (defined by
the $\beta$-labels), which are  arranged in the fifteen Tables
\ref{alpha1}-\ref{alpha15}, one for each $G^{\mathbb{C}}$-orbit
($\alpha$-label). Within each table, each $G$-orbit, represented by
a column,  splits into distinct  $H^*$-orbits, which are
distinguished either by the $\gamma$-labels (rows in the table), or,
for a same $\gamma$-label, by the signatures of certain tensor
classifiers (further horizontal splitting of the corresponding
$\gamma,\beta$- entry of the table). This further splitting is
labeled  by $\delta_1,\,\delta_2$.\par Solutions describing regular
static black holes fall in the  first four $G^{\mathbb{C}}$-orbits.
The other $G^{\mathbb{C}}$-nilpotent orbits have degree of
nilpotency higher than 3 (we work in the fundamental representation
of ${\rm F}_{4(4)}$). We shall give examples of single-center static
solutions in these orbits, which however all lift to singular
four-dimensional space-times, consistently with the regularity
condition  (\ref{regcondnil}).
\par
Let us now discuss the general structure of $H^{little}_c[h]$. It
can be represented as the semidirect product of a continuous group
in the identity sector of $H^*$ and the discrete \emph{stabilizer}
$\mathcal{HW}$ of the Cartan subalgebra $\mathcal{C}_{H^*}$:
\begin{equation}
H^{little}_c[h]=H^{little}_{c,(0)}[h]\ltimes \mathcal{HW}\,.
\end{equation}
The groups $H^{little}_{c,(0)}[h]$, for each standard triple, are
listed in Table \ref{Hlittle0} in  Appendix \ref{AA}. The group $\mathcal{HW}$ is a new
object first introduced, to our knowledge, in the physics literature
in \cite{Fre:2011ns}, and is defined as follows:
\begin{equation}
\mathcal{HW}=\{g\in H^*\,\vert\,\,\forall h\in
\mathcal{C}_{H^*}\,:\,\,g^{-1}\,h\,g=h\}\,.
\end{equation}
A simple way of characterizing $\mathcal{HW}$ is as a normal subgroup of the \emph{generalized Weyl group} $\mathcal{GW}$ \cite{Fre:2011ns} of $\mathfrak{g}$. Let us briefly review the definition of the latter.
Given a positive root $\alpha$ of $\mathfrak{g}$ defined with respect to a Cartan subalgebra $\mathcal{C}$ of $\mathfrak{g}$, it is known that the  Weyl group $\mathcal{W}$ of $\mathfrak{g}$ is generated by the reflections in the positive roots $\alpha$ of $\mathfrak{g}$ which are effected by means of the adjoint action of a $G$-elements $O_\alpha$ of the form:
\begin{equation}
O_\alpha\equiv e^{\frac{\pi}{\sqrt{2}\,|\alpha|}\,(E_\alpha-E_{-\alpha})}\,.
\end{equation}
It is indeed straightforward to prove that
\begin{align}
O_\alpha^{-1} H_\beta O_\alpha &=H_{\sigma_\alpha(\beta)}\,\,;\,\,\,
\sigma_\alpha(\beta)\equiv \beta-2\,\frac{(\beta,\alpha)}{|\alpha|^2}\,\alpha\,.
\end{align}
We shall choose the Cartan subalgebra $\mathcal{C}=\mathcal{C}_{H^*}$ of $\mathfrak{g}=\mathfrak{f}_{4(4)}$ to consist of non-compact  (i.e. represented by symmetric matrices) in $H^*$. This is a Cartan subalgebra of $\mathfrak{H}^*$ as well. Diagonalizing the adjoint action of $\mathcal{C}_{H^*}$ over $\mathfrak{g}$, we define shift generators $E_{\pm \alpha}$, some of which will lie in $\mathfrak{H}^*$ and some in $\mathfrak{K}^*$. We then divide the root system $\Delta$ of $\mathfrak{g}$ correspondingly in the following disjoint sets:
\begin{align}
\Delta_+ &=\Delta[\mathfrak{H}^*]\oplus \Delta[\mathfrak{K}^*]\,\nonumber\\
\alpha &\in  \Delta[\mathfrak{H}^*] \,\leftrightarrow \,\,E_{\alpha}\in \mathfrak{H}^*\,,\nonumber\\
\alpha &\in  \Delta[\mathfrak{K}^*] \,\leftrightarrow \,\,E_{\alpha}\in \mathfrak{K}^*\,.\nonumber
\end{align}
The orthogonal matrices $O_\alpha$ (or even just those corresponding to the simple roots $\alpha_i$ of $\mathfrak{g}$) generate themselves a discrete group $\mathcal{GW}$ which is larger than the Weyl group $\mathcal{W}$. It is the largest subgroup of $G$ whose adjoint action leaves $\mathcal{C}$ stable. A generic element of $\mathcal{GW}$ can indeed be written as the product of an element of $\mathcal{W}$ times an element of the stabilizer $\mathcal{HW}$ of the Cartan subalgebra $\mathcal{C}$, which  is a normal subgroup of $\mathcal{GW}$,  so that we can write:
\begin{equation}
\mathcal{W}=\frac{\mathcal{GW}}{\mathcal{HW}}\,.
\end{equation}
A simple way of characterizing $\mathcal{HW}$ is as the subgroup of $\mathcal{GW}$ generated by `` squared reflections'' $O_\alpha^2$ (or simply by $O_{\alpha_i}^2$), whose adjoint action on a generic element of $\mathcal{C}$ clearly leaves it invariant.
Notice that if $O_\alpha^4={\bf 1}$, as it is the case for $\mathfrak{f}_{4(4)}$ or the models considered in \cite{Fre:2011ns}, then  $\mathcal{HW}\subset H^*$, even if $E_\alpha\in \mathfrak{K}^*$, i.e. $\alpha\in \Delta[\mathfrak{K}^*]$. In this case indeed we have:
\begin{equation}
(O_\alpha^T)^2\eta O_\alpha^2=\eta  O_\alpha^4=\eta\,\,\Rightarrow\,\,\,O_\alpha^2\in H^*\,.
\end{equation}
These transformations in $\mathcal{HW}$ do not belong to the identity sector of $H^*$, but are nevertheless important since they relate, just as any other transformation in $H^{little}_c[h]$, different solutions to eq.s (\ref{eq1},\ref{eq2}).\par
Following \cite{Fre:2011ns}, we also define a subgroup $\mathcal{GW}_H$ of
$\mathcal{GW}$ as its intersection with $H^*$: $\mathcal{GW}_H=\mathcal{GW}\cap H^*$. Clearly $\mathcal{HW}\subset \mathcal{GW}_H$ and we can consider the coset
\begin{equation}
\mathcal{W}_H=\frac{\mathcal{GW}_H}{\mathcal{HW}}\subset \mathcal{W}\,,
\end{equation}
which can be characterized as the subgroup of the Weyl group whose action leaves the two root subspaces $\Delta[\mathfrak{H}^*]$ and $\Delta[\mathfrak{K}^*]$ invariant. This analysis provides us with a useful alternative way of finding representatives in $\mathcal{C}_{H^*}$ of the various $H^*$-orbits of $h$, identified by the $\gamma$-labels. Such representatives could either be constructed directly using the $\gamma$-labels, or we can start from representatives in $\mathcal{C}_{H^*}$ of $G^{\mathbb{C}}$-orbits of $h$ within $\mathfrak{g}^{\mathbb{C}}$, each defined by a set of  $\alpha$-labels. If we act on this representative by means of $\mathcal{W}/\mathcal{W}_H$ we find different representatives of the same orbit in $\mathcal{C}_{H^*}$ which are not related by $H^*$, namely representatives of distinct $H^*$-orbits. Not all these representatives are neutral elements of triples with $E$ and $F=E^T$ in $\mathfrak{K}^*$. If we impose this further condition, we end up with a set of $H^*$-orbits for the given $\alpha$-label which precisely correspond to the allowed $\gamma$-labels. They coincide for each $\alpha$-label with the $\beta$-labels listed in Table \ref{f44orbitar}. Then we take a representative neutral element $h$ for each $\gamma$-label and proceed with the solution of eq.s (\ref{eq1},\ref{eq2}).\par
For the  ${\rm F}_{4(4)}$-model the Weyl group has $1152$ elements, of which only $96$ belong to $H^*$ and thus close the subgroup $\mathcal{W}_H$. The stabilizer $\mathcal{HW}$ has order 16. We summarize below these data:
\begin{align}
|\mathcal{W}|&=1152\,\,;\,\,\,\left\vert\mathcal{W}_H\right\vert=96\,\,;\,\,\,\vert\frac{\mathcal{W}}{\mathcal{W}_H}\vert=12\,\,;\,\,\,
|\mathcal{HW}|=16\,\,;\,\,\,|\mathcal{GW}|=16\times 1152=18432\,.
\end{align}
\subsection{The Orbits}
Here we discuss the explicit construction of the orbit in the model under consideration.
As pointed out in the previous section, Given a standard triple $\{E,\,F,\,h\}$, whose neutral element $h$ is in the fundamental domain of the simple roots $\alpha_i$ of $\mathfrak{g}^{\mathbb{C}}=\mathfrak{f}_4^\mathbb{C}$, the $G^{\mathbb{C}}$-orbit of the nilpositive element $E$
is uniquely defined by the $\alpha$-labels $\alpha_{i}(h)$ which  take value in $\{0,1,2\}$:
\begin{equation}\alpha\textrm{-\emph{labels}};\qquad (\alpha_{1}(h),\alpha_{2}(h),\alpha_{3}(h),\alpha_{4}(h)).\end{equation}
For each $\alpha$-label we choose a representative $E,$ and it may happen that two different representatives are conjugated
by an element $X\in F_{4(4)},$ that is
\begin{equation}X^{-1}E\,X=E',\qquad X\in F_{4(4)}.\end{equation} In this case $E$ and $E'$ lie in the same nilpotent orbit, and therefore one obtains a single $F_{4(4)}$-orbit. We present the $F_{4(4)}$ single orbits in Table \ref{f44orbitar}. If this is not the case, then one can distinguish two or three different $F_{4(4)}$-orbits through what we have called the $\beta$-labels which provide a complete classification of the $F_{4(4)}$-orbits.
As mentioned in the previous section, the nilpotent $F_{4(4)}$-orbits are in one-to-one correspondence with the nilpotent $H^{\mathbb{C}}$-orbits in $\mathfrak{K}^{\mathbb{C}}$, complexification of $\mathfrak{K}$,\footnote{We recall that $H^{\mathbb{C}}$ is the complexification of the maximal compact subgroup $H={\rm SU}(2)\times {\rm USp}(6)$ of $G={\rm F}_{4(4)}$, whose algebra is denoted by $\mathfrak{H}^{\mathbb{C}}\subset \mathfrak{f}_4^\mathbb{C}$, not to be confused with the complexification $H^{*\mathbb{C}}$ of $H^*={\rm SL}(2)\times {\rm Sp}'(6)$ subgroup of ${\rm F}_{4(4)}$, whose Lie algebra is denoted by $\mathfrak{H}^{*\mathbb{C}}\subset \mathfrak{f}_4^\mathbb{C}$. ${H}^{\mathbb{C}}$ and $H^{*\mathbb{C}}$ are clearly isomorphic in $G^{\mathbb{C}}$ and  so are their Lie algebras $\mathfrak{H}^{\mathbb{C}}$, $\mathfrak{H}^{*\mathbb{C}}$, though the latter are described by different generators.} which in turn are classified by the $\beta$-labels. To define the latter we need to refer to a suitable Cartan subalgebra $\mathcal{C}_H$ of $i\,\mathfrak{H}$ which the element $i\,(E-F)$ should belong to. Since $E-F$ is also an element of $\mathfrak{K}^*$, we choose $\mathcal{C}_H$ to lie in the intersection  $i(\mathfrak{H}\cap \mathfrak{K}^*)$. A possible choice of basis for $\mathcal{C}_H$ is:
\begin{align}
\hat{H}_{\beta_1}&=2i\,(K_4+K_{11})\,\,;\,\,\,\hat{H}_{\beta_2}=2i\,(K_3-K_{11})\,\,;\,\,\,\hat{H}_{\beta_3}=-2i \left(K_3+K_4-K_{11}-K_{14}\right)\,,\nonumber\\\hat{H}_{\beta_4}&=i(-K_3+K_4-K_{11}+K_{14})\,,
\end{align}
where $\beta_1,\,\beta_2,\,\beta_3$ are the simple roots of $\mathfrak{sp}(6,\mathbb{C})$ in $\mathfrak{H}^\mathbb{C}$, while $\beta_4$ is the simple root of the $\mathfrak{sl}(2,\mathbb{C})$ subalgebra commuting with it. The roots $\beta_1,\beta_2,\beta_4$ have squared length equal to $2$, while $\beta_3$ has  squared length equal to $4$.
The corresponding Dynkin diagram is
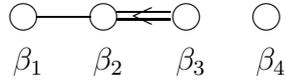
\vspace{0mm}
\begin{center}
\begin{picture}(110,30)
\put (10,10){\circle {10}}
\put (6,-10){$\beta_{1}$}
\put (15,10){\line (1,0){20}}
\put (40,10){\circle {10}}
\put (36,-10){$\beta_{2}$}
\put (45,12){\line (1,0){20}}
\put (50,7.3){{$<$}}
\put (45,9){\line (1,0){20}}
\put (71,10){\circle {10}}
\put (67,-10){$\beta_{3}$}
\put (101,10){\circle {10}}
\put (97,-10){$\beta_{4}$}

\end{picture}

\end{center}
\vspace{5mm}

The $\beta$-labels associated with a triple $\{E,F,h\}$ are then computed as
\begin{equation}\beta\textrm{-\emph{labels}};\qquad (\beta_{1}(i(E-F)),\beta_{2}(i(E-F)),\beta_{3}(i(E-F)),\beta_{4}(i(E-F))).\end{equation}
If we define the simple weights $\lambda^k$ associated with $\beta_k$ as usual by the property that:
\begin{equation}
\langle \lambda^i,\,\beta_j\rangle=2 \frac{(\lambda^i,\,\beta_j)}{(\beta_j,\,\beta_j)}=\delta^i_j\,,
\end{equation}
we can write the corresponding basis of $\mathcal{C}_H$ as:
\begin{equation}
\hat{H}_{\lambda^i}=C^{ij}\,\hat{H}_{\beta_j}\,,
\end{equation}
where $C^{ij}$ is the inverse of the Cartan matrix $C_{ij}$:
\begin{equation}
(C_{ij})=\left(2 \frac{(\beta_i,\,\beta_j)}{(\beta_j,\,\beta_j)}\right)=\left (\begin{array}{cccc}
  2 & -1&0&0\\
  -1 & 2&-1&0\\
0&-2&2&0\\
0&0&0&2
\end{array} \right)\,.
\end{equation}
If we denote by $n_i$ the $\beta$-labels, knowing $n_i$ we can construct the corresponding matrix $E-F$ as follows:
\begin{equation}
i(E-F)=\sum_{k=1}^4\,2\,n_k\frac{\hat{H}_{\lambda^k}}{(\beta_k,\,\beta_k)}\,.
\end{equation}
$H^*$-orbits of the neutral element $h$ of a triple are classified by the $\gamma$-labels defined in the previous section as the values $\beta'_k(h)$ on it of the simple roots $\beta_k'$ associated with the complexification $\mathfrak{H}^{*\mathbb{C}}$ of $\mathfrak{H}^{*}$. The corresponding Dynkin diagram is the same as for $\beta_k$, though these roots are now referred to a non-compact Cartan subalgebra $\mathcal{C}_{H^*}$ in $\mathfrak{H}^*\cap\mathfrak{K}$. We can choose as a basis of $\mathcal{C}_{H^*}$ the following matrices:
\begin{align}
\tilde{H}_{\beta'_1}&=2\,(J_4+J_{11})\,\,;\,\,\,\tilde{H}_{\beta'_2}=2\,(J_3-J_{11})\,\,;\,\,\,\tilde{H}_{\beta'_3}=-2 \left(J_3+J_4-J_{11}-J_{14}\right)\,,\nonumber\\
\tilde{H}_{\beta'_4}&=(-J_3+J_4-J_{11}+J_{14})\,,\label{CHstarb}
\end{align}
Notice that the $\tilde{H}_{\beta'_k}$ and the $\hat{H}_{\beta_k}$ are mapped into one another by replacing $J_\alpha$ with $iK_\alpha$. The corresponding Cartan subalgebras are isomorphic in $\mathfrak{g}^\mathbb{C}$ through the action of $G^\mathbb{C}$. The same is true for $h$ and $i(E-F)$. We construct the simple weights $\lambda_k'$ associated with $\beta_k'$ and the corresponding basis of matrices $\tilde{H}_{\lambda^{\prime k}}$. Given the $\gamma$-labels $n_k=\beta'_k(h)$, we can construct the corresponding $h$ as follows:
\begin{equation}
h=\sum_{k=1}^4\,2\,n_k\frac{\tilde{H}_{\lambda^{\prime k}}}{(\beta'_k,\,\beta'_k)}\,.
\end{equation}
In Table \ref{f44orbitar} the $G={\rm F}_{4(4)}$-nilpotent orbits are listed with the corresponding $\alpha$- and $\beta$- labels. There are $15$ $\alpha$-labels defining $15$ distinct nilpotent orbits of $F_4^\mathbb{C}$, and are denoted by $\alpha^{(s)}$. For a same $\alpha$-label we can  have more $\beta$-ones signalling that the corresponding ${\rm F}_4^\mathbb{C}$-orbit branches with respect to ${\rm F}_{4(4)}$. When this occurs, we denote the $\beta$-labels by $\beta^{(1)},\,\beta^{(2)},$.. in the order in which they are listed in Table \ref{f44orbitar}. The possible $\gamma$-labels for a same $\alpha$- one are the same as the $\beta$-labels and thus are not listed.

\begin{table}[h!]
\begin{center}
\begin{equation}
\begin{array}{|c|c|c|c|}
\hline
\text{$\mathrm{F}_{4(4)}$-orbit} & %(\alpha_1(h),\alpha_2(h))
\text{$\alpha$-labels}
& %(\beta_1(h),\beta_2(h))
\text{$\beta$-labels}&\text{Degree of nilpotency }\nn\\ \hline
\mathcal{O}_1 & (1,\,0,\,0,\,0)  &(0,\,0,\,1,\,1)& 2\nn\\ \hline

\mathcal{O}_2 & (0,\,0,\,0,\,1)  &(1,\,0,\,0,\,2)& 3 \nn\\

\mathcal{O}_3 & (0,\,0,\,0,\,1)  &(0,\,1,\,0,\,0)& 3 \nn\\ \hline

\mathcal{O}_4 & (0,\,1,\,0,\,0)  &(0,\,0,\,1,\,3)& 3 \nn\\
\mathcal{O}_5 & (0,\,1,\,0,\,0)  &(1,\,0,\,1,\,1) & 3\nn\\ \hline

\mathcal{O}_6 & (2,\,0,\,0,\,0)  &(0,\,0,\,0,\,4) & 3\nn\\
\mathcal{O}_7 & (2,\,0,\,0,\,0)  &(2,\,0,\,0,\,0)& 3 \nn\\
\mathcal{O}_8 & (2,\,0,\,0,\,0)  &(0,\,0,\,2,\,2) & 3\nn\\ \hline

\mathcal{O}_9 & (0,\,0,\,0,\,2)  &(0,\,2,\,0,\,0) &5\nn\\ \hline
\mathcal{O}_{10} & (0,\,0,\,1,\,0)  &(1,\,1,\,0,\,2)&4 \nn\\ \hline

\mathcal{O}_{11} & (2,\,0,\,0,\,1)  &(1,\,0,\,2,\,4) &5\nn\\
\mathcal{O}_{12} & (2,\,0,\,0,\,1)  &(0,\,1,\,2,\,2) &5\nn\\ \hline

\mathcal{O}_{13} & (0,\,1,\,0,\,1)  &(1,\,1,\,1,\,1)&5 \nn\\ \hline

\mathcal{O}_{14} & (1,\,0,\,1,\,0)  &(1,\,0,\,3,\,1)&5 \nn\\
\mathcal{O}_{15} & (1,\,0,\,1,\,0)  &(1,\,1,\,1,\,3) &5\nn\\ \hline

\mathcal{O}_{16} & (0,\,2,\,0,\,0)  &(0,\,0,\,4,\,0) &5\nn\\
\mathcal{O}_{17} & (0,\,2,\,0,\,0)  &(0,\,2,\,0,\,4)&5 \nn\\
\mathcal{O}_{18} & (0,\,2,\,0,\,0)  &(2,\,0,\,2,\,2) &5\nn\\ \hline

\mathcal{O}_{19} & (2,\,2,\,0,\,0)  &(0,\,0,\,4,\,8) &7\nn\\
\mathcal{O}_{20} & (2,\,2,\,0,\,0)  &(2,\,0,\,4,\,4) &7\nn\\ \hline

\mathcal{O}_{21} & (1,\,0,\,1,\,2)  &(1,\,3,\,1,\,3)&9 \nn\\ \hline

\mathcal{O}_{22} & (0,\,2,\,0,\,2)  &(0,\,4,\,0,\,4)&9 \nn\\
\mathcal{O}_{23} & (0,\,2,\,0,\,2)  &(2,\,2,\,2,\,2) &9\nn\\ \hline

\mathcal{O}_{24} & (2,\,2,\,0,\,2)  &(2,\,2,\,4,\,4) &11\nn\\
\mathcal{O}_{25} & (2,\,2,\,0,\,2)  &(4,\,0,\,4,\,8)&11 \nn\\ \hline

\mathcal{O}_{26} & (2,\,2,\,2,\,2)  &(4,\,4,\,4,\,8) &17\nn
\\ \hline
%\end{align}
\end{array}
\end{equation}
\caption{\small The
 six nonzero $F_{4(4)}$-orbits and their degree of nilpotency.}
\label{f44orbitar}
\end{center}
\end{table}
Below we list all the labels, giving the corresponding spectrum of the adjoint action of
 $h$ over $\mathfrak{g}$ for the $\alpha$-labels, of the adjoint action of
 $h$ over $\mathfrak{H}^*$ for the $\gamma$-labels and of the adjoint action of
 $i\,(E-F)$ over $\mathfrak{H}$ for the $\beta$-ones. Moreover, for each $\alpha$-label we give the ``angular momentum'' decomposition of the adjoint of $\mathfrak{g}^{\mathbb{C}}$ with respect to the ${\rm SL}(2,\mathbb{C})$ subgroup of $G^{\mathbb{C}}$ generated by the standard triple.\par
 We apply to the classification of the $H^*$-nilpotent orbits in $\mathfrak{K}^*$ the systematic method
 defined in the previous section: We start from a representative $h$ for each $\gamma$-label, we solve eq.s (\ref{eq1}), (\ref{eq2}) in $E\in \mathfrak{K}^*$, and group the solutions under the action of $H_c^{little}[h]$. We find that solutions which are not connected through $H_c^{little}[h]$ can be distinguished by the signatures of tensor classifiers, and thus belong to distinct $H^*$-orbits. The result of this classification is summarized in Tables \ref{alpha1}-\ref{alpha15}. In the next subsection we list the $\alpha,\,\beta,\,\gamma $-labels. In Subsection \ref{tcs} we review the construction of the tensor classifiers.

\subsection{$\alpha,\,\beta,\,\gamma $-labels}
%\alpha1
\paragraph{$\alpha^{(1)}$-label}:
\begin{eqnarray}\label{ealpha1}\alpha^{(1)}&=&1\times (\pm 2)+14\times(\pm1)+22\times (0)\\
&=&1\times (J=1)\oplus 14\times \Big(J=\frac{1}{2}\Big)\oplus 21\times (J=0)\end{eqnarray}
and
\begin{equation}\gamma=7\times(\pm1)+10\times (0)=\beta\end{equation}
corresponding to $(0,0,1,1)$ label. The associated orbit and its representative are given in Table \ref{alpha1}.
%table(alpha1)
\begin{table}[htbp]
\scalefont{1}
\begin{center}
\begin{tabular}{|c|c|c|c|}\hline
\backslashbox{$\gamma$-label}{$\beta$-label}
&(0,\,0,\,1,\,1)&\null\\ \hline
(0,\,0,\,1,\,1)&
$\renewcommand{\arraystretch}{1}
\begin{array}{c}
\\-\frac{1}{2}(H_{1}-H_{4})- K_{4}\end{array}$&$\mathcal{O}_{\cdot H^*}$\\ \hline

\null  & $\mathcal{O}_{1}$&\null\\ \hline
\end{tabular}
\caption{\small The $H^*$-orbit $\mathcal{O}_{1 H^*}$ within the  $\rm{F}_{4(4)}$-orbits $\mathcal{O}_{1}.$}
\label{alpha1}
\end{center}
\end{table}

%alpha2
\paragraph{$\alpha^{(2)}$-label}:
\begin{eqnarray}\label{ealpha2}\alpha^{(2)}&=&7\times(\pm2)+8\times(\pm 1)+22\times (0)\\
&=&7\times (J=1)\oplus8\times\Big(J=\frac{1}{2}\Big)\oplus15\times (J=0).\end{eqnarray}
The $\gamma-\beta$-labels are given by
\begin{eqnarray}\gamma^{(1)}&=&2\times (\pm 2)+4\times(\pm1 )+12\times(0)=\beta^{(1)}\\
\gamma^{(2)}&=&3\times (\pm 2)+4\times (\pm 1)+10\times(0)=\beta^{(2)}\end{eqnarray}
corresponding, respectively, to $(1,0,0,2)$ and $(0,1,0,0)$ labels. The associated orbits and their representatives are presented in Table \ref{alpha2}.

%table(alpha2)
\begin{table}[htbp]

\scalefont{.9}
\begin{tabular}{|c|c|c|c|}\hline
\backslashbox{$\gamma$-label}{$\beta$-label}
&(1,\,0,\,0,\,2)&(0,\,1,\,0,\,0)&\null\\ \hline
(1,\,0,\,0,\,2)&
$\renewcommand{\arraystretch}{1}
\begin{array}{c}
\\\frac{1}{2}(-H_{1}-H_{2}-H_{3}+H_{4})\\- K_{14}- K_{4}\end{array}$&$\renewcommand{\arraystretch}{1}
\begin{array}{c}
\\ \frac{1}{2}(-H_{1}+H_{2}+H_{3}+H_{4})\\+ K_{14}- K_{4}\end{array}$ &$\mathcal{O}_{\cdot H^*}$\\ \hline
(0,\,1,\,0,\,0)  & $- K_{2}+ K_{15}+ K_{6}+ K_{13}$ & $- K_{16}- K_{12}- K_{5}- K_{1}$&$\mathcal{O'}_{\cdot H^*}$\\ \hline
\null  & $\mathcal{O}_{2}$& $\mathcal{O}_{3}$&\null\\ \hline
\end{tabular}
\caption{\small The four $H^*$-orbits $\mathcal{O}_{2 H^*},\,\mathcal{O}_{2H^*}',\,\mathcal{O}_{3H^*},\,\mathcal{O}_{3H^*}'$ within the two $\rm{F}_{4(4)}$-orbits $\mathcal{O}_{2}$ and $\mathcal{O}_{3}$.
Our labeling of the orbits is indicated by the last row and the rightmost column.}
\label{alpha2}

\end{table}

%alpha3

\paragraph{$\alpha^{(3)}$-label}:
\begin{eqnarray}\label{ealpha3}\alpha^{(3)}&=&2\times (\pm 3)+6\times (\pm 2)+12\times (\pm 1)+12\times (0)\\
&=&2\times \Big(J=\frac{3}{2}\Big)\oplus 6\times(J=1)\oplus 10\times \Big(J=\frac{1}{2}\Big)\oplus 6 \times (J=0),\end{eqnarray}
and
\begin{eqnarray}\gamma^{(1)}&=&1\times (\pm 3)+6 \times(\pm 1) +10\times (0)=\beta^{(1)}\\
\gamma^{(2)}&=&1\times (\pm 3)+2 \times (\pm 2)+6 \times (\pm)+6\times (0)=\beta^{(2)},\end{eqnarray}
corresponding, respectively, to $(0,0,1,3)$ and $(1,0,1,1)$ labels. The associated orbits and their representatives are presented in Table \ref{alpha3}.

%table(alpha3)
\begin{table}[htbp]
\scalefont{.75}
\begin{tabular}{|c|c|c|c|}\hline
\backslashbox{$\gamma$-label}{$\beta$-label}
&(0,\,0,\,1,\,3)&(1,\,0,\,1,\,1)&\null\\ \hline
(0,\,0,\,1,\,3)&
$\renewcommand{\arraystretch}{1}
\begin{array}{c}
\frac{1}{2}(3 K_{16}+3 K_{12}- K_{5}- K_{2}\\ +K_{15}- K_{1}+ K_{6}+ K_{13})\end{array}$&$\renewcommand{\arraystretch}{1}
\begin{array}{c}
\frac{1}{2}( K_{16}+ K_{12}-3 K_{5}+ K_{2}\\- K_{15}-3 K_{1}- K_{6}- K_{13})\end{array}$ &$\mathcal{O}_{\cdot H^*}$\\ \hline
(1,\,0,\,1,\,1)  & $\renewcommand{\arraystretch}{1}
\begin{array}{c}
\frac{1}{2}( K_{16}+ K_{12}- K_{5}-3 K_{2}\\+ K_{15}- K_{1}+3 K_{6}+ K_{13})\end{array}$&

\begin{tabular}{|c|c|c|}\hline
  $\delta_1$& $\renewcommand{\arraystretch}{1}
\begin{array}{c} \frac{1}{2}(-3 K_{16}-3 K_{12}- K_{5}+ K_{2}\\+ K_{15}- K_{1}- K_{6}+ K_{13})
 \end{array}$ &$\bar{ \mathcal{O}}'_{.H^*}$\\ \hline
$\delta_2$ & $\renewcommand{\arraystretch}{1}
\begin{array}{c}\frac{1}{2}(- K_{16}- K_{12}- K_{5}- K_{2}\\+ K_{15}- K_{1}+ K_{6}+ K_{13})\\
+\sqrt{2}\,( K_{24}+ K_{20})
 \end{array}$ & $\hat{ \mathcal{O}}'_{.H^*}$ \\ \hline
\end{tabular}

  &
$ \mathcal{O'}_{\cdot H^*}$ \\ \hline
\null  & $\mathcal{O}_{4}$& $\mathcal{O}_{5}$&\null\\ \hline
\end{tabular}
\caption{\small The four $H^*$-orbits $\mathcal{O}_{4 H^*},\,\mathcal{O}_{4H^*}',\,\mathcal{O}_{5H^*},\,\mathcal{O}_{5H^*}'$ within the two $\rm{F}_{4(4)}$-orbits $\mathcal{O}_{4}$ and $\mathcal{O}_{5}$. The two $H^*$-suborbits $\bar{\mathcal{O}}'_{5 H^*},\,\hat{\mathcal{O}}'_{5H^*}$ within $\mathcal{O}_{5H^*}'.$}
\label{alpha3}
\end{table}

%alpha4

\paragraph{$\alpha^{(4)}$-label}:
\begin{eqnarray}\alpha^{(4)}&=&1\times (\pm 4)+14\times(\pm 2)+22\times (0)\\
&=&1\times(J=2)\oplus 13\times (J=1)\oplus 8\times (J=0).\end{eqnarray} The $\gamma-\beta$ labels read
\begin{eqnarray} \gamma^{(1)}&=&1\times (\pm 4)+22\times (0)=\beta^{(1)}\\
\gamma^{(2)}&=&1\times (\pm 4)+4\times (\pm 2)+14\times (0)=\beta^{(2)}\\
\gamma^{(3)}&=&7\times (\pm 2)+10\times (0)=\beta^{(3)}\end{eqnarray}
corresponding, respectively, to $(0,0,0,4),$ $(2,0,0,0)$ and $(0,0,2,2)$ labels. The associated orbits and their representatives are listed in Table \ref{alpha4}.

%table(alpha4)

\begin{sidewaystable}

\scalefont{1}
\begin{tabular}{|c|c|c|c|c|}\hline
\backslashbox{$\gamma$-label}{$\beta$-label}
&(0,\,0,\,0,\,4)&(2,\,0,\,0,\,0)&(0,\,0,\,2,\,2)&\null\\ \hline

(0,\,0,\,0,\,4)
& $\renewcommand{\arraystretch}{0.5}
\begin{array}{c}
-H_{1}-H_{2}- K_{14}\\ \\+ K_{3}- K_{4}+ K_{11}\end{array}$
&
$\renewcommand{\arraystretch}{0.5}
\begin{array}{c}
 K_{12}+ K_{2}- K_{1}+\\ \\  K_{13}+\sqrt{2}( K_{24}+ K_{20})\end{array}$
&$\renewcommand{\arraystretch}{0.5}
\begin{array}{c}
\frac{1}{4}(-3H_{1}-H_{2}+H_{3}+H_{4})\\ \\ +\frac{1}{2}( K_{3}-2 K_{12}- K_{5}+ K_{15}\\ \\-2 K_{4}+ K_{11}+ K_{1}-2 K_{6}\\ \\- K_{13})\\ \\+\frac{1}{\sqrt{2}}( K_{21}+ K_{24}+ K_{17}+ K_{20})\end{array}$
&$\mathcal{O}_{\cdot H^*}$\\ \hline

(2,\,0,\,0,\,0)
&$\renewcommand{\arraystretch}{0.5}
\begin{array}{c}
 K_{12}- K_{5}- K_{2}\\ \\+ K_{15}+2  K_{6}\end{array}$
&

\begin{tabular}{|c|c|c|}\hline
  $\delta_1$& $\renewcommand{\arraystretch}{1}
\begin{array}{c} 2 K_{16}+ K_{12}+\\  K_{5}- K_{2}- K_{15}
 \end{array}$ &$\bar{ \mathcal{O}}'_{.H^*}$\\ \hline
$\delta_2$ & $\renewcommand{\arraystretch}{1}
\begin{array}{c}- K_{12}- K_{2}- K_{1}\\ + K_{13}+\sqrt{2}( K_{24}+ K_{20})
 \end{array}$ & $\hat{ \mathcal{O}}'_{.H^*}$ \\ \hline
\end{tabular}

&$\renewcommand{\arraystretch}{0.5}
\begin{array}{c}
 K_{16}+ K_{5}- K_{15}\\ \\- K_{6}+\sqrt{2}( K_{24}+ K_{20})\end{array}$
&$\mathcal{O'}_{\cdot H^*}$\\ \hline

(0,\,0,\,2,\,2)
& $\renewcommand{\arraystretch}{1}
\begin{array}{c} K_{12}- K_{5}- K_{2}\\ \\+ K_{15}- K_{1}+ K_{6}\\ \\+K_{16}+ K_{13}
\end{array}$
& $\renewcommand{\arraystretch}{0.5}
\begin{array}{c}
- K_{16}- K_{12}- K_{5}\\ \\- K_{2}- K_{15}- K_{1}\\  \\+ K_{6}- K_{13}\end{array}$
&

\begin{tabular}{|c|c|c|}\hline
  $\delta_1$& $\renewcommand{\arraystretch}{1}
\begin{array}{c} -2( K_{16}+ K_{12})
 \end{array}$ &$\bar{ \mathcal{O}}''_{.H^*}$\\ \hline
$\delta_2$ & $\renewcommand{\arraystretch}{1}
\begin{array}{c}-2( K_{5}+ K_{1})
 \end{array}$ & $\hat{ \mathcal{O}}''_{.H^*}$ \\ \hline
\end{tabular}

&$\mathcal{O''}_{\cdot H^*}$\\ \hline
\null  & $\mathcal{O}_{6}$& $\mathcal{O}_{7}$&$\mathcal{O}_{8}$&\null\\ \hline
\end{tabular}
\caption{\small The nine $H^*$-orbits $\mathcal{O}_{6 H^*},\,\mathcal{O}_{6H^*}',\,\mathcal{O}_{6H^*}'' ,\,\mathcal{O}_{7H^*},\,\mathcal{O}_{7H^*}',\, \mathcal{O}_{7H^*}'' \,\mathcal{O}_{8 H^*},\,\mathcal{O}_{8H^*}',\, \mathcal{O}_{8H^*}'' $ within the three $\rm{F}_{4(4)}$-orbits $\mathcal{O}_{6},$ $\mathcal{O}_{7}$  and $\mathcal{O}_{8}$.The four $H^*$-suborbits $\bar{\mathcal{O}}'_{7 H^*},\,\hat{\mathcal{O}}'_{7 H^*},\,\bar{\mathcal{O}}''_{8 H^*} ,\,\hat{\mathcal{O}}''_{8 H^*}$  within the two $H^*$-orbits $\mathcal{O}'_{7H^*}$ and $\mathcal{O}''_{8H^*}$.}
\label{alpha4}
\end{sidewaystable}

%alpha5

\paragraph{$\alpha^{(5)}$-label}:
\begin{eqnarray}\alpha^{(5)}&=&7\times (\pm 4)+8\times (\pm2)+22\times (0)\\
&=&7\times (J=2)\oplus 1\times (J=1)\oplus 14\times (J=0),\end{eqnarray}
and
\begin{equation}\gamma=3\times (\pm 4)+4\times (\pm 2)+10\times (0)=\beta,\end{equation}
corresponding to $(0,2,0,0)$ label. The associated orbit  and its representative is listed in Table \ref{alpha5}.

%table(alpha5)
\begin{table}[htbp]
\scalefont{1}
\begin{center}
\begin{tabular}{|c|c|c|c|}\hline
\backslashbox{$\gamma$-label}{$\beta$-label}
&(0,\,2,\,0,\,0)&\null\\ \hline
(0,\,2,\,0,\,0)&
$\renewcommand{\arraystretch}{1}
\begin{array}{c}
-2( K_{21}+ K_{22}- K_{17}+ K_{18})\end{array}$&$\mathcal{O}_{\cdot H^*}$\\ \hline

\null  & $\mathcal{O}_{9}$&\null\\ \hline
\end{tabular}
\caption{\small The $H^*$-orbit $\mathcal{O}_{9 H^*}$ within the  $\rm{F}_{4(4)}$-orbits $\mathcal{O}_{9}.$}
\label{alpha5}
\end{center}
\end{table}

%alpha6

\paragraph{$\alpha^{(6)}$-label}:
\begin{eqnarray}\alpha^{(6)}&=&3\times (\pm 4)+2\times (\pm 3)+9\times (\pm 2)+6\times (\pm 1)+12\times(0)\\
&=&3\times (J=2)\oplus 2 \times \Big(J=\frac{3}{2}\Big)\oplus 6\times (J=1)\oplus 4\times \Big(J=\frac{1}{2}\Big)\oplus 3\times (J=0),\end{eqnarray}
and
\begin{equation}\gamma=1\times (\pm 4)+1 \times (\pm 3)+4 \times (\pm 2)+3\times (\pm 1)+6\times (0)=\beta,\end{equation}
corresponding to $(1,1,0,2)$ label. The associated orbit  and its representative is listed in Table \ref{alpha6}.

%table(alpha6)
\begin{table}[htbp]
\scalefont{1}
\begin{center}
\begin{tabular}{|c|c|c|c|}\hline
\backslashbox{$\gamma$-label}{$\beta$-label}
&(1,\,1,\,0,\,2)&\null\\ \hline
(1,\,1,\,0,\,2)&
$\renewcommand{\arraystretch}{1}
\begin{array}{c}
-\frac{1}{\sqrt{2}}(H_{2}+H_{3}+ K_{16}+2 K_{14}\\+ K_{12}+ K_{5}+ K_{2}- K_{15}\\+2 K_{22}+ K_{1}- K_{6}- K_{13}+2 K_{18})\end{array}$&$\mathcal{O}_{\cdot H^*}$\\ \hline

\null  & $\mathcal{O}_{10}$&\null\\ \hline
\end{tabular}
\caption{\small The $H^*$-orbit $\mathcal{O}_{10 H^*}$ within the  $\rm{F}_{4(4)}$-orbits $\mathcal{O}_{10}.$}
\label{alpha6}
\end{center}
\end{table}

%\alpha7
\paragraph{$\alpha^{(7)}$-label}:
\begin{eqnarray}\alpha^{(7)}&=&1\times (\pm 6)+5\times (\pm 4)+4\times (\pm 3)+6 \times (\pm 2)+4\times (\pm 1)+12 \times (0)\\
&=&1\times (J=3)\oplus 4\times (J=2)\oplus 4\times \Big(J=\frac{3}{2}\Big)\oplus 1\times (J=1)\oplus 6\times (J=0).\end{eqnarray}
The $\gamma-\beta$-labels are written as
\begin{eqnarray}\gamma^{(1)}&=&2\times (\pm 4)+2\times (\pm 3)+2 \times (\pm 2)+2\times (\pm 1)+6\times (0)=\beta^{(1)}\\
\gamma^{(2)}&=&3\times (\pm 4)+2 \times (\pm 3)+2 \times (\pm 2)+2\times (\pm 1)+6\times (0)=\beta^{(2)}\end{eqnarray}
corresponding, respectively, to $(1,0,2,4)$ and $(0,1,2,2).$ The associated orbits and their representatives are listed in Table \ref{alpha7}.

%table(alpha7)

\begin{table}[htbp]
\scalefont{.9}
\begin{tabular}{|c|c|c|c|}\hline
\backslashbox{$\gamma$-label}{$\beta$-label}
&(1,\,0,\,2,\,4)&(0,\,1,\,2,\,2)&\null\\ \hline

(1,\,0,\,2,\,4)
& $\renewcommand{\arraystretch}{1}
\begin{array}{c}
-H_{2}-H_{3}-2 K_{14}-\\\sqrt{3}( K_{5}- K_{15}+ K_{1}- K_{13})\end{array}$
&$\renewcommand{\arraystretch}{1}
\begin{array}{c}
-H_{2}-H_{3}-2  K_{14}-\\ \sqrt{3}( K_{16}+ K_{12}+ K_{2}- K_{6})\end{array}$
 &$\mathcal{O}_{\cdot H^*}$\\ \hline

(0,\,1,\,2,\,2)
& $\renewcommand{\arraystretch}{1}
\begin{array}{c}
-H_{1}-H_{4}-2 K_{11}+\\ \sqrt{3}( K_{16}+ K_{12}- K_{5}- K_{1})\end{array}$
& $\renewcommand{\arraystretch}{1}
\begin{array}{c}
-H_{1}-H_{4}-2 K_{11}\\-\sqrt{3}( K_{2}+ K_{15}- K_{6}+ K_{13})\end{array}$
&$\mathcal{O'}_{\cdot H^*}$\\ \hline

\null  & $\mathcal{O}_{11}$& $\mathcal{O}_{12}$&\null\\ \hline
\end{tabular}
\caption{\small The four $H^*$-orbits $\mathcal{O}_{11 H^*},\,\mathcal{O}_{11 H^*}',\,\mathcal{O}_{12 H^*},\,\mathcal{O}_{12 H^*}'$ within the two $\rm{F}_{4(4)}$-orbits $\mathcal{O}_{11}$ and $\mathcal{O}_{12}$.}
\label{alpha7}
\end{table}

%alpha8
\paragraph{$\alpha^{(8)}$-label}:
\begin{eqnarray}\alpha^{(8)}&=&2\times (\pm 5)+3 \times (\pm 4)+4\times (\pm 3)+5 \times (\pm 2)+8\times (\pm 1)+8\times (0)\\
&=&2\times \Big(J=\frac{5}{2}\Big)\oplus (J=2)\oplus 2\times \Big(J=\frac{3}{2}\Big)\oplus 2 \times (J=1)\oplus 4\times \Big(J=\frac{1}{2}\Big)\oplus 3\times (J=0),\nonumber\\\end{eqnarray}
and
\begin{equation}\gamma=1\times (\pm 5)+1\times (\pm 4)+2 \times (\pm 3)+2\times (\pm 2)+4\times (\pm 1)+4\times (0)=\beta\end{equation}
corresponding to $(1,1,1,1)$ label. The associated orbit and its representative is presented in Table \ref{alpha8}.

%table(alpha8)
\begin{table}[htbp]
\scalefont{1}
\begin{center}
\begin{tabular}{|c|c|c|c|}\hline
\backslashbox{$\gamma$-label}{$\beta$-label}
&(1,\,1,\,1,\,1)&\null\\ \hline
(1,\,1,\,1,\,1)&
$\renewcommand{\arraystretch}{1}
\begin{array}{c}
\frac{1}{2}( K_{16}+ K_{12}- K_{2}- K_{15}- K_{1}+ K_{6}- K_{13}-K_{5})\\-2( K_{21}+ K_{22}- K_{17}+ K_{18})\end{array}$&$\mathcal{O}_{\cdot H^*}$\\ \hline

\null  & $\mathcal{O}_{13}$&\null\\ \hline
\end{tabular}
\caption{\small The $H^*$-orbit $\mathcal{O}_{13 H^*}$ within the  $\rm{F}_{4(4)}$-orbits $\mathcal{O}_{13}.$}
\label{alpha8}
\end{center}
\end{table}

%alpha9
\paragraph{$\alpha^{(9)}$-label}:
\begin{eqnarray}\alpha^{(9)}&=&1\times (\pm 6)+2\times (\pm 5)+2\times (\pm 4)+6\times (\pm 3)+5\times (\pm 2)+6\times (\pm 1)+8\times (0)\\
&=&1\times (J=3)\oplus 2\times \Big(J=\frac{5}{2}\Big)\oplus 1\times (J=2)\oplus 4\times \Big(J=\frac{3}{2}\Big)\oplus 3\times (J=1)\oplus 3\times (J=0),\nonumber\\ \end{eqnarray}
and
\begin{eqnarray}\gamma^{(1)}&=&1\times (\pm 5)+2\times (\pm 4)+3\times (\pm 3)+3\times (\pm 1)+6\times (0)=\beta^{(1)}\\
\gamma^{(2)}&=&1\times (\pm 5)+1\times (\pm 4)+3 \times (\pm 3)+2 \times (\pm 2)+3 \times (\pm 1)+4\times (0)=\beta^{(2)}\end{eqnarray}
corresponding, respectively, to $(1,0,3,1)$ and $(1,1,1,3)$ labels. The associated orbits and their representatives are listed in Table \ref{alpha9}.

%table(alpha9)

\begin{table}[htbp]
\scalefont{.9}
\begin{tabular}{|c|c|c|c|}\hline
\backslashbox{$\gamma$-label}{$\beta$-label}
&(1,\,0,\,3,\,1)&(1,\,1,\,1,\,3)&\null\\ \hline

(1,\,0,\,3,\,1)
& $\renewcommand{\arraystretch}{0.5}
\begin{array}{c}
-\frac{1}{2}(H_{1}-2H_{2}+2H_{3}+H_{4})\\+2 K_{3}- K_{11}-\sqrt{6}( K_{22}+ K_{18})\end{array}$
&$\renewcommand{\arraystretch}{0.5}
\begin{array}{c}
-\frac{1}{2}(H_{1}+2H_{2}-2H_{3}+H_{4})\\-2 K_{3}- K_{11}-\sqrt{6}( K_{22}+ K_{18})\end{array}$
 &$\mathcal{O}_{\cdot H^*}$\\ \hline

(1,\,1,\,1,\,3)
& $\renewcommand{\arraystretch}{0.5}
\begin{array}{c}
-\frac{1}{2}(2H_{1}+H_{2}+H_{3}+2H_{4})\\- K_{14}-2 K_{11}-\sqrt{6}( K_{7}- K_{10})\end{array}$
& $\renewcommand{\arraystretch}{0.5}
\begin{array}{c}
-\frac{1}{2}(2H_{1}-H_{2}-H_{3}+2H_{4})\\+ K_{14}-2 K_{11}-\sqrt{6}( K_{7}- K_{10})\end{array}$
&$\mathcal{O'}_{\cdot H^*}$\\ \hline

\null  & $\mathcal{O}_{14}$& $\mathcal{O}_{15}$&\null\\ \hline
\end{tabular}
\caption{\small The four $H^*$-orbits $\mathcal{O}_{14 H^*},\,\mathcal{O}_{14 H^*}',\,\mathcal{O}_{15 H^*},\,\mathcal{O}_{15 H^*}'$ within the two $\rm{F}_{4(4)}$-orbits $\mathcal{O}_{14}$ and $\mathcal{O}_{15}$.}
\label{alpha9}
\end{table}

%\alpha10
\paragraph{$\alpha^{(10)}$-label}:
\begin{eqnarray}\alpha^{(10)}&=&2\times (\pm 6)+6\times (\pm 4)+12 \times (\pm 2)+12\times (0)\\
&=&2\times (J=3)\oplus 4\times (J=2)\oplus 6\times (J=1),\end{eqnarray}
The $\gamma-\beta$ labels are given by
\begin{eqnarray}\gamma^{(1)}&=&6\times (\pm 4)+12\times (0)=\beta^{(1)}\\
\gamma^{(2)}&=&4\times (\pm 4)+4 \times (\pm 2)+8\times (0)=\beta^{(2)}\\
\gamma^{(3)}&=&1\times (\pm 6)+2\times (\pm 4)+6\times (\pm 2)+6 \times (0)=\beta^{(3)},\end{eqnarray}
corresponding, respectively, to  $(0,0,4,0),$ $(0,2,0,4)$ and $(2,0,2,2).$ The associated orbits and their representatives are listed in Table \ref{alpha10}.

%table{alpha10}
\begin{sidewaystable}
\scalefont{.83}

%\begin{table}[htbp]

\begin{tabular}{|c|c|c|c|c|}\hline
\backslashbox{$\gamma$-label}{$\beta$-label}
&(0,\,0,\,4,\,0)&(0,\,2,\,0,\,4)&(2,\,0,\,2,\,2)&\null\\ \hline

(0,\,0,\,4,\,0)
& $\renewcommand{\arraystretch}{0.5}
\begin{array}{c}
-H_{1}+H_{2}-H_{4}\\- K_{14}+ K_{3}-2 K_{11}\\-\sqrt{3}( K_{2}+ K_{15}- K_{6}+ K_{13})\end{array}$
&
$\renewcommand{\arraystretch}{0.5}
\begin{array}{c}
-\sqrt{3}H_{3}+2 K_{16}+2 K_{12}\\+\sqrt{3}( K_{14}+ K_{3})- K_{2}\\+ K_{15}+ K_{6}+ K_{13}\end{array}$
&$\renewcommand{\arraystretch}{0.5}
\begin{array}{c}
-\sqrt{3}H_{3}+\sqrt{3}( K_{14}+ K_{3})\\- K_{16}- K_{12}- K_{5}+ K_{2}\\
- K_{15}- K_{1}- K_{6}- K_{13}\\-\sqrt{2}( K_{7}- K_{10})\end{array}$
&$\mathcal{O}_{\cdot H^*}$\\ \hline

(0,\,2,\,0,\,4)
&$\renewcommand{\arraystretch}{0.5}
\begin{array}{c}
-H_{1}-H_{2}-H_{4}\\- K_{14}+ K_{3}-2 K_{11}\\-\sqrt{3}( K_{2}+ K_{15}- K_{6}+ K_{13})\end{array}$

& \begin{tabular}{|c|c|c|}\hline
  $\delta_1$& $\renewcommand{\arraystretch}{1}
\begin{array}{c} -H_{1}-H_{4}-2 K_{11}\\+\sqrt{3}( K_{16}+ K_{12}- K_{5}\\- K_{1})+\sqrt{2}( K_{8}+ K_{9})
 \end{array}$ &$\bar{ \mathcal{O}}'_{.H^*}$\\ \hline
$\delta_2$ & $\renewcommand{\arraystretch}{1}
\begin{array}{c}-H_{1}+H_{2}-H_{4}\\+ K_{14}- K_{3}-2 K_{11}-\\ \sqrt{3}( K_{2}+ K_{15}- K_{6}+ K_{13})
 \end{array}$ & $\hat{ \mathcal{O}}'_{.H^*}$ \\ \hline
\end{tabular}

&$\renewcommand{\arraystretch}{0.5}
\begin{array}{c}
-H_{1}-H_{4}-2H_{12}\\-\sqrt{3}( K_{2}+ K_{15}- K_{6}\\+ K_{13})+\sqrt{2}( K_{8}+ K_{9})\end{array}$
&$\mathcal{O'}_{\cdot H^*}$\\ \hline

(2,\,0,\,2,\,2)
& $\renewcommand{\arraystretch}{0.5}
\begin{array}{c}
-\frac{1}{2}(H_{1}-3H_{2}+H_{3}\\ +H_{4})+ K_{14}+2 K_{3}\\- K_{11}-\sqrt{6}( K_{22}+ K_{18})\end{array}$
&$\renewcommand{\arraystretch}{0.5}
\begin{array}{c}
-\frac{1}{2}(H_{1}+H_{2}-3H_{3}\\+H_{4})+ K_{14}-2 K_{3}\\- K_{11}-\sqrt{6}( K_{22}+ K_{18})\end{array}$
&

\begin{tabular}{|c|c|c|}\hline
  $\delta_1$& $\renewcommand{\arraystretch}{1}
\begin{array}{c} -\frac{1}{2}(H_{1}+3H_{2}-H_{3}\\+H_{4})- K_{14}-2 K_{3}\\- K_{11}-\sqrt{6}( K_{22}+ K_{18})
 \end{array}$ &$\bar{ \mathcal{O}}''_{.H^*}$\\ \hline
$\delta_2$ & $\renewcommand{\arraystretch}{1}
\begin{array}{c}-\frac{1}{2}(H_{1}-H_{2}+3H_{3}\\+H_{4})- K_{14}+2 K_{3}\\- K_{11}-\sqrt{6}( K_{22}+ K_{18})
 \end{array}$ & $\hat{ \mathcal{O}}''_{.H^*}$ \\ \hline
\end{tabular}

&$\mathcal{O''}_{\cdot H^*}$\\ \hline

\null  & $\mathcal{O}_{16}$& $\mathcal{O}_{17}$&$\mathcal{O}_{18}$&\null\\ \hline
\end{tabular}
\caption{\small The nine $H^*$-orbits $\mathcal{O}_{16 H^*},\,\mathcal{O}_{16H^*}',\,\mathcal{O}_{16H^*}'' ,\,\mathcal{O}_{17H^*},\,\mathcal{O}_{17H^*}',\, \mathcal{O}_{17H^*}'' \,\mathcal{O}_{18 H^*},\,\mathcal{O}_{18H^*}',\, \mathcal{O}_{18H^*}'' $ within the three $\rm{F}_{4(4)}$-orbits $\mathcal{O}_{16},$ $\mathcal{O}_{17}$  and $\mathcal{O}_{18}$.The four $H^*$-suborbits $\bar{\mathcal{O}}'_{17 H^*},\,\hat{\mathcal{O}}'_{17 H^*},\,\bar{\mathcal{O}}''_{18 H^*} ,\,\hat{\mathcal{O}}''_{18 H^*}$  within the two $H^*$-orbits $\mathcal{O}'_{17H^*}$ and $\mathcal{O}''_{18H^*}$}
\label{alpha10}
%\end{table}
\end{sidewaystable}

%alpha11
\paragraph{$\alpha^{(11)}$-label}:
\begin{eqnarray}\alpha^{(11)}&=&1\times (\pm 10)+1\times (\pm 8)+6\times (\pm6)+6 \times (\pm 4)+7\times (\pm 2)+10\times (0)\\
&=&1\times (J=5)\oplus 5\times (J=3)\oplus 1\times (J=1)\oplus 3\times (J=0).\end{eqnarray}
The $\gamma-\beta$ labels are
\begin{eqnarray}\gamma^{(1)}&=&1\times (\pm 8)+6\times (\pm 4)+10\times (0)=\beta^{(1)}\\
\gamma^{(2)}&=&1\times (\pm 8)+2 \times (\pm 6)+4\times (\pm 4)+2 \times (\pm 2)+6\times (0)=\beta^{(2)},\end{eqnarray}
corresponding, respectively, to  $(0,0,4,8),$ and $(2,0,4,4).$ The associated orbits and their representatives are listed in Table \ref{alpha11}.

%table(alpha11)

\begin{table}[htbp]
\scalefont{0.75}
\begin{tabular}{|c|c|c|c|}\hline
\backslashbox{$\gamma$-label}{$\beta$-label}
&(0,\,0,\,4,\,8)&(2,\,0,\,4,\,4)&\null\\ \hline

(0,\,0,\,4,\,8)
& $\renewcommand{\arraystretch}{0.5}
\begin{array}{c}
-\sqrt{\frac{3}{2}}(H_1+2H_{2}+H_{4})\\-\sqrt{6}( K_{14}- K_{3}- K_{11})\\-\sqrt{\frac{5}{2}}( K_{16}+ K_{12}+ K_{5}+ K_{2}\\- K_{15}+ K_{1}- K_{6}- K_{13})\end{array}$
&$\renewcommand{\arraystretch}{0.5}
\begin{array}{c}
-\sqrt{\frac{3}{2}}(H_1-2H_{2}+H_{4})\\+\sqrt{6}( K_{14}- K_{3}+ K_{11})\\-\sqrt{\frac{5}{2}}( K_{16}+ K_{12}+ K_{5}+ K_{2}\\- K_{15}+ K_{1}- K_{6}- K_{13})\end{array}$
 &$\mathcal{O}_{\cdot H^*}$\\ \hline

(2,\,0,\,4,\,4)
& $\renewcommand{\arraystretch}{0.5}
\begin{array}{c}
-\sqrt{\frac{3}{2}}(H_1-2H_{3}+H_{4})\\+\sqrt{6}( K_{14}- K_{3}- K_{11})\\+\sqrt{\frac{5}{2}}( K_{16}+ K_{12}- K_{5}+ K_{2}\\+ K_{15}- K_{1}- K_{6}+ K_{13})\end{array}$
&

\begin{tabular}{|c|c|c|}\hline
  $\delta_1$& $\renewcommand{\arraystretch}{1}
\begin{array}{c} -\sqrt{\frac{3}{2}}(H_{2}+H_{3})-\sqrt{6} K_{14}\\-2\sqrt{3}( K_{21}- K_{17})+\\\sqrt{\frac{5}{2}}( K_{16}+ K_{12}- K_{5}+ K_{2}\\+ K_{15}- K_{1}- K_{6}+ K_{13})
 \end{array}$ &$\bar{ \mathcal{O}}'_{.H^*}$\\ \hline
$\delta_2$ & $\renewcommand{\arraystretch}{1}
\begin{array}{c}-\sqrt{\frac{3}{2}}(H_{1}+2H_{2}+H_{4})\\-\sqrt{6}( K_{14}+ K_{3}+ K_{11})+\\\sqrt{\frac{5}{2}}( K_{16}+ K_{12}- K_{5}+ K_{2}\\+ K_{15}- K_{1}- K_{6}+ K_{13})
 \end{array}$ & $\hat{ \mathcal{O}}'_{.H^*}$ \\ \hline
\end{tabular}

&$\mathcal{O'}_{\cdot H^*}$\\ \hline

\null  & $\mathcal{O}_{19}$& $\mathcal{O}_{20}$&\null\\ \hline
\end{tabular}
\caption{\small The four $H^*$-orbits $\mathcal{O}_{19 H^*},\,\mathcal{O}_{19 H^*}',\,\mathcal{O}_{20 H^*},\,\mathcal{O}_{20 H^*}'$ within the two $\rm{F}_{4(4)}$-orbits $\mathcal{O}_{19}$ and $\mathcal{O}_{20}$.The two $H^*$-suborbits $\bar{\mathcal{O}}'_{20 H^*},\,\hat{\mathcal{O}}'_{20H^*}$ within $\mathcal{O}_{20H^*}'.$}
\label{alpha11}
\end{table}

%alpha12
\paragraph{$\alpha^{(12)}$-label}:
\begin{eqnarray}\alpha^{(12)}&=&1\times (\pm 10)+2\times (\pm 9)+1\times (\pm 8)+2\times (\pm 7)+2\times (\pm 6)\nonumber\\&&+2\times (\pm 5)+2\times (\pm 4))+4 \times (\pm 3)+
3\times (\pm 2)+4\times (\pm 1)+6\times (0)\nonumber\\
&=&1\times (J=5)\oplus 2\times \Big(J=\frac{9}{2}\Big)\oplus 1\times (J=3)\oplus 2\times \Big(J=\frac{3}{2}\Big)\oplus 1\times (J=1)\oplus 3\times (J=0)\nonumber\\\end{eqnarray}
and
\begin{equation}\gamma=1\times (\pm 9)+1\times (\pm 8)+1 \times (\pm 7)+1\times (\pm 5)+2\times (\pm 4)+2 \times (\pm 3)+2\times (\pm 1)+4\times (0)=\beta\end{equation}
corresponding to $(1,3,1,3)$ label. The associated orbit and its representative is given in Table \ref{alpha12}.

%table(alpha12)

\begin{table}[htbp]
\scalefont{1}
\begin{center}
\begin{tabular}{|c|c|c|c|}\hline
\backslashbox{$\gamma$-label}{$\beta$-label}
&(1,\,3,\,1,\,3)&\null\\ \hline
(1,\,3,\,1,\,3)&
$\renewcommand{\arraystretch}{1}
\begin{array}{c}
-\frac{3}{2}(H_{2}+H_{3})-3 K_{14}+\\
4( K_{23}- K_{19})-\sqrt{10}( K_{7}- K_{10})\end{array}$&$\mathcal{O}_{\cdot H^*}$\\ \hline

\null  & $\mathcal{O}_{21}$&\null\\ \hline
\end{tabular}
\caption{\small The $H^*$-orbit $\mathcal{O}_{21 H^*}$ within the  $\rm{F}_{4(4)}$-orbits $\mathcal{O}_{21}.$}
\label{alpha12}
\end{center}
\end{table}

%alpha13
\paragraph{$\alpha^{(13)}$-label}:
\begin{eqnarray}\alpha^{(13)}&=&2\times (\pm 10)+3\times (\pm 8)+4\times (\pm 6)+5\times (\pm 4)+8\times (\pm 2)+8\times (0)\\
&=&2\times (J=5)\oplus 1\times (J=4)\oplus 1\times (J=3)\oplus 1\times (J=2)\oplus 3\times (J=1)\nonumber\\ \end{eqnarray}
The $\gamma-\beta$ take the values
\begin{eqnarray}\gamma^{(1)}&=&3\times (\pm 8)+5\times (\pm 4)+8\times (0)=\beta^{(1)}\\
\gamma^{(2)}&=&1\times (\pm 10)+1\times (\pm 8)+2 \times (\pm 6)+2 \times (\pm 4)+4\times (\pm 2)+4\times (0)=\beta^{(2)},\end{eqnarray}
corresponding, respectively, to $(0,4,0,4)$ and $(2,2,2,2)$ labels. The associated orbits and their representatives are given in Table \ref{alpha13}.

%\alpha(table13)

\begin{table}[htbp]
\scalefont{.85}
\begin{tabular}{|c|c|c|c|}\hline
\backslashbox{$\gamma$-label}{$\beta$-label}
&(0,\,4,\,0,\,4)&(2,\,2,\,2,\,2)&\null\\ \hline

(0,\,4,\,0,\,4)
& $\renewcommand{\arraystretch}{0.5}
\begin{array}{c}
2( K_{16}+ K_{12}- K_{5}+2 K_{23}\\- K_{1}-2 K_{19})- K_{2}- K_{15}+\\ K_{6}- K_{13}+\sqrt{10}( K_{8}+ K_{9})\end{array}$
&$\renewcommand{\arraystretch}{0.5}
\begin{array}{c}
 K_{16}+ K_{12}- K_{5}- K_{1}+\\+2(2 K_{23}- K_{2}- K_{15}+ K_{6}+\\- K_{13}-2 K_{19})+\sqrt{10}( K_{8}+ K_{9})\end{array}$
 &$\mathcal{O}_{\cdot H^*}$\\ \hline

(2,\,2,\,2,\,2)
& $\renewcommand{\arraystretch}{0.5}
\begin{array}{c}
-H_{2}-2H_{3}-3 K_{14}\\+ K_{3}+4( K_{23}- K_{19})\\-\sqrt{10}( K_{7}- K_{10})\end{array}$
& $\renewcommand{\arraystretch}{0.5}
\begin{array}{c}
-2H_{2}-H_{3}-3 K_{14}\\- K_{3}+4( K_{23}- K_{19})\\-\sqrt{10}( K_{7}- K_{10})\end{array}$
&$\mathcal{O'}_{\cdot H^*}$\\ \hline

\null  & $\mathcal{O}_{22}$& $\mathcal{O}_{23}$&\null\\ \hline
\end{tabular}
\caption{\small The four $H^*$-orbits $\mathcal{O}_{22 H^*},\,\mathcal{O}_{22 H^*}',\,\mathcal{O}_{23 H^*},\,\mathcal{O}_{23 H^*}'$ within the two $\rm{F}_{4(4)}$-orbits $\mathcal{O}_{22}$ and $\mathcal{O}_{23}$.}
\label{alpha13}
\end{table}

%alpa14
\paragraph{$\alpha^{(14)}$-label}:
\begin{eqnarray}\alpha^{(14)}&=&1\times (\pm 14)+1\times (\pm 12)+3\times (\pm 10)+3\times (\pm 8)+4\times (\pm 6)+5\times (\pm 4)\nonumber\\&&+6\times (\pm2)+6\times (0)\\
&=&1\times (J=7)\oplus 2\times (J=5)\oplus 1\times (J=3)\oplus 1\times (J=2)\oplus 1\times (J=1).\end{eqnarray}
The $\gamma-\beta$ labels read
\begin{eqnarray}\gamma^{(1)}&=&1\times (\pm 12)+1\times (\pm 10)+2\times (\pm 8)+1\times (\pm 6)+3\times (\pm 4)\nonumber\\&&+2\times (\pm 2)+4\times (0)=\beta^{(1)}\\
\gamma^{(2)}&=&1\times (\pm 12)+3\times (\pm 8)+5\times (\pm 4)+6\times (0)=\beta^{(2)},\end{eqnarray}
corresponding, respectively, to $(2,2,4,4)$ and $(4,0,4,8)$ labels. The associated orbits and their representatives are given in Table \ref{alpha14}.

%table(alpha14)
\begin{table}[htbp]
\scalefont{.85}
\begin{tabular}{|c|c|c|c|}\hline
\backslashbox{$\gamma$-label}{$\beta$-label}
&(2,\,2,\,4,\,4)&(4,\,0,\,4,\,8)&\null\\ \hline

(2,\,2,\,4,\,4)
& $\renewcommand{\arraystretch}{0.5}
\begin{array}{c}
-\frac{1}{\sqrt{2}}(5H_{2}+H_{3})-\sqrt{2}(3 K_{14}+\\2 K_{3})+\sqrt{\frac{7}{2}}( K_{16}+ K_{12}- K_{5}+\\
 K_{2}+ K_{15}- K_{1}- K_{6}\\+ K_{13})+2\sqrt{5}( K_{23}- K_{19})\end{array}$
&$\renewcommand{\arraystretch}{0.5}
\begin{array}{c}
-\frac{1}{\sqrt{2}}(H_{2}+5H_{3})-\sqrt{2}(3 K_{14}-\\2 K_{3})+\sqrt{\frac{7}{2}}( K_{16}+ K_{12}- K_{5}+\\
 K_{2}+ K_{15}- K_{1}- K_{6}\\+ K_{13})+2\sqrt{5}( K_{23}- K_{19})\end{array}$
 &$\mathcal{O}_{\cdot H^*}$\\ \hline

(4,\,0,\,4,\,8)
& $\renewcommand{\arraystretch}{0.5}
\begin{array}{c}
-\frac{1}{2\sqrt{2}}(H_{1}+H_{2}-H_{3}+H_{4})\\-5( K_{21}- K_{17})+\sqrt{\frac{7}{2}}( K_{16}+ K_{12}\\- K_{5}+ K_{2}+ K_{15}- K_{1}- K_{6}\\+ K_{13})-\frac{1}{\sqrt{2}}( K_{3}+ K_{11})-\\
\sqrt{10}( K_{8}+ K_{23}+ K_{9}+ K_{19})\end{array}$
& $\renewcommand{\arraystretch}{0.5}
\begin{array}{c}
-\frac{5}{2\sqrt{2}}(H_{1}+H_{2}-H_{3}+H_{4})\\-( K_{21}- K_{17})+\sqrt{\frac{7}{2}}( K_{16}+ K_{12}\\- K_{5}+ K_{2}+ K_{15}- K_{1}- K_{6}\\+ K_{13})-\frac{5}{\sqrt{2}}( K_{3}+ K_{11})-\\
\sqrt{10}( K_{8}+ K_{23}+ K_{9}+ K_{19})\end{array}$
&$\mathcal{O'}_{\cdot H^*}$\\ \hline

\null  & $\mathcal{O}_{24}$& $\mathcal{O}_{25}$&\null\\ \hline
\end{tabular}
\caption{\small The four $H^*$-orbits $\mathcal{O}_{24 H^*},\,\mathcal{O}_{24 H^*}',\,\mathcal{O}_{25 H^*},\,\mathcal{O}_{25 H^*}'$ within the two $\rm{F}_{4(4)}$-orbits $\mathcal{O}_{24}$ and $\mathcal{O}_{25}$.}
\label{alpha14}
\end{table}

%alpha15
\paragraph{$\alpha^{(15)}$-label}:
\begin{eqnarray}\alpha^{(15)}&=&1\times (\pm 22)+1\times (\pm 20)+1\times (\pm 18)+1\times (\pm 16)+2 \times (\pm 14)+2 \times (\pm 12)\nonumber\\&&
+3\times (\pm 10)+3\times (\pm 8)+3\times (\pm 6)+3\times (\pm 4)+4\times (\pm 2)+4\times (0)\\
&=&1\times (J=11)\oplus 1\times (J=7)\oplus 1\times (J=5)\oplus 1\times (J=1),\end{eqnarray}
and
\begin{equation}\gamma=1\times (\pm 20)+1\times (\pm 16)+2\times (\pm 12)+3\times (\pm 8)+3\times (\pm 4)+4\times (0)=\beta,\end{equation}
corresponding to $(4,4,4,8)$. The associated orbit and its representative is presented in Table \ref{alpha15}.

%table(alpha15)
\begin{table}[htbp]
\scalefont{1}
\begin{center}
\begin{tabular}{|c|c|c|c|}\hline
\backslashbox{$\gamma$-label}{$\beta$-label}
&(4,\,4,\,4,\,8)&\null\\ \hline
(4,\,4,\,4,\,8)&
$\renewcommand{\arraystretch}{1}
\begin{array}{c}
-\sqrt{\frac{21}{2}}(H_{2}-H_{3})+4\sqrt{2}( K_{23}- K_{19})\\-\sqrt{42} K_{3}+2\sqrt{15}( K_{8}+ K_{9})+\sqrt{\frac{11}{2}}( K_{16}+ K_{12}\\- K_{5}
+ K_{2}+ K_{15}- K_{1}- K_{6}+ K_{13})\end{array}$&$\mathcal{O}_{\cdot H^*}$\\ \hline

\null  & $\mathcal{O}_{26}$&\null\\ \hline
\end{tabular}
\caption{\small The $H^*$-orbit $\mathcal{O}_{26 H^*}$ within the  $\rm{F}_{4(4)}$-orbits $\mathcal{O}_{26}.$}
\label{alpha15}
\end{center}
\end{table}

\subsubsection{Tensor Classifier Analysis}\label{tcs}
Let us introduce a set of \emph{tensor classifiers} (TC) which
proves to be a valuable tool for the classification. These are
rank-two symmetric $H^*$-tensors, constructed out of the Lax
components $\Delta^{\alpha,A}$ at radial infinity, whose signature
is used as an $H^*$-invariant feature.\par Let us introduce the
relevant quantities. We denote by $s_\alpha,\,t_x$, $\alpha=1,2,3$,
$x=1,\,\dots, 21$ the generators of the $\mathfrak{sl}(2,\mathbb{R})$
and $\mathfrak{sp}'(6,\mathbb{R})$ subalgebras of $\mathfrak{H}^*$.
Their adjoint action on the generators $K_{a,A}$, $a=1,2$,
$A=1,\dots, 14$, of $\mathfrak{K}^*$ in the ${\bf (2,{\bf 14}')}$ of
$H^*$ is defined  by the following commutation relations:
\begin{equation}
[s_\alpha,K_{a,A}]=-s_{\alpha
a}{}^b\,K_{b,A}\,\,,\,\,\,[t_x,K_{a,A}]=-t_{x A}{}^B\,K_{a,B}\,,
\end{equation}
where the matrices $(s_{\alpha})_ a{}^b,\,(t_{x})_A{}^B$, describe
the generators $s_{\alpha},\,t_x$ in the ${\bf 2}$ and ${\bf 14}'$
representations  respectively. Using the symplectic property of
these matrices, we can construct the following symmetric tensors
\begin{equation}
s_{\alpha\, ab}=s_{\alpha
a}{}^c\,\epsilon_{cb}\,\,,\,\,\,t_{x\,AB}\equiv t_{x
A}{}^C\,\mathbb{C}_{CB}\,.
\end{equation}
 We start defining now a set of tensors which are of second order in
$\Delta^{\alpha,A}$. Using the general decompositions
\begin{eqnarray} ({\bf 14}' \times {\bf 14}')_{sym.}&=& {\bf 21}+{\bf 84},\label{decs}\\
 ({\bf 14}' \times {\bf 14}')_{antisym.} &=&{\bf 1}+{\bf 90}\,,\end{eqnarray}
we see that the ${\rm SL}(2,\mathbb{R})$-singlets in the product of
two Lax components $\Delta^{a A} \Delta^{b B}$can only fall in the
representations ${\bf (1,1)}+{\bf (1,90)}$, so that we may write:
\begin{equation}
\Delta^{a A} \Delta^{b B}\,\epsilon_{ab}=
\mathcal{T}^{AB}+\mathcal{T}\,\mathbb{C}^{AB}\,.
\end{equation}
The singlet $\mathcal{T}$ is zero for all matrices $L_0$ associated
with extremal solutions, since
\begin{equation}
0=c^2\propto{\rm Tr}(L_0^2)\propto \Delta^{a A} \Delta^{b
B}\,\epsilon_{ab}\mathbb{C}_{AB}=14\,\mathcal{T}\,,
\end{equation}
where $c$ is the extremality parameter of the four dimensional
solution. From the antisymmetric tensor $\mathcal{T}^{AB}$ in the
${\bf (1,90)}$ we can construct a symmetric tensor classifier
$\mathcal{T}_{xy}$ as follows:
\begin{equation}
\mathcal{T}_{xy}=\frac{1}{2}\,t_{x\,AC}\,t_{y\,BD}\,\mathbb{C}^{CD}\,\mathcal{T}^{AB}=
\frac{1}{2}\,t_{x\,AC}\,t_{y\,BD}\,\mathbb{C}^{CD}\, \Delta^{a A}
\Delta^{b B}\epsilon_{ab}\,.
\end{equation}
The signature of this tensor, i.e. the number of positive, negative
and null eigenvalues, is an $H^*$-invariant feature which is useful
for distinguishing different orbits. This tensor has moreover an
other relevance to the study of black holes: It vanishes if and only
if the extremal solution is BPS. To show this we recall that  the
$D=3$ theory under consideration is characterized by $14$ fermionic
fields $\lambda^A$ whose supersymmetry variation on the geodesic
background is expressed in terms of the Lax components by eq.
(\ref{susytrans}). As shown in the last paragraph of Sect.
\ref{sbhg}, the existence of a residual supersymmetry is equivalent
to the property of $\Delta^{a\,A}$ to factorize:
$\Delta^{a\,A}=\epsilon^a\,\Delta^A$. This feature is in turn
equivalent to the vanishing of
$\Delta^{a,A}\Delta^{b,B}\epsilon_{ab}$ and thus of
$\mathcal{T}_{xy}$:
\begin{eqnarray}
\mathrm{SUSY}\,\,\Leftrightarrow\,\,\,\Delta^{a,A}\Delta^{b,B}\epsilon_{ab}&=&0\,\,\,\Leftrightarrow
\,\,\, \mathcal{T}_{xy}\equiv 0\,.\label{SUSY}
\end{eqnarray}
This is consistent with our last statement of Sect. \ref{sbhg}:
residual supersymmetry is a $G$-invariant feature of the geodesic
or, equivalently, an $H^*$-invariant feature of $L_0$. We have
indeed related it to the vanishing of an  $H^*$-covariant
tensor.\par We can construct other symmetric covariant matrices
which are of second order in the Lax tensor $\Delta^{a,A}$, like the
following four tensors which are symmetric in the couples $(a,A),
(b, B)$:
 \begin{equation}T_{(21)}^{a A,b B}\equiv \Delta^{a C}  \Delta^{b D}P_{(21)\,CD}{}^{AB}\,,\\
 T_{(84)}^{a A,b B}\equiv \Delta^{a C}  \Delta^{b D} P_{(84)\,CD}{}^{AB}\,,\end{equation}
 where
 \begin{equation} P_{(21)\,AB}{}^{CD}\equiv-t_{x\,AB}t^{x\,CD}\,\,;\,\,
 P_{(84)\,AB}{}^{CD}\equiv\delta_{(AB)}{}^{CD}+t_{x\,AB}t^{x\,CD}\,,\end{equation}
are the projectors onto the ${\bf 21}$ and ${\bf 84}$, respectively,
and  the adjoint indices $x,y$ of ${\rm Sp}(6,\mathbb{R})$ are
lowered and raised  using the metric $\eta_{xy}\equiv {\rm Tr}(t_x
t_y)$, proportional to the Cartan-Killing metric of the algebra.\par
Next we introduce a set of quartic tensor classifiers. To this end
we define the following quantity:
\begin{equation}
W_\alpha{}^{(AB)}\equiv s_{\alpha\,ab}\,\Delta^{a,A}\Delta^{b,B}\,,
\end{equation}
By virtue of (\ref{decs}), the representation labeled by symmetric
couple $(AB)$ can be decomposed into the ${\bf 21}+{\bf 84}$:
\begin{align}
W_{\alpha\,
x}&=W_\alpha{}^{(AB)}\,t_{x\,AB}\,\,,\,\,\,W^{(84)}{}_\alpha{}^{(AB)}=W_\alpha{}^{(CD)}\,P_{(84)\,CD}{}^{AB}\,,
\end{align}
and the following $21\times 21$, $105\times 105$ and $3\times 3$
symmetric tensors can be constructed:
\begin{align}
\mathfrak{T}_{xy}&\equiv W_{\alpha\, x}\,W_{\beta\,
y}\,\eta^{\alpha\beta}\,\,;\,\,\,\mathfrak{T}_{(84)}{}^{(AB),(CD)}\equiv
W^{(84)}{}_\alpha{}^{(AB)}\,W^{(84)}{}_\beta{}^{(CD)}\,\eta^{\alpha\beta}\,\,;\,\,\,
\mathbb{T}_{\alpha\beta}\equiv W_{\alpha\, x}\,W_{\beta\,
y}\,\eta^{xy}\,,
\end{align}
where $\eta_{\alpha\beta}\equiv {\rm Tr}(s_\alpha s_\beta)$. Let us
now define the tensor
\begin{equation}
\Gamma_{aC,\,bD}\equiv
\epsilon_{ac}\epsilon_{bd}\epsilon_{a'b'}K_{ABA'B'}K_{C'CDD'}\mathbb{C}^{B'C'}\,\Delta^{c,A}\,\,\Delta^{d,B}\,
\Delta^{a',D'}\,\Delta^{b',A'}\,,
\end{equation}
where $K_{ABCD}$ is the rank-4 totally symmetric invariant tensor in
the four-fold product of the ${\bf 14}'$:
\begin{equation}
K_{ABCD}=t_{x\,AB}t^x{}_{CD}-\frac{1}{5}\,\mathbb{C}_{A(C}\mathbb{C}_{D)B}=K_{(ABCD)}\,,
\end{equation}
in terms of which quartic ${\rm Sp}'(6,\mathbb{R})$-invariant $I_4(Q)$ of a generic vector $Q=(Q^A)$ in the ${\bf 14}'$ reads:
\begin{equation}
I_4(Q)\equiv -\frac{5}{144}\, K_{ABCD}\,Q^AQ^BQ^CQ^D\,.
\end{equation}
Using the above definitions, we introduce the following $28\times
28$  tensor classifiers:
\begin{equation}
\mathbb{T}^{(3,21)}{}_{aA,\,bB}\equiv
\Gamma_{aC,\,bD}\,P_{(21)\,AB}{}^{CD}\,\,;\,\,\,\mathbb{T}^{(3,84)}{}_{aA,\,bB}\equiv
\Gamma_{aC,\,bD}\,P_{(84)\,AB}{}^{CD}\,.
\end{equation}
The signatures of the tensors $\mathcal{T}_{xy}$, $T_{(21)}^{a A,b
B}$, $T_{(84)}^{a A,b
B}$, $\mathfrak{T}_{xy}$, $\mathbb{T}_{\alpha\beta}$, $\mathfrak{T}_{(84)}{}^{(AB),(CD)},\,
\mathbb{T}^{(3,21)}{}_{aA,\,bB}$, $\mathbb{T}^{(3,84)}{}_{aA,\,bB}$
provide a valuable tool to discriminate between the various orbits.
Although they do not represent a complete set of symmetric tensors,
they are sufficient, together with the $\gamma,\,\beta$ labels, to
classify the orbits. The order-2 tensors $T_{(21)}^{a A,b
B},\,T_{(84)}^{a A,b B}$ and the order-4 ones
$\mathbb{T}^{(3,21)}{}_{aA,\,bB},\,\mathbb{T}^{(3,84)}{}_{aA,\,bB}$
are quite important in this respect, since they allow to distinguish
distinct orbits which share the same $\gamma-\beta$ labels. They
occur in the  third, fourth, tenth and eleventh
$G^{\mathbb{C}}$-orbits. In Table \ref{orbitstc} of Appendix \ref{AA} we list for each $H^*$-orbit the signatures of the tensor classifiers.

\section{Generating Solutions}\label{generatingsol}
In \cite{Bergshoeff:2008be} and \cite{Chemissany:2009hq}  representatives of the (regular and small) single center black hole orbits
with the least number of parameters (generating solutions) were explicitly constructed in symmetric supergravities. In particular it  was shown that  these were dilatonic solutions described by null geodesics in
a characteristic submanifold $\mathcal{M}_N$ of the form\footnote{In the presence of hypermultiplets in the $\mathcal{N}=2,\,D=4$ theory, additional ${\rm SO}(1,1)$ factors will appear in the definition of $\mathcal{M}_N$, in number equal to the rank of the corresponding quaternionic manifold.}:
\begin{eqnarray}\label{genman}
\mathcal{M}_N=\left(\frac{{\rm SL}(2,\mathbb{R})}{{\rm SO}(1,1)}\right)^p=(dS_2)^p\subset \frac{G}{H^*}\,,\label{MN}
\end{eqnarray}
where $p$ is the non-compact rank of the coset $H^\star/H_c$, $H_c=\mathrm{H}'_4\times {\rm U}(1)$ being the maximal compact subgroup of $H^\star$. For our model
\begin{equation}
p=\mbox{rank}\left(\frac{H^*}{H_c}\right)=\mbox{rank}\left(\frac{{\rm SL}(2,\mathbb{R})}{{\rm U}(1)}\times \frac{G_4'}{H'_4}\right)=\mbox{rank}\left(\frac{{\rm SL}(2,\mathbb{R})}{{\rm U}(1)}\times \frac{{\rm Sp}'(6,\mathbb{R})}{{\rm U}'(3)}\right)=4\,.
\end{equation}
 The generating geodesic will be the product of geodesics inside the four $dS_2$ factors of $\mathcal{M}_N$.
  One can show \cite{Bergshoeff:2008be} that $p$ is related to the electric and magnetic charges:
In fact the normal form of the electric-magnetic charge vector with respect to the action of $H_c={\rm U}'(3)\times {\rm U}(1) $ is a $p$-charge vector. This means that, by acting by means of $H_c$ on a generic combination $Y_0^A\,K_A$ in the representation ${\bf R}'$, we can always rotate it into a subspace of dimension  $p=4$ in the coset  (\ref{MN}):
\begin{equation}
Y_0^A\,K_A \,\stackrel{H_c}{\longrightarrow }\,\, Y_0^\ell\,\mathcal{K}_\ell\,,
\end{equation}
where $\mathcal{K}_\ell=(\mathcal{T}_\ell+\eta\,\mathcal{T}_\ell^T\eta)/2$ and $\mathcal{T}_\ell$ are $p=4$ out of the $T_M$ generators. This means that, by beans of $H_c$, the central- and matter-charge vector at infinity can always be reduced to $p$ real parameters.
The generators $\mathcal{T}_\ell$ can be identified with the $\mathfrak{f}_{4(4)}$-shift generators (with respect to the non-compact Cartan subalgebra in the coset $G/H^*$) corresponding to $p=4$ of the $\gamma_M$ roots which, in the basis $\{T_0,\,{\bf h}_i\}$ of the non-compact Cartan subalgebra of the coset $G/H^*$, are described by mutually orthogonal 4-vectors. These generators define the $p=4$ $\mathfrak{sl}(2)_\ell$  isometry algebras of the four $dS_2$ factors of $\mathcal{M}_N$. The $\mathfrak{sl}(2)_\ell={\rm Span}(\mathcal{J}_\ell, \mathcal{H}_\ell,\,\mathcal{K}_\ell)$ algebras are constructed as follows:
\begin{equation}
\mathcal{J}_\ell=\frac{1}{2}\,(\mathcal{T}_\ell+\mathcal{T}_\ell^T)\,\,;\,\,\,\mathcal{K}_\ell=
\frac{1}{2}\,(\mathcal{T}_\ell-\mathcal{T}_\ell^T)\,\,;\,\,\,\mathcal{H}_\ell=
\frac{1}{2}[\mathcal{T}_\ell,\,\mathcal{T}_\ell^T]\,.
\end{equation}
where the above generators satisfy the following relation:
\begin{align}
[\mathcal{H}_\ell, \,\mathcal{J}_{\ell'}]&=\delta_{\ell\ell'}\,\mathcal{K}_{\ell'}\,\,,\,\,\,[\mathcal{H}_\ell, \,\mathcal{K}_{\ell'}]=\delta_{\ell\ell'}\,\mathcal{J}_{\ell'}\,\,,\,\,\,[\mathcal{J}_\ell, \,\mathcal{K}_{\ell'}]=-\delta_{\ell\ell'}\,\mathcal{H}_{\ell'}\,.
\end{align}
The matrices $\mathcal{J}_\ell$ generate the four $\mathrm{SO}(1,1)_\ell$ groups in the denominator of $\mathcal{M}_N$.
The normal manifold $\mathcal{M}_N$ is parametrized by the three dilatons $\varphi_i$, the scalar $U$ and four of the 14 $\mathcal{Z}^M$. The corresponding generating solution will therefore be a four-charge dilatonic one.
We can make two choices for the four mutually orthogonal roots $\tilde{\gamma}_\ell$ among the $\gamma_M$, which we give below as vectors $(\gamma_M(T_0),\gamma_M({\bf h}_i))$ corresponding to different sets of scalars $\mathcal{Z}^M$:\footnote{For later convenience we choose as range of $\ell$ the values $0,1,4,6$, in light of the truncation of the 14 charges to the eight of the STU model.}
\begin{align}
(\mathcal{Z}_0,\mathcal{Z}^1,\mathcal{Z}^4,\mathcal{Z}^6)&:\,\tilde{\gamma}_0=\gamma_1=\frac{1}{2}(1,-1,-1,-1)
\,,\,\,\tilde{\gamma}_1=\gamma_9=\frac{1}{2}(1,-1,1,1)
\,,\nonumber\\&\tilde{\gamma}_4=\gamma_{12}=\frac{1}{2}(1,1,-1,1)\,\,,\,\,\tilde{\gamma}_6=\gamma_{14}=\frac{1}{2}(1,1,1,-1)\,,\label{choice1}\\
(\mathcal{Z}^0,\mathcal{Z}_1,\mathcal{Z}_4,\mathcal{Z}_6)&:\,\,\tilde{\gamma}_0=
\gamma_8=\frac{1}{2}(1,1,1,1)\,,\,\,\tilde{\gamma}_1=\gamma_2=
\frac{1}{2}(1,1,-1,-1)
\,,\nonumber\\&\tilde{\gamma}_4=\gamma_{5}=\frac{1}{2}(1,-1,1,-1)\,\,,\,\,\,\tilde{\gamma}_6=\gamma_{7}=\frac{1}{2}(1,-1,-1,1)\,,\label{choice2}
\end{align}
Given the general relation between $\mathcal{Z}^M$ and the quantized charges $\Gamma^M$ of the solution \cite{Bergshoeff:2008be}:
\begin{equation}
\dot{\mathcal{Z}}^M=\mathbb{F}^M_{\tau 0}=-e^{2U}\,\mathbb{C}^{MN}\,\mathcal{M}_{4\,NP}(\phi^r)\,\Gamma^P\,,
\end{equation}
we can say that the set (\ref{choice1}) corresponds to a dilatonic solution with charges $q_0,p^1,p^4,p^6$, interpreted as originating from a set of $D0,\,D4$ branes, while the choice  (\ref{choice2}) yields  a solution with charges $p^0,q_1,q_4,q_6$, originating from $D6,\,D2$ branes. We shall choose the normal form corresponding to the first choice, so that the $\mathcal{T}_\ell$, $\ell=0,1,4,6$, can be identified with the following $T_M$:
\begin{equation}
\mathcal{T}_0=T_1\,\,;\,\,\,\mathcal{T}_1=T_9\,\,;\,\,\,\mathcal{T}_4=T_{12}\,\,;\,\,\,\mathcal{T}_6=T_{14}\,.
\end{equation}
Note that both choices define a truncation of the STU model, i.e. $\mathcal{M}_N$ is a totally geodesic submanifold of $\mathcal{M}^{*(STU)}_{QK}\subset G/H^*$ of eq. (\ref{STUQK}).
We wish to emphasize here that while  the STU truncation exists  for all symmetric models with a rank 3 $\mathcal{M}_{SK}$, the construction of $\mathcal{M}_N$ is universal for symmetric models and allows to construct representatives of the $H^*$ orbits corresponding to regular and small black holes.  \par
 A diagonalizable $L_0$ can be rotated by means of $H^*$ into a Cartan subalgebra in the coset (\ref{genman}).
In particular the component of $L_0$ in the tangent space of some of the $dS_2$ factors may be a compact  (i.e. anti-symmetric in the chosen real representation) matrix. This is the case if $L_0$ has imaginary eigenvalues. The corresponding geodesic will have a projection onto some of the $dS_2$ subspaces, which hits the boundary of the solvable (i.e. physical) patch and, as a consequence, $e^{-U}$ will vanish at finite $\tau$, signalling a true singularity of the four-dimensional space-time metric. In order for the solution generated by a diagonalizable $L_0$ to be regular, it must have real eigenvalues only (i.e. $L_0$ must be symmetric). We shall deal with such solutions in a next section.\par
Let us restrict to extremal solutions in $\mathcal{M}_N$ generated by nilpotent $L_0$.
Having defined the normal form according to (\ref{choice1}) we proceed in defining the nilpotent elements in the coset $\mathcal{M}_N$:
\begin{equation}
N_\ell^{(\epsilon_\ell)}=\mathcal{H}_\ell-\epsilon_\ell\,\mathcal{K}_\ell\,\,,\,\,\,[\mathcal{J}_\ell, N_\ell^{(\epsilon_\ell)}]=\epsilon_\ell\, N_\ell^{(\epsilon_\ell)}\,,\,\,\,\ell=0,1,4,6\,,
\end{equation}
where $\epsilon_\ell=\pm 1$.
Consider now the geodesic originating in the origin $O$ of $\mathcal{M}_N$ at radial infinity, corresponding to $\varphi_i=U=0$, with initial velocity $L_0$ which, being $\mathbb{L}(O)={\bf 1}$, coincides with the Noether charge matrix $Q$ in eq. (\ref{noe}). A generic nilpotent $L_0$ on the tangent space to $\mathcal{M}_N$ will be a combination of $N_\ell^{(\epsilon_\ell)}$ of the form:
\begin{equation}
Q=L_0=\sum_{\ell=0,1,4,6} k_\ell \,N_\ell^{(\epsilon_\ell)}=\sum_{\ell=0,1,4,6}^4 k_\ell \,(\mathcal{H}_\ell-\epsilon_\ell\,\mathcal{K}_\ell )\,.\label{L0G}
\end{equation}
All these combinations have vanishing NUT charge: ${\rm Tr}(Q T_\bullet)=0$. The coefficients $k_\ell$ of $\mathcal{H}_\ell$ define the scalar charges and ADM mass, while the coefficients of  $\mathcal{K}_\ell $ define the electric and magnetic charges, which can be computed using  eq. (\ref{GQ}) to be:
\begin{equation}
q_0=-\epsilon_0\,k_0/\sqrt{2}\,\,,\,\,\,p^\ell=\epsilon_\ell\,k_\ell/\sqrt{2}\,\,,\ell=1,4,6\,.\label{chG}
\end{equation}
The ADM mass is computed by tracing $Q$ with $T_0$, as in eq. (\ref{MADM}) and reads
\begin{equation}
M_{ADM}=\lim_{\tau\rightarrow 0^-}\dot{U}=\frac{1}{4}\sum_{\ell}
k_\ell\,.\label{ADMass}
\end{equation}
Solving (\ref{geosi1}) or, equivalently, (\ref{geosi2}), we find the following solution \cite{Bergshoeff:2008be,Chemissany:2009hq}:
\begin{align}
e^{-2U}&=\sqrt{{\bf H}_0 {\bf H}_1 {\bf H}_4 {\bf H}_6}\,,\,\,e^{\varphi_1}=\sqrt{\frac{{\bf H}_0 {\bf H}_1}{{\bf H}_4 {\bf H}_6}}\,,\,\,e^{\varphi_2}=\sqrt{\frac{{\bf H}_0 {\bf H}_4}{{\bf H}_1 {\bf H}_6}}\,,\,\,e^{\varphi_3}=\sqrt{\frac{{\bf H}_0 {\bf H}_6}{{\bf H}_1 {\bf H}_4}}\,,\label{gsol1}\\
\mathcal{Z}^0&= \frac{q_0\,\tau}{{\bf H}_0}\,\,,\,\,\,\mathcal{Z}_k=-\frac{p^k\,\tau}{{\bf H}_k}\,,\,\,k=1,4,6\,,\label{gsol2}
\end{align}
where we have introduced the harmonic functions:
\begin{equation}
{\bf H}_0=1-k_0\,\tau=1+\sqrt{2}\,\epsilon_0\,q_0\,\tau\,\,;\,\,\,{\bf H}_\ell=1-k_\ell\,\tau=1-\sqrt{2}\,\epsilon_\ell\,p^\ell\,\tau\,\,,\,\,\,\ell=1,4,6\,.
\end{equation}
We see that, if one of the $k_\ell$ $(\ell=0,1,4,6)$ is negative, the corresponding ${\bf H}_\ell$ vanishes at finite $\tau=1/k_\ell<0$ and so does  $e^{-2U}$, signalling a true space-time singularity. \emph{ Regular solutions therefore correspond to positive, non vanishing $k_\ell$}.
%Having characterized in a precise algebraic way the space of
%nilpotent orbits of possible Lax operator, we consider the solutions
%to the smaller model based on the manifold $\mathcal{M}_{N}.$ The
%manifold $\mathcal{M}_{N}$ will be referred to as the \emph{normal
%form} of $G/H^*$ for (regular and small) single center solutions with (non)-diagnolizable Lax operator
%$L=\Delta^{\alpha|A}\Lambda_{\alpha|A}.$ This manifold is the
%product of $p=4$
%$\left(\frac{\SL(2,\mathbbm{R})}{\SO(2)}\right)$-factors times $r-p$
%$\textrm{O}(1,1)$ factors. The dilatons in the
%$\SO(1,1)^{r-p}=\SO(1,1)^4$ factor of $\mathcal{M}_{N},$ from $N=2$
%viewpoint, are hyperscalars in the $N=2$ theory and will not be
%therefore relevant for our discussion. $p=4$ is related to the
%electric and magnetic charges; the 4-charge vector is virtually the
%normal form of the electric-magnetic charge vector with respect to
%the action of $H_{4}\times U(1)_{\textrm{E}}.$ The normal space
%$\mathcal{M}_{N}$ of the $F_{4}$model coincides with the one of the
%$N=2$ STU model. In the $F_{4(4)}$-model $H_{4}=\textrm{U}(3)$ and
%the normal form with respect to $\textrm{U}(3)\times \textrm{U}(1)$
%has indeed four parameters which can be chosen as
%\begin{itemize}
% \item The $D_0-D_4-D_4-D_4$ (anti-)brane charges $q_{0},\,p^{1},\,p^2,\,p^3.$
%\item  The $D_{6}-D_{2}-D_{2}-D_{2}$ (anti-)brane charges $p^0,\,q_{1},\,q_{2},\,q_{3}.$
%\end{itemize}
%The corresponding scalar manifold is described by the special coordinate frame  o  originating from the %reduction of a $D=5$ theory to $D=4 $ STU model.
In this case the solution has a finite horizon area given by:
\begin{equation}
A_H=4\pi\,\lim_{\tau\rightarrow -\infty} \frac{e^{-2U}}{\tau^2}=4\pi\,\sqrt{k_0 k_1 k_4 k_6}=4\pi\,\sqrt{
\epsilon I_4(p,q)}=4\pi\,\sqrt{
|I_4(p,q)|}\,,
\end{equation}
where $I_4(p,q)=4 q_0 p^1 p^4 p^6 $ is the quartic $G_4$-invariant function of the electric and magnetic charges  expressed in the charges of the solution, and $\epsilon\equiv -\prod_{\ell}\epsilon_\ell$. The near horizon geometry is $AdS_2\times S^2$ and, in approaching it,  the scalar fields evolve towards values which are fixed solely in terms of the quantized charges, consistently with the \emph{attractor phenomenon} \cite{Ferrara:1995ih,Ferrara:1996um,Ferrara:1997tw,Andrianopoli:1997wi} (see also \cite{Andrianopoli:2006ub} for a review of extremal black holes):
\begin{equation}
\lim_{\tau\rightarrow -\infty} e^{\varphi_1}=\sqrt{\epsilon\,\frac{q_0 p^1}{p^4 p^6}}\,,\,\,\lim_{\tau\rightarrow -\infty} e^{\varphi_2}=\sqrt{\epsilon\,\frac{q_0 p^4}{p^1 p^6}}\,,\,\,\lim_{\tau\rightarrow -\infty} e^{\varphi_3}=\sqrt{\epsilon\,\frac{q_0 p^6}{p^1 p^4}}\,.
\end{equation}
If some of the $k_\ell$ vanish we end up with solutions having a vanishing horizon area, namely a naked singularity at $\tau=-\infty$. Such solutions are called \emph{small} black holes.\par
Let us elaborate now on the $H^*$-orbit of $L_0$. We can easily see that $L_0$, as defined in (\ref{L0G}), does not satisfy eq.s (\ref{eq1}), (\ref{eq2}).
It can however be mapped into one which satisfies (\ref{eq1}), (\ref{eq2}) by means of an ${\rm SO}(1,1)^p$ transformation, generated by the $\mathcal{J}_\ell$, whose effect is to rescale each $k_\ell$ by a positive number and bring them to: $k_\ell=0,\pm 1$. Such transformation clearly cannot affect the signs of $k_\ell$.
Let us consider then an $L_0$ given by (\ref{L0G}), with $k_\ell^2=0,1$. We see that, if we identify the nilpositive element $E$ of the standard triple with $L_0$, the nilnegative $F$ with $L_0^T$ and $h$ with $[L_0,L_0^T]$ we have:
 \begin{equation}
 h=[L_0,L_0^T]=-2\,\sum_\ell\epsilon_\ell k_\ell^2\,\mathcal{J}_\ell\,\,;\,\,\,i\,(E-F)=i\,(L_0-L_0^T)=-2\,\sum_\ell\epsilon_\ell k_\ell (i\mathcal{K}_\ell)\,.
 \end{equation}
Within $\mathfrak{g}^{\mathbb{C}}$, the elements $i\mathcal{K}_\ell$ and $\mathcal{J}_\ell$ are $G^{\mathbb{C}}$-conjugate, just as the complexifications $\mathfrak{H}^{*\,\mathbb{C}}$ and $\mathfrak{H}^{\mathbb{C}}$, of $\mathfrak{H}^*$ and $\mathfrak{H}$ respectively,  are in  $\mathfrak{g}^{\mathbb{C}}$. In particular $\{i\mathcal{K}_\ell \}$ and $\{\mathcal{J}_\ell\}$ are bases of Cartan subalgebras $\mathcal{C}_H$ and $\mathcal{C}_{H^*}$, respectively  in $\mathfrak{H}^{\mathbb{C}}$ and $\mathfrak{H}^{*\,\mathbb{C}}$. If $\beta_k$ and $\beta_k^\prime$ are the
$\mathfrak{H}^{\mathbb{C}}$ and $\mathfrak{H}^{*\,\mathbb{C}}$ simple roots referred to $\mathcal{C}_H$ and $\mathcal{C}_{H^*}$, respectively, we have that:
\begin{equation}
\beta_k(i\mathcal{K}_\ell)=\beta'_k(\mathcal{J}_\ell)\,.
\end{equation}
Since by definition the $\beta$-labels associated with $E$ are $\{\beta_k(i\,(E-F))\}$ and the $\gamma$-labels are $\{\beta'_k(h)\}$, we have:
\begin{equation}
\beta-\mbox{label}\,=\,\{ -2\,\sum_\ell\epsilon_\ell k_\ell
\beta_k(i\mathcal{K}_\ell)\}\,\,;\,\,\,\gamma-\mbox{label}\,=\,\{
-2\,\sum_\ell\epsilon_\ell k_\ell^2
\beta'_k(\mathcal{J}_\ell)\}\,.\label{gabela}
\end{equation}
We see that the regularity condition $k_\ell=0,1$ implies the coincidence of $\gamma$-and $\beta$-labels. This is a formal proof, using the generating solution, of the property:
\begin{equation}
\mbox{Regularity}\,\Rightarrow\,\,\gamma-\mbox{label}\,=\,\beta-\mbox{label}\,,
\end{equation}
first conjectured in \cite{Bossard:2009we}. A similar proof was given in \cite{Fre:2011uy} for the ${\rm G}_{2(2)}$-model.
We stress here that this proof applies to all $\mathcal{N}=2,\,D=4$ theories with symmetric rank-3 special K\"ahler manifold, since for all of them $p=4$, the normal manifold is given by (\ref{MN}) and the generating solution by eq.s (\ref{gsol1}), (\ref{gsol2}).\par
From (\ref{gabela}) we see that the $\beta$-labels only depend on the normal form $(\epsilon_\ell k_\ell)$ of the central and matter charges. In fact  it was shown on general grounds in \cite{Bossard:2009we} that the $\beta$-labels only depend on the $G_4$-orbit of the quantized charges $\Gamma$.

\subsection{Regular Black Holes and the $\alpha^{(4)}$-Orbit}
Using the generating solution we can obtain representatives of all the $H^*$-suborbits in the first four $G^{\mathbb{C}}$-orbits. These are precisely the nilpotent orbits whose step of nilpotency does not exceed 3 and are classified in Tables \ref{alpha1}, \ref{alpha2}, \ref{alpha3}, \ref{alpha4}.
If all $k_\ell\neq 0$, we are in the  fourth $G^{\mathbb{C}}$-orbit, defined by the $\alpha^{(4)}$-label $(2,0,0,0)$, with nilpotency step 3. Let us consider these orbits one by one in light of the known classification of $D=4$ extremal black holes \cite{Bellucci:2006xz}.
\paragraph{Orbit $O_{6H^*}$: The regular BPS solution.} The representative is
\begin{equation}
L_0=N_0^-+N_1^++N_4^++N_6^+\,.\label{lo6h}
\end{equation}
From eq. (\ref{chG}) we see that all charges are positive and equal to $1/\sqrt{2}$. The quartic invariant is positive. The tensor classifier $\mathcal{T}_{xy}$ vanishes, signalling that the solution in BPS. The corresponding $H^*$ orbit is denoted by $O_{6H^*}$ and is identified in Table \ref{alpha4} by the $\beta$ and $\gamma$ labels both coinciding with $(0,0,0,4)$.\par
The signatures of the relevant tensor classifiers are:
\begin{align}
{\rm Sign}(\mathcal{T}_{xy})&=(0_+,\,0_-)\,,\nonumber\\
{\rm Sign}(T_{(21)}^{aA,bB})&=(2_+,\,12_-)\,,\nonumber\\
{\rm Sign}(T_{(84)}^{aA,bB})&=(13_+,\,1_-)\,,\nonumber\\
{\rm Sign}(\mathfrak{T}_{xy})&=(0_+,\,0_-)\,,\nonumber\\
{\rm Sign}(\mathbb{T}_{\alpha \beta})&=(0_+,\,1_-)\,,\nonumber\\
{\rm Sign}(\mathfrak{T}_{(84)}^{(AB),(CD)})&=(0_+,\,0_-)\,,\nonumber\\
% {\rm Sign}(\mathbb{T}^{(3,21)}_{aA,bB})&=(0_+,\,0_-)\,,\nonumber\\
% {\rm Sign}(\mathbb{T}^{(3,84)}_{aA,bB})&=(0_+,\,0_-)\,.\nonumber\\
\end{align}
\paragraph{Orbit $O_{7H^*}$: singular BPS solution.}
The representative is
\begin{equation}
L_0=-N_0^--N_1^++N_4^++N_6^+\,.\label{lo7h}
\end{equation}
The $\beta$-label is $(2,0,0,0)$ while the $\gamma$-label is $(0,0,0,4)$.
Since $k_0=k_1=-1$ the solution is singular. It is BPS though since $\mathcal{T}_{xy}=0$.
The signatures of the relevant tensor classifiers are:
\begin{align}
{\rm Sign}(\mathcal{T}_{xy})&=(0_+,\,0_-)\,,\nonumber\\
{\rm Sign}(T_{(21)}^{aA,bB})&=(6_+,\,8_-)\,,\nonumber\\
{\rm Sign}(T_{(84)}^{aA,bB})&=(9_+,\,5_-)\,,\nonumber\\
{\rm Sign}(\mathfrak{T}_{xy})&=(0_+,\,0_-)\,,\nonumber\\
{\rm Sign}(\mathbb{T}_{\alpha \beta})&=(0_+,\,1_-)\,,\nonumber\\
{\rm Sign}(\mathfrak{T}_{(84)}^{(AB),(CD)})&=(0_+,\,0_-)\,,\nonumber\\
% {\rm Sign}(\mathbb{T}^{(3,21)}_{aA,bB})&=(0_+,\,0_-)\,,\nonumber\\
% {\rm Sign}(\mathbb{T}^{(3,84)}_{aA,bB})&=(0_+,\,0_-)\,.\nonumber\\
\end{align}
Since they differ by the $\beta$-label, the orbits $O_{6H^*}$ and $O_{7H^*}$ belong to different $F_{4(4)}$-orbits.
 \paragraph{Orbit $O_{8H^*}$: singular BPS solution.}
 The representative is
\begin{equation}
L_0=-N_0^-+N_1^++N_4^++N_6^+\,.\label{lo8h}
\end{equation}
The $\beta$-label is $(0,0,2,2)$ while the $\gamma$-label is $(0,0,0,4)$.
Since $k_0=-1$ the solution is singular. It is BPS though since $\mathcal{T}_{xy}=0$.
The signatures of the relevant tensor classifiers are:
\begin{align}
{\rm Sign}(\mathcal{T}_{xy})&=(0_+,\,0_-)\,,\nonumber\\
{\rm Sign}(T_{(21)}^{aA,bB})&=(7_+,\,7_-)\,,\nonumber\\
{\rm Sign}(T_{(84)}^{aA,bB})&=(8_+,\,6_-)\,,\nonumber\\
{\rm Sign}(\mathfrak{T}_{xy})&=(0_+,\,0_-)\,,\nonumber\\
{\rm Sign}(\mathbb{T}_{\alpha \beta})&=(1_+,\,0_-)\,,\nonumber\\
{\rm Sign}(\mathfrak{T}_{(84)}^{(AB),(CD)})&=(0_+,\,0_-)\,.\nonumber\\
%{\rm Sign}(\mathbb{T}^{(3,21)}_{aA,bB})&=(0_+,\,0_-)\,,\nonumber\\
% {\rm Sign}(\mathbb{T}^{(3,84)}_{aA,bB})&=(0_+,\,0_-)\,.\nonumber\\
\end{align}
The orbits $O_{6H^*}$, $O_{7H^*}$ and $O_{8H^*}$ belong to three different $F_{4(4)}$-orbits.
\paragraph{Orbit $O'_{6H^*}$: singular non-BPS solution.} The representative is
\begin{equation}
L_0=-N_0^+-N_1^-+N_4^++N_6^+\,.\label{lo6ph}
\end{equation}
Since some $k_\ell$ are negative, the solution is singular.
 The tensor classifier $\mathcal{T}_{xy}$ does not vanish, signalling that the solution in non-BPS. The corresponding $H^*$ orbit is denoted by $O'_{6H^*}$ and is identified in Table \ref{alpha4} by the $\beta-$ and $\gamma-$ labels given by  $(0,0,0,4)$, $(2,0,0,0)$ respectively.\par
The signatures of the relevant tensor classifiers are:
\begin{align}
{\rm Sign}(\mathcal{T}_{xy})&=(5_+,\,1_-)\,,\nonumber\\
{\rm Sign}(T_{(21)}^{aA,bB})&=(4_+,\,6_-)\,,\nonumber\\
{\rm Sign}(T_{(84)}^{aA,bB})&=(9_+,\,1_-)\,,\nonumber\\
{\rm Sign}(\mathfrak{T}_{xy})&=(0_+,\,1_-)\,,\nonumber\\
{\rm Sign}(\mathbb{T}_{\alpha \beta})&=(0_+,\,0_-)\,,\nonumber\\
{\rm Sign}(\mathfrak{T}_{(84)}^{(AB),(CD)})&=(2_+,\,1_-)\,,\nonumber\\
{\rm Sign}(\mathbb{T}^{(3,21)}_{aA,bB})&=(0_+,\,0_-)\,,\nonumber\\
{\rm Sign}(\mathbb{T}^{(3,84)}_{aA,bB})&=(0_+,\,0_-)\,.\nonumber\\
\end{align}
\paragraph{Orbit $\bar{O}'_{7H^*}$: Regular non-BPS solution.} The representative is
\begin{equation}
L_0=+N_0^++N_1^-+N_4^++N_6^+\,.\label{lo7ph}
\end{equation}
From eq. (\ref{chG}) we see that the charges $q_0$ and $p^1$ are
$-1/\sqrt{2}$, while $p^4$ and $p^6$ are  $1/\sqrt{2}$. The quartic
invariant is positive and the solution is regular. The tensor
classifier $\mathcal{T}_{xy}$ is non-vanishes, signalling that the
solution in non-BPS. The corresponding $H^*$ orbit is denoted by
$\bar{O}'_{7H^*}$ and is identified in Table \ref{alpha4}  by the
$\beta-$ and $\gamma-$ labels both given by  $(2,0,0,0)$.\par The
signatures of the relevant tensor classifiers are:
\begin{align}
{\rm Sign}(\mathcal{T}_{xy})&=(1_+,\,5_-)\,,\nonumber\\
{\rm Sign}(T_{(21)}^{aA,bB})&=(4_+,\,6_-)\,,\nonumber\\
{\rm Sign}(T_{(84)}^{aA,bB})&=(9_+,\,1_-)\,,\nonumber\\
{\rm Sign}(\mathfrak{T}_{xy})&=(0_+,\,1_-)\,,\nonumber\\
{\rm Sign}(\mathbb{T}_{\alpha \beta})&=(0_+,\,0_-)\,,\nonumber\\
{\rm Sign}(\mathfrak{T}_{(84)}^{(AB),(CD)})&=(2_+,\,1_-)\,,\nonumber\\
{\rm Sign}(\mathbb{T}^{(3,21)}_{aA,bB})&=(0_+,\,0_-)\,,\nonumber\\
{\rm Sign}(\mathbb{T}^{(3,84)}_{aA,bB})&=(0_+,\,0_-)\,.\nonumber\\
\end{align}
Note that $\bar{O}'_{7H^*}$ and $O'_{6H^*}$ are only distinguished
by the signature of $\mathcal{T}_{xy}$.
\paragraph{Orbit $\hat{O}'_{7H^*}$: singular non-BPS solution.} The representative is
\begin{equation}
L_0=+N_0^+-N_1^--N_4^++N_6^+\,.\label{lho7ph}
\end{equation}
Since some $k_\ell$ are negative, the solution is singular.
 The tensor classifier $\mathcal{T}_{xy}$ does not vanish, signalling that the solution in non-BPS. The corresponding $H^*$ orbit is
  denoted by $\hat{O}'_{7H^*}$ and is identified in Table \ref{alpha4} by the same $\beta-$ and $\gamma-$
   labels as the orbit $\bar{O}'_{7H^*}$. The two orbits are also distinguished by a further $\delta$-label.\par
The signatures of the relevant tensor classifiers are:
\begin{align}
{\rm Sign}(\mathcal{T}_{xy})&=(3_+,\,3_-)\,,\nonumber\\
{\rm Sign}(T_{(21)}^{aA,bB})&=(6_+,\,4_-)\,,\nonumber\\
{\rm Sign}(T_{(84)}^{aA,bB})&=(7_+,\,3_-)\,,\nonumber\\
{\rm Sign}(\mathfrak{T}_{xy})&=(0_+,\,1_-)\,,\nonumber\\
{\rm Sign}(\mathbb{T}_{\alpha \beta})&=(0_+,\,0_-)\,,\nonumber\\
{\rm Sign}(\mathfrak{T}_{(84)}^{(AB),(CD)})&=(2_+,\,1_-)\,,\nonumber\\
{\rm Sign}(\mathbb{T}^{(3,21)}_{aA,bB})&=(0_+,\,0_-)\,,\nonumber\\
{\rm Sign}(\mathbb{T}^{(3,84)}_{aA,bB})&=(0_+,\,0_-)\,,\nonumber\\
\end{align}
and clearly show that the orbits $\hat{O}'_{7H^*}$ and
$\bar{O}'_{7H^*}$ are different.
\paragraph{Orbit $O'_{8H^*}$: singular non-BPS solution.} The representative is
\begin{equation}
L_0=+N_0^+-N_1^-+N_4^++N_6^+\,.\label{lo8ph}
\end{equation}
Since some $k_\ell$ are negative, the solution is singular.
 The tensor classifier $\mathcal{T}_{xy}$ does not vanish, signalling that the solution in non-BPS. The corresponding $H^*$ orbit is denoted by ${O}'_{8H^*}$ and is identified in Table \ref{alpha4}  by the $\beta-$ and $\gamma-$ labels given by  $(0,0,2,2)$, $(2,0,0,0)$ respectively.\par
The signatures of the relevant tensor classifiers are:
\begin{align}
{\rm Sign}(\mathcal{T}_{xy})&=(3_+,\,3_-)\,,\nonumber\\
{\rm Sign}(T_{(21)}^{aA,bB})&=(5_+,\,5_-)\,,\nonumber\\
{\rm Sign}(T_{(84)}^{aA,bB})&=(6_+,\,4_-)\,,\nonumber\\
{\rm Sign}(\mathfrak{T}_{xy})&=(1_+,\,0_-)\,,\nonumber\\
{\rm Sign}(\mathbb{T}_{\alpha \beta})&=(0_+,\,0_-)\,,\nonumber\\
{\rm Sign}(\mathfrak{T}_{(84)}^{(AB),(CD)})&=(2_+,\,1_-)\,,\nonumber\\
{\rm Sign}(\mathbb{T}^{(3,21)}_{aA,bB})&=(0_+,\,0_-)\,,\nonumber\\
{\rm Sign}(\mathbb{T}^{(3,84)}_{aA,bB})&=(0_+,\,0_-)\,.\nonumber\\
\end{align}
\paragraph{Orbit $O''_{6H^*}$: singular non-BPS solution.} The representative is
\begin{equation}
L_0=N_0^+-N_1^+-N_4^+-N_6^+\,.\label{lo6pph}
\end{equation}
Since  some $k_\ell$ are negative,  the solution is singular.
 The tensor classifier $\mathcal{T}_{xy}$ does not vanish, signalling that the solution in non-BPS. The corresponding $H^*$ orbit is denoted by ${O}''_{6H^*}$ and is identified in Table \ref{alpha4}  by the $\beta-$ and $\gamma-$ labels given by  $(0,0,0,4)$, $(0,0,2,2)$ respectively.\par
The signatures of the relevant tensor classifiers are:
\begin{align}
{\rm Sign}(\mathcal{T}_{xy})&=(5_+,\,1_-)\,,\nonumber\\
{\rm Sign}(T_{(21)}^{aA,bB})&=(2_+,\,6_-)\,,\nonumber\\
{\rm Sign}(T_{(84)}^{aA,bB})&=(8_+,\,1_-)\,,\nonumber\\
{\rm Sign}(\mathfrak{T}_{xy})&=(0_+,\,0_-)\,,\nonumber\\
{\rm Sign}(\mathbb{T}_{\alpha \beta})&=(0_+,\,0_-)\,,\nonumber\\
{\rm Sign}(\mathfrak{T}_{(84)}^{(AB),(CD)})&=(2_+,\,1_-)\,,\nonumber\\
{\rm Sign}(\mathbb{T}^{(3,21)}_{aA,bB})&=(0_+,\,0_-)\,,\nonumber\\
{\rm Sign}(\mathbb{T}^{(3,84)}_{aA,bB})&=(0_+,\,1_-)\,.\nonumber\\
\end{align}
\paragraph{Orbit $O''_{7H^*}$: singular non-BPS solution.} The representative is
\begin{equation}
L_0=N_0^++N_1^++N_4^+-N_6^+\,.\label{lo7pph}
\end{equation}
Since  some $k_\ell$ are negative,  the solution is singular.
 The tensor classifier $\mathcal{T}_{xy}$ does not vanish, signalling that the solution in non-BPS. The corresponding $H^*$ orbit is denoted by ${O}''_{7H^*}$ and is identified in Table \ref{alpha4}  by the $\beta-$ and $\gamma-$ labels given by  $(2,0,0,0)$, $(0,0,2,2)$ respectively.\par
The signatures of the relevant tensor classifiers are:
\begin{align}
{\rm Sign}(\mathcal{T}_{xy})&=(3_+,\,3_-)\,,\nonumber\\
{\rm Sign}(T_{(21)}^{aA,bB})&=(4_+,\,4_-)\,,\nonumber\\
{\rm Sign}(T_{(84)}^{aA,bB})&=(6_+,\,3_-)\,,\nonumber\\
{\rm Sign}(\mathfrak{T}_{xy})&=(0_+,\,0_-)\,,\nonumber\\
{\rm Sign}(\mathbb{T}_{\alpha \beta})&=(0_+,\,0_-)\,,\nonumber\\
{\rm Sign}(\mathfrak{T}_{(84)}^{(AB),(CD)})&=(2_+,\,1_-)\,,\nonumber\\
{\rm Sign}(\mathbb{T}^{(3,21)}_{aA,bB})&=(0_+,\,0_-)\,,\nonumber\\
{\rm Sign}(\mathbb{T}^{(3,84)}_{aA,bB})&=(0_+,\,1_-)\,.\nonumber\\
\end{align}
\paragraph{Orbit $\bar{O}''_{8H^*}$ ($\delta_1$): regular non-BPS solution with $I_4<0$.} The representative is
\begin{equation}
L_0=N_0^++N_1^++N_4^++N_6^+\,.\label{lo8pph}
\end{equation}
All the $k_\ell$ are positive and   the solution is regular. The charges are read off eq. (\ref{chG}) to be $q_0=-1/\sqrt{2}$, $p^1=p^4=p^6=1/\sqrt{2}$, and the quartic invariant is negative.
 The tensor classifier $\mathcal{T}_{xy}$ does not vanish, signalling that the solution in non-BPS. The corresponding $H^*$ orbit is denoted by ${O}''_{8H^*}$ and is identified in Table \ref{alpha4}  by the $\beta-$ and $\gamma-$ labels given by  $(0,0,2,2)$, $(0,0,2,2)$ respectively.\par
The signatures of the relevant tensor classifiers are:
\begin{align}
{\rm Sign}(\mathcal{T}_{xy})&=(1_+,\,5_-)\,,\nonumber\\
{\rm Sign}(T_{(21)}^{aA,bB})&=(2_+,\,6_-)\,,\nonumber\\
{\rm Sign}(T_{(84)}^{aA,bB})&=(8_+,\,1_-)\,,\nonumber\\
{\rm Sign}(\mathfrak{T}_{xy})&=(0_+,\,0_-)\,,\nonumber\\
{\rm Sign}(\mathbb{T}_{\alpha \beta})&=(0_+,\,0_-)\,,\nonumber\\
{\rm Sign}(\mathfrak{T}_{(84)}^{(AB),(CD)})&=(2_+,\,1_-)\,,\nonumber\\
{\rm Sign}(\mathbb{T}^{(3,21)}_{aA,bB})&=(0_+,\,0_-)\,,\nonumber\\
{\rm Sign}(\mathbb{T}^{(3,84)}_{aA,bB})&=(1_+,\,0_-)\,.\nonumber\\
\end{align}
\paragraph{Orbit $\hat{O}''_{8H^*}$ ($\delta_2$): singular non-BPS solution.} The representative is
\begin{equation}
L_0=-N_0^+-N_1^++N_4^++N_6^+\,.\label{lho8pph}
\end{equation}
Since  some $k_\ell$ are negative,  the solution is singular.
 The tensor classifier $\mathcal{T}_{xy}$ does not vanish, signalling that the solution in non-BPS. The corresponding $H^*$ orbit is denoted in Table \ref{alpha4}  by $\hat{O}''_{8H^*}$ to distinguish it from ${O}''_{8H^*}$ with which it shares the same  $\beta-$ and $\gamma-$ labels.\par
The signatures of the relevant tensor classifiers are:
\begin{align}
{\rm Sign}(\mathcal{T}_{xy})&=(3_+,\,3_-)\,,\nonumber\\
{\rm Sign}(T_{(21)}^{aA,bB})&=(4_+,\,4_-)\,,\nonumber\\
{\rm Sign}(T_{(84)}^{aA,bB})&=(6_+,\,3_-)\,,\nonumber\\
{\rm Sign}(\mathfrak{T}_{xy})&=(0_+,\,0_-)\,,\nonumber\\
{\rm Sign}(\mathbb{T}_{\alpha \beta})&=(0_+,\,0_-)\,,\nonumber\\
{\rm Sign}(\mathfrak{T}_{(84)}^{(AB),(CD)})&=(2_+,\,1_-)\,,\nonumber\\
{\rm Sign}(\mathbb{T}^{(3,21)}_{aA,bB})&=(0_+,\,0_-)\,,\nonumber\\
{\rm Sign}(\mathbb{T}^{(3,84)}_{aA,bB})&=(1_+,\,0_-)\,.\nonumber\\
\end{align}
Thus, in summary, the regular BPS and non-BPS solutions are all
described in the diagonal orbits
$O_{6H^*},\,\bar{O}'_{7H^*},\,\bar{O}''_{8H^*}$. Their
representatives are characterized by having the same scalar charges
(same $k_\ell$) but different electric and magnetic charges
(different $\epsilon_\ell k_\ell$). In general, if we define
$\epsilon_0'=-\epsilon_0, \epsilon_\ell'=\epsilon_\ell$,
$\ell=1,4,6$, one can verify
\cite{Bergshoeff:2008be,Chemissany:2009hq} that representatives
$L_0$ of the first $O_{6H^*}$ orbit in the form (\ref{L0G}) (thus
with  all $k_\ell$ positive) are all characterized by having
$\epsilon'_0=\epsilon'_1=\epsilon'_4=\epsilon'_6$ (so that
$\epsilon=+1$ and $I_4>0$) and generate regular BPS solutions. If,
on the other hand, all $k_\ell$ are positive, $\epsilon=1$ but the
$\epsilon'_\ell$ are not all equal (there are 6 possibilities) the
solution lies in the $\bar{O}'_{7H^*}$ orbit and is regular non-BPS
with $I_4>0$ \cite{Bergshoeff:2008be,Chemissany:2009hq}. Finally if
all $k_\ell$ are positive but  $\epsilon=-1$ the solution lies in
the $\bar{O}''_{8H^*}$ orbit and is regular non-BPS with $I_4<0$
\cite{Bergshoeff:2008be,Chemissany:2009hq}. What we have shown here
is that the generating solution allows to derive representatives of
all the $H^*$-orbits with degree of nilpotency not exceeding 3. Let
us observe that the gradings $\epsilon_\ell$ of the nilpotent
generators entering  $L_0$ are related to a particular
$\gamma$-label. We can move from one orbit to an other in Table
\ref{alpha4} by observing that:
\begin{align}
e^{-i\pi \mathcal{J}_\ell}\, N^{\epsilon}_\ell \,e^{i\pi \mathcal{J}_\ell}&=-N^{\epsilon}_\ell\,,\nonumber\\
e^{-\pi \mathcal{K}_\ell }\, N^{\epsilon}_\ell \,e^{\pi \mathcal{K}_\ell }&=-N^{-\epsilon}_\ell\,,\nonumber\\
e^{-\pi \mathcal{K}_\ell }e^{-i\pi \mathcal{J}_\ell }N^{\epsilon}_\ell\,e^{i\pi \mathcal{J}_\ell }e^{\pi \mathcal{K}_\ell }&=N^{-\epsilon}_\ell\,,
\end{align}
so that the adjoint action on the representative $L^{(BPS)}_0$, in
the form (\ref{L0G}), of the BPS orbit $O_{6H^*}$, of an even number
of transformations $e^{i\pi \mathcal{J}_\ell }e^{\pi
\mathcal{K}_\ell }$, have the effect of switching an even number of
$\epsilon_\ell$, yielding a representative of the non-BPS orbit
$\bar{O}'_{7H^*}$. By the same token the action of an odd number of
such transformations will map $L^{(BPS)}_0$ into the non-BPS
$\bar{O}''_{8H^*}$ orbit:
\begin{align}
\left(\prod_{even}e^{-\pi \mathcal{K}_\ell } e^{-i\pi
\mathcal{J}_\ell }\right)L^{(BPS)}_0 \left(\prod_{even} e^{i\pi
\mathcal{J}_\ell }e^{\pi \mathcal{K}_\ell }\right)&\in
\bar{O}'_{7H^*}\,,\nonumber\\\left(\prod_{odd}e^{-\pi
\mathcal{K}_\ell } e^{-i\pi \mathcal{J}_\ell }\right)L^{(BPS)}_0
\left(\prod_{odd} e^{i\pi \mathcal{J}_\ell }e^{\pi \mathcal{K}_\ell
}\right)&\in \bar{O}''_{8H^*}\,.
\end{align}
In general the adjoint action of $e^{\pi \mathcal{K}_\ell }$ will
not alter the $G$-orbit (and thus the $\beta$-label)  since the
transformation belongs to $G$. It may alter the $\gamma$-label and
thus make us move vertically in Table \ref{alpha4}. Moreover its
action changes the sign of $\epsilon_\ell$ and of $k_\ell$, keeping
the sign of the corresponding electric-magnetic charge fixed. The
transformation $e^{i\pi \mathcal{J}_\ell }$, on the other hand,
belongs to $H^{\prime\,\mathbb{C}}$, complexification of $H^*$. Its
action will therefore not alter the $\gamma$-labels and the
supersymmetry property of the solution (since $\mathcal{T}_{xy}$ is
also a $H^{\prime\,\mathbb{C}}$-covariant tensor) but, being it in
$G^{\mathbb{C}}/G$, it may affect the $G$-orbit of the solution and
thus the corresponding $\beta$-label. It will in other words make us
move horizontally in Table \ref{alpha4} or, for fixed $\beta$- and
$\gamma$-labels, vertically from the $\delta_1$ to the $\delta_2$
orbits, whenever this further splitting exists.\par The relations
discussed above among the various $H^*$-orbits, clearly extend to
any of their representatives.

%begintable
\begin{sidewaystable}

\scalefont{1}
\begin{tabular}{|c|c|c|c|c|}\hline
\backslashbox{$\gamma$-label}{$\beta$-label}
&(0,\,0,\,0,\,4)&(2,\,0,\,0,\,0)&(0,\,0,\,2,\,2)&\null\\ \hline

(0,\,0,\,0,\,4)
& $\renewcommand{\arraystretch}{0.5}
\begin{array}{c}
L_0=N_0^-+N_1^++\\N_4^++N_6^+\end{array}$
&
$\renewcommand{\arraystretch}{0.5}
\begin{array}{c}
L_0=-N_0^--N_1^++N_4^++N_6^+\end{array}$
&$\renewcommand{\arraystretch}{0.5}
\begin{array}{c}
L_0=-N_0^-+N_1^++N_4^++N_6^+\end{array}$
&$\mathcal{O}_{\cdot H^*}$\\ \hline

(2,\,0,\,0,\,0)
&$\renewcommand{\arraystretch}{0.5}
\begin{array}{c}
L_0=-N_0^+-N_1^-+\\N_4^++N_6^+\end{array}$
&

\begin{tabular}{|c|c|c|}\hline
  $\delta_1$& $\renewcommand{\arraystretch}{1}
\begin{array}{c} L_0=N_0^++N_1^-+\\N_4^++N_6^+
 \end{array}$ &$\bar{ \mathcal{O}}'_{.H^*}$\\ \hline
$\delta_2$ & $\renewcommand{\arraystretch}{1}
\begin{array}{c}L_0=+N_0^+-N_1^--\\N_4^++N_6^+
 \end{array}$ & $\hat{ \mathcal{O}}'_{.H^*}$ \\ \hline
\end{tabular}

&$\renewcommand{\arraystretch}{0.5}
\begin{array}{c}
L_0=+N_0^+-N_1^-+N_4^++N_6^+\end{array}$
&$\mathcal{O'}_{\cdot H^*}$\\ \hline

(0,\,0,\,2,\,2)
& $\renewcommand{\arraystretch}{1}
\begin{array}{c}L_0=N_0^+-N_1^+-\\N_4^+-N_6^+
\end{array}$
& $\renewcommand{\arraystretch}{0.5}
\begin{array}{c}
L_0=N_0^++N_1^++N_4^+-N_6^+\end{array}$
&

\begin{tabular}{|c|c|c|}\hline
  $\delta_1$& $\renewcommand{\arraystretch}{1}
\begin{array}{c}L_0=N_0^++\\N_1^++N_4^++N_6^+
 \end{array}$ &$\bar{ \mathcal{O}}''_{.H^*}$\\ \hline
$\delta_2$ & $\renewcommand{\arraystretch}{1}
\begin{array}{c}L_0=-N_0^+-N_1^++\\N_4^++N_6^+
 \end{array}$ & $\hat{ \mathcal{O}}''_{.H^*}$ \\ \hline
\end{tabular}

&$\mathcal{O''}_{\cdot H^*}$\\ \hline
\null  & $\mathcal{O}_{6}$& $\mathcal{O}_{7}$&$\mathcal{O}_{8}$&\null\\ \hline
\end{tabular}
\caption{\small The representatives of $\alpha^{(4)}$- orbtis
in terms of the generating solution.}
\label{alpha4g}
\end{sidewaystable}
%endtable

\subsection{Small Black Holes and the $\alpha^{(1)},\,\alpha^{(2)},\,\alpha^{(3)}$ Orbits}
\subsubsection{Small Lightlike Black Holes ($\alpha^{(3)}$)}
Let us first consider the case in which one of the $k_\ell$
vanishes. In this case $I_4=0$, though $\partial I_4/\partial k_\ell
\neq 0$. The corresponding $G^{\mathbb{C}}$-orbit is the one defined
by  the $\alpha^{(3)}$-label $(0,1,0,0)$. Let us consider these
orbits one by one in light of the known classification of $D=4$ small black holes \cite{Ferrara:1997uz,Ceresole:2010nm}:
\paragraph{Orbit $O_{4H^*}$: small, light-like, BPS black hole.}
A representative of this orbit, in the form (\ref{L0G}) can be
obtained by setting one parameters $k_\ell$ in the representative of
$O_{6H^*}$ (i.e. regular BPS black hole, $k_\ell>0$,
$\epsilon_1=\epsilon_4=\epsilon_6=-\epsilon_0$) to zero. This
amounts to setting in the generating solution for regular BPS black
holes one of the charges to zero. For example we can choose
\begin{equation}
L_0=N_0^-+N_1^++N_4^+\,,
\end{equation}
obtained by setting $p^6\rightarrow 0$ in the generating solution for regular BPS black holes.
Since $\mathcal{T}_{xy}=0$, the solution is BPS.
The $\beta$- and $\gamma$-labels are both $(0,0,1,3)$. This  solution has vanishing horizon area and thus a naked singularity at $\tau\rightarrow -\infty$ but no singularity at finite $\tau$. Though $I_4=0$, its gradient with respect to the electric-magnetic charges is non-vanishing. These small black holes are named \emph{lightlike}.
The signatures of the relevant tensor classifiers are:
\begin{align}
{\rm Sign}(\mathcal{T}_{xy})&=(0_+,\,0_-)\,,\nonumber\\
{\rm Sign}(T_{(21)}^{aA,bB})&=(2_+,\,6_-)\,,\nonumber\\
{\rm Sign}(T_{(84)}^{aA,bB})&=(7_+,\,1_-)\,,\nonumber\\
{\rm Sign}(\mathfrak{T}_{xy})&=(0_+,\,0_-)\,,\nonumber\\
{\rm Sign}(\mathbb{T}_{\alpha \beta})&=(0_+,\,0_-)\,,\nonumber\\
{\rm Sign}(\mathfrak{T}_{(84)}^{(AB),(CD)})&=(0_+,\,0_-)\,,\nonumber\\
%{\rm Sign}(\mathbb{T}^{(3,21)}_{aA,bB})&=(0_+,\,0_-)\,,\nonumber\\
%{\rm Sign}(\mathbb{T}^{(3,84)}_{aA,bB})&=(0_+,\,0_-)\,.\nonumber\\
\end{align}
\paragraph{Orbit $O_{5H^*}$: singular BPS black hole.}
A representative of this orbit is obtained by setting to zero one of the charges in the generating solution of $O_{7H^*}$ or $O_{8H^*}$. We can choose for instance
\begin{equation}
L_0=-N_0^--N_1^++N_4^+\,,
\end{equation}
One of the $k_\ell$ is negative, implying a singularity at finite $\tau$. The solution is still BPS since $\mathcal{T}_{xy}=0$.
The signatures of the relevant tensor classifiers are:
\begin{align}
{\rm Sign}(\mathcal{T}_{xy})&=(0_+,\,0_-)\,,\nonumber\\
{\rm Sign}(T_{(21)}^{aA,bB})&=(4_+,\,4_-)\,,\nonumber\\
{\rm Sign}(T_{(84)}^{aA,bB})&=(5_+,\,3_-)\,,\nonumber\\
{\rm Sign}(\mathfrak{T}_{xy})&=(0_+,\,0_-)\,,\nonumber\\
{\rm Sign}(\mathbb{T}_{\alpha \beta})&=(0_+,\,0_-)\,,\nonumber\\
{\rm Sign}(\mathfrak{T}_{(84)}^{(AB),(CD)})&=(0_+,\,0_-)\,,\nonumber\\
%{\rm Sign}(\mathbb{T}^{(3,21)}_{aA,bB})&=(0_+,\,0_-)\,,\nonumber\\
%{\rm Sign}(\mathbb{T}^{(3,84)}_{aA,bB})&=(0_+,\,0_-)\,.\nonumber\\
\end{align}
The $\beta$- and $\gamma$- labels of the orbit are $(1,0,1,1)$ and
$(0,0,1,3)$ respectively.
\paragraph{Orbit $O'_{4H^*}$: singular BPS black hole.}
This orbit is obtained as a singular limit (implemented by setting some of the $k_\ell$ to zero) of the off-diagonal orbits in the $\alpha^{(4)}$-class.
We can choose for instance
\begin{equation}
L_0=N_0^+-N_1^+-N_4^+\,,
\end{equation}
Some of the   $k_\ell$ are negative, implying a singularity at finite $\tau$. The solution is non-BPS since $\mathcal{T}_{xy}\neq 0$.
The $\beta$- and $\gamma$- labels of the orbit are $(0,0,1,3)$ and $(1,0,1,1)$ respectively.
The signatures of the relevant tensor classifiers are:
\begin{align}
{\rm Sign}(\mathcal{T}_{xy})&=(3_+,\,1_-)\,,\nonumber\\
{\rm Sign}(T_{(21)}^{aA,bB})&=(2_+,\,3_-)\,,\nonumber\\
{\rm Sign}(T_{(84)}^{aA,bB})&=(5_+,\,1_-)\,,\nonumber\\
{\rm Sign}(\mathfrak{T}_{xy})&=(0_+,\,0_-)\,,\nonumber\\
{\rm Sign}(\mathbb{T}_{\alpha \beta})&=(0_+,\,0_-)\,,\nonumber\\
{\rm Sign}(\mathfrak{T}_{(84)}^{(AB),(CD)})&=(2_+,\,1_-)\,,\nonumber\\
{\rm Sign}(\mathbb{T}^{(3,21)}_{aA,bB})&=(0_+,\,0_-)\,,\nonumber\\
{\rm Sign}(\mathbb{T}^{(3,84)}_{aA,bB})&=(0_+,\,0_-)\,.\nonumber\\
\end{align}
\paragraph{Orbit $\bar{O}'_{5H^*}$ ($\delta_1$): small, lightlike, non-BPS black hole.}
This orbit is obtained as a singular limit (implemented by setting
some of the $k_\ell$ to zero) of the off-diagonal orbits in the
$\alpha^{(4)}$-class.
We can choose for instance
\begin{equation}
L_0=N_0^++N_1^++N_4^+\,,
\end{equation}
All   $k_\ell$ are positive, implying a singularity only at $\tau\rightarrow -\infty$. The solution is non-BPS since $\mathcal{T}_{xy}\neq 0$. The $\beta$- and $\gamma$- labels of the orbit are both $(1,0,1,1)$. This  solution generates the small, lightlike , non-BPS black holes.
The signatures of the relevant tensor classifiers are:
\begin{align}
{\rm Sign}(\mathcal{T}_{xy})&=(1_+,\,3_-)\,,\nonumber\\
{\rm Sign}(T_{(21)}^{aA,bB})&=(2_+,\,3_-)\,,\nonumber\\
{\rm Sign}(T_{(84)}^{aA,bB})&=(5_+,\,1_-)\,,\nonumber\\
{\rm Sign}(\mathfrak{T}_{xy})&=(0_+,\,0_-)\,,\nonumber\\
{\rm Sign}(\mathbb{T}_{\alpha \beta})&=(0_+,\,0_-)\,,\nonumber\\
{\rm Sign}(\mathfrak{T}_{(84)}^{(AB),(CD)})&=(2_+,\,1_-)\,,\nonumber\\
{\rm Sign}(\mathbb{T}^{(3,21)}_{aA,bB})&=(0_+,\,0_-)\,,\nonumber\\
{\rm Sign}(\mathbb{T}^{(3,84)}_{aA,bB})&=(0_+,\,0_-)\,.\nonumber\\
\end{align}
\paragraph{Orbit $\hat{O}'_{5H^*}$ ($\delta_2$): singular non-BPS solution.}
 We can choose for this orbit the following representative
\begin{equation}
L_0=N_0^++N_1^+-N_4^+\,,
\end{equation}
One of the $k_\ell$ is negative, implying a singularity at finite $\tau$. The solution is non-BPS since $\mathcal{T}_{xy}\neq 0$. The $\beta$- and $\gamma$- labels of the orbit are still both equal to $(1,0,1,1)$.
The signatures of the relevant tensor classifiers are:
\begin{align}
{\rm Sign}(\mathcal{T}_{xy})&=(2_+,\,2_-)\,,\nonumber\\
{\rm Sign}(T_{(21)}^{aA,bB})&=(3_+,\,2_-)\,,\nonumber\\
{\rm Sign}(T_{(84)}^{aA,bB})&=(4_+,\,2_-)\,,\nonumber\\
{\rm Sign}(\mathfrak{T}_{xy})&=(0_+,\,0_-)\,,\nonumber\\
{\rm Sign}(\mathbb{T}_{\alpha \beta})&=(0_+,\,0_-)\,,\nonumber\\
{\rm Sign}(\mathfrak{T}_{(84)}^{(AB),(CD)})&=(2_+,\,1_-)\,,\nonumber\\
{\rm Sign}(\mathbb{T}^{(3,21)}_{aA,bB})&=(0_+,\,0_-)\,,\nonumber\\
{\rm Sign}(\mathbb{T}^{(3,84)}_{aA,bB})&=(0_+,\,0_-)\,.\nonumber\\
\end{align}
This orbit is distinguished from $O'_{5H^*}$ by the tensor classifiers.

%table(alpha3)
\begin{table}[htbp]
\scalefont{.75}
\begin{tabular}{|c|c|c|c|}\hline
\backslashbox{$\gamma$-label}{$\beta$-label}
&(0,\,0,\,1,\,3)&(1,\,0,\,1,\,1)&\null\\ \hline
(0,\,0,\,1,\,3)&
$\renewcommand{\arraystretch}{1}
\begin{array}{c}
L_0=N_0^-+N_1^++N_4^+\end{array}$&$\renewcommand{\arraystretch}{1}
\begin{array}{c}
L_0=-N_0^--N_1^++N_4^+\end{array}$ &$\mathcal{O}_{\cdot H^*}$\\ \hline
(1,\,0,\,1,\,1)  & $\renewcommand{\arraystretch}{1}
\begin{array}{c}
L_0=N_0^+-N_1^+-N_4^+\end{array}$&

\begin{tabular}{|c|c|c|}\hline
  $\delta_1$& $\renewcommand{\arraystretch}{1}
\begin{array}{c} L_0=N_0^++N_1^++N_4^+
 \end{array}$ &$\bar{ \mathcal{O}}'_{.H^*}$\\ \hline
$\delta_2$ & $\renewcommand{\arraystretch}{1}
\begin{array}{c}L_0=N_0^++N_1^+-N_4^+
 \end{array}$ & $\hat{ \mathcal{O}}'_{.H^*}$ \\ \hline
\end{tabular}

 &
$ \mathcal{O'}_{\cdot H^*}$ \\ \hline
\null  & $\mathcal{O}_{4}$& $\mathcal{O}_{5}$&\null\\ \hline
\end{tabular}
\caption{\small  The representatives of $\alpha^{(3)}$- orbtis
in terms of the generating solution.}
\label{alpha3g}
\end{table}
\subsubsection{Small Critical Black Holes ($\alpha^{(2)}$)}
Now we  consider the case in which two of the $k_\ell$ vanish. In
this case the following properties, which can be easily verified on
the generating solution, extend to the whole orbit of the
electric-magnetic charges in the representation ${\bf R}$ of $G_4$:
\begin{equation}
I_4(p,q)=0\,\,;\,\,\,\frac{\partial I_4}{\partial \Gamma^M}\equiv
0\,,\label{pqcritical}
\end{equation}
 The
corresponding $G^{\mathbb{C}}$-orbit is the one defined by  the
$\alpha^{(2)}$-label $(0,0,0,1)$. Let us consider these orbits one
by one:
\paragraph{Orbit $O_{2H^*}$: small, critical, BPS black hole.}
A representative of this orbit, in the form (\ref{L0G}) can be
obtained by setting one parameters $k_\ell$ in the representative of
$O_{4H^*}$ (i.e. lightlike BPS black hole) to zero. This amounts to
setting in the generating solution for regular BPS black holes two
of the charges to zero. For example we can choose
\begin{equation}
L_0=N_0^-+N_1^+\,,\label{lo2h}
\end{equation}
obtained by setting $p^4,p^6\rightarrow 0$ in the generating
solution for regular BPS black holes. Since $\mathcal{T}_{xy}=0$,
the solution is BPS. The $\beta$- and $\gamma$-labels are both
$(1,0,0,2)$. This  solution has vanishing horizon area and thus a
naked singularity at $\tau\rightarrow -\infty$ but no singularity at
finite $\tau$. Since both  $I_4$ and its gradient with respect to
the charges vanish, these small black holes are named
\emph{critical}. This orbit could also be reached from
$\bar{O}''_{8H^*}$ by setting the $q_0$ and one of the $p^\ell$,
$\ell=1,4,6$, charges to zero in the non-BPS regular generating
solution with $I_4<0$, or from $\bar{O}'_{7H^*}$ by setting in the
generating solution of regular non-BPS black holes with $I_4>0$ two
charges to zero.\par The signatures of the relevant tensor
classifiers are:
\begin{align}
{\rm Sign}(\mathcal{T}_{xy})&=(0_+,\,0_-)\,,\nonumber\\
{\rm Sign}(T_{(21)}^{aA,bB})&=(2_+,\,3_-)\,,\nonumber\\
{\rm Sign}(T_{(84)}^{aA,bB})&=(4_+,\,1_-)\,,\nonumber\\
{\rm Sign}(\mathfrak{T}_{xy})&=(0_+,\,0_-)\,,\nonumber\\
{\rm Sign}(\mathbb{T}_{\alpha \beta})&=(0_+,\,0_-)\,,\nonumber\\
{\rm Sign}(\mathfrak{T}_{(84)}^{(AB),(CD)})&=(0_+,\,0_-)\,,\nonumber\\
%{\rm Sign}(\mathbb{T}^{(3,21)}_{aA,bB})&=(0_+,\,0_-)\,,\nonumber\\
%{\rm Sign}(\mathbb{T}^{(3,84)}_{aA,bB})&=(0_+,\,0_-)\,.\nonumber\\
\end{align}
\paragraph{Orbit $O_{3H^*}$: singular BPS black hole.}
We can choose as representative of this orbit the matrix:
\begin{equation}
L_0=N_0^--N_1^+\,,
\end{equation}
This can be obtained by acting on  the representative (\ref{lo2h})
of $O_{2H^*}$ with $e^{i\pi \mathcal{J}_1}$. This has the effect of changing
the $\beta$-label to $(0,1,0,0)$ while keeping the $\gamma$ one
unaltered and equal to $(1,0,0,2)$. The corresponding solution
exhibits a $D=4$ true space-time singularity at finite $\tau$, since some of the $k_\ell$ are negative.\par
The signatures of the relevant tensor classifiers are:
\begin{align}
{\rm Sign}(\mathcal{T}_{xy})&=(0_+,\,0_-)\,,\nonumber\\
{\rm Sign}(T_{(21)}^{aA,bB})&=(3_+,\,2_-)\,,\nonumber\\
{\rm Sign}(T_{(84)}^{aA,bB})&=(3_+,\,2_-)\,,\nonumber\\
{\rm Sign}(\mathfrak{T}_{xy})&=(0_+,\,0_-)\,,\nonumber\\
{\rm Sign}(\mathbb{T}_{\alpha \beta})&=(0_+,\,0_-)\,,\nonumber\\
{\rm Sign}(\mathfrak{T}_{(84)}^{(AB),(CD)})&=(0_+,\,0_-)\,,\nonumber\\
%{\rm Sign}(\mathbb{T}^{(3,21)}_{aA,bB})&=(0_+,\,0_-)\,,\nonumber\\
%{\rm Sign}(\mathbb{T}^{(3,84)}_{aA,bB})&=(0_+,\,0_-)\,.\nonumber\\
\end{align}
\paragraph{Orbit $O'_{2H^*}$: singular BPS black hole.}
We can choose as representative of this orbit the matrix:
\begin{equation}
L_0=N_0^+-N_1^+\,,
\end{equation}
The $\gamma$- and $\beta$-labels are $(0,1,0,0)$ and $(1,0,0,2)$
respectively and the orbit is denoted by $O'_{2H^*}$. Since
$\mathcal{T}_{xy}$ is non-vanishing, the solution is non-BPS. Some of the $k\ell$ are negative implying a singularity at finite $\tau$.\par The
signatures of the relevant tensor classifiers are:
\begin{align}
{\rm Sign}(\mathcal{T}_{xy})&=(2_+,\,1_-)\,,\nonumber\\
{\rm Sign}(T_{(21)}^{aA,bB})&=(0_+,\,0_-)\,,\nonumber\\
{\rm Sign}(T_{(84)}^{aA,bB})&=(3_+,\,1_-)\,,\nonumber\\
{\rm Sign}(\mathfrak{T}_{xy})&=(0_+,\,0_-)\,,\nonumber\\
{\rm Sign}(\mathbb{T}_{\alpha \beta})&=(0_+,\,0_-)\,,\nonumber\\
{\rm Sign}(\mathfrak{T}_{(84)}^{(AB),(CD)})&=(2_+,\,1_-)\,,\nonumber\\
{\rm Sign}(\mathbb{T}^{(3,21)}_{aA,bB})&=(0_+,\,0_-)\,,\nonumber\\
{\rm Sign}(\mathbb{T}^{(3,84)}_{aA,bB})&=(0_+,\,0_-)\,.\nonumber\\
\end{align}
\paragraph{Orbit $O'_{3H^*}$: small, critical non-BPS black hole.}
We can choose as representative of this orbit the matrix:
\begin{equation}
L_0=N_0^++N_1^+\,,
\end{equation}
Both $\gamma$- and $\beta$-labels are $(0,1,0,0)$. The solution is
non-BPS and describes a small black hole whose charges are in the
$G_4$-orbit characterized by the properties (\ref{pqcritical}). It
is therefore a small critical non-BPS black hole. Its $D=4$
space-time geometry exhibits a singularity at $\tau\rightarrow
-\infty$. The signatures of the relevant tensor classifiers are:
\begin{align}
{\rm Sign}(\mathcal{T}_{xy})&=(1_+,\,2_-)\,,\nonumber\\
{\rm Sign}(T_{(21)}^{aA,bB})&=(0_+,\,0_-)\,,\nonumber\\
{\rm Sign}(T_{(84)}^{aA,bB})&=(3_+,\,1_-)\,,\nonumber\\
{\rm Sign}(\mathfrak{T}_{xy})&=(0_+,\,0_-)\,,\nonumber\\
{\rm Sign}(\mathbb{T}_{\alpha \beta})&=(0_+,\,0_-)\,,\nonumber\\
{\rm Sign}(\mathfrak{T}_{(84)}^{(AB),(CD)})&=(2_+,\,1_-)\,,\nonumber\\
{\rm Sign}(\mathbb{T}^{(3,21)}_{aA,bB})&=(0_+,\,0_-)\,,\nonumber\\
{\rm Sign}(\mathbb{T}^{(3,84)}_{aA,bB})&=(0_+,\,0_-)\,.\nonumber\\
\end{align}

\begin{table}[htbp]
\begin{center}
\scalefont{.9}
\begin{tabular}{|c|c|c|c|}\hline
\backslashbox{$\gamma$-label}{$\beta$-label}
&(1,\,0,\,0,\,2)&(0,\,1,\,0,\,0)&\null\\ \hline
(1,\,0,\,0,\,2)&
$\renewcommand{\arraystretch}{1}
\begin{array}{c}
L_0=N_0^-+N_1^+\end{array}$&$\renewcommand{\arraystretch}{1}
\begin{array}{c}
L_0=N_0^--N_1^+\end{array}$ &$\mathcal{O}_{\cdot H^*}$\\ \hline
(0,\,1,\,0,\,0)  & $L_0=N_0^+-N_1^+$ & $L_0=N_0^++N_1^+$&$\mathcal{O'}_{\cdot H^*}$\\ \hline
\null  & $\mathcal{O}_{2}$& $\mathcal{O}_{3}$&\null\\ \hline
\end{tabular}
\caption{\small The representatives of $\alpha^{(2)}$- orbits in terms of the generating solution.}
\label{alpha2g}
\end{center}
\end{table}

\subsubsection{$O_{1H^*}$: Small Doubly-Critical Black Holes ($\alpha^{(1)}$)}
By setting any of the three charges of the generating solution to
zero we end up in the orbit $O_{1H^*}$. We can take as
representative the matrices:
\begin{equation}
L_0=N_\ell^\pm\,,
\end{equation}
The electric-magnetic charges satisfy the   following general
properties
\begin{equation}
I_4(p,q)=0\,\,;\,\,\,\frac{\partial I_4}{\partial \Gamma^M}\equiv
0\,\,;\,\,\,\left.\frac{\partial^2 I_4}{\partial \Gamma^M\partial
\Gamma^N}\right\vert_{{\rm Adj}\mathfrak{g}_4}\equiv
0\,,\label{pqdcritical}
\end{equation}
which define a specific $G_4$-orbit of the representation ${\bf R}$
of the electric-magnetic charges. The solution is called
\emph{doubly critical small black hole}. It is BPS since
$\mathcal{T}_{xy}$ vanishes and has a singularity for
$\tau\rightarrow -\infty$. The signatures of the relevant tensor
classifiers are:
\begin{align}
{\rm Sign}(\mathcal{T}_{xy})&=(0_+,\,0_-)\,,\nonumber\\
{\rm Sign}(T_{(21)}^{aA,bB})&=(0_+,\,0_-)\,,\nonumber\\
{\rm Sign}(T_{(84)}^{aA,bB})&=(1_+,\,0_-)\,,\nonumber\\
{\rm Sign}(\mathfrak{T}_{xy})&=(0_+,\,0_-)\,,\nonumber\\
{\rm Sign}(\mathbb{T}_{\alpha \beta})&=(0_+,\,0_-)\,,\nonumber\\
{\rm Sign}(\mathfrak{T}_{(84)}^{(AB),(CD)})&=(0_+,\,0_-)\,,\nonumber\\
%{\rm Sign}(\mathbb{T}^{(3,21)}_{aA,bB})&=(0_+,\,0_-)\,,\nonumber\\
%{\rm Sign}(\mathbb{T}^{(3,84)}_{aA,bB})&=(0_+,\,0_-)\,.\nonumber\\
\end{align}

\begin{table}[htbp]
\scalefont{1}
\begin{center}
\begin{tabular}{|c|c|c|c|}\hline
\backslashbox{$\gamma$-label}{$\beta$-label}
&(0,\,0,\,1,\,1)&\null\\ \hline
(0,\,0,\,1,\,1)&
$\renewcommand{\arraystretch}{1}
\begin{array}{c}
L_0=N_\ell^\pm\end{array}$&$\mathcal{O}_{\cdot H^*}$\\ \hline

\null  & $\mathcal{O}_{1}$&\null\\ \hline
\end{tabular}
\caption{\small The representative of $\alpha^{(1)}$- orbit in terms of the generating solution.}
\label{alpha1g}
\end{center}
\end{table}
The various representatives of the $H^*$-orbits within $\alpha^{(1)}-\alpha^{(4)}$ $G^{\mathbb{C}}$-orbits, discussed above are listed in Tables \ref{alpha4g}-\ref{alpha1g}.

\subsection{Sum Rules}
From the explicit expression of the representatives of the various
orbits we can deduce \emph{sum rules}, namely express
representatives of each orbit as sum of representatives of orbits
with lower degree of nilpotency. This is relevant if we wish to
apply the orbit analysis to the study of black hole composites.\par
For instance the Lax representative (\ref{lo6h}) of the regular BPS
black hole orbit $O_{6H^*}$ is the sum of $N_0^-+N_1^+$ and
$N_4^-+N_6^+$, which both belong to the orbits $O_{2H^*}$ pertaining
to small critical BPS black holes. \par The representative
(\ref{lo7ph}) of the regular non-BPS orbit $\bar{O}'_{7H^*}$, on the
other hand,  is the sum either  of $N_0^++N_1^-$ and $N_4^++N_6^+$,
still both belonging to $O_{2H^*}$, or of $N_0^++N_4^+$  and
$N_1^-+N_6^+$, both in $O'_{3H^*}$ (critical non-BPS black hole).
\par The representative (\ref{lo8pph}) of the regular non-BPS orbit $\bar{O}''_{8H^*}$
can be written as $N_0^++N_1^+$ and $N_4^++N_6^+$ in $O'_{3H^*}$ and
$O_{2H^*}$, respectively. \par It is interesting to analyze the
representatives of the off-diagonal or the $\delta_2$-orbits, where
some of the $k_\ell $ are negative. In this case the Lax matrix can
still be written as sum of matrices belonging to
$\alpha^{(2)}$-orbits which contain small non singular solutions. In
this combination however the generator of a small black hole
component appears multiplied by $-1$. As a result, some of the
scalar fields start their flow at radial infinity with the
\emph{wrong} derivative and  a singularity is produced at finite
$\tau$. Take for instance the representative (\ref{lho7ph}) of
$\hat{O}'_{7H^*}$. It is the sum of the two Laxes: $N_0^++N_6^+$ and
$-(N_1^-+N_4^+)$ both in $O'_{3H^*}$. While $N_0^++N_6^+$ generates
a genuine small, critical non-BPS black hole, $-(N_1^-+N_4^+)$ would
generate one if we were to redefine $\tau\rightarrow -\tau$. For the
same $\tau<0$ the second matrix produces a solution with a
singularity at finite $\tau$. The same applies to the orbit
$\hat{O}''_{8H^*}$. \subsection{Asymptotics of the Generating
Solution} In this section we wish to comment on the behavior of the
generating solution at radial infinity and give a further
characterization of those $H^*$-orbits which were discarded as being
associated with singular solutions (here we refer to solutions
exhibiting a singularity at finite $\tau$).
 To this end we introduce a first order description of the
 generating solution in terms of a \emph{fake-superpotential} $W$
 \cite{Ceresole:2007wx,Andrianopoli:2007gt,Andrianopoli:2009je,Bossard:2009we,Ceresole:2009vp,Andrianopoli:2010bj}.
 This amounts to writing the four-dimensional fields in the generating solution as solutions to a
 first order ``gradient-flow'' system of equations of the form:
\begin{align}
\dot{U}=e^U\,W\,\,;\,\,\,\dot{\phi}^r=2\,e^U\,G^{rs}\,\frac{\partial
W}{\partial \phi^s}\,,\label{1ord}
\end{align}
 defined by a duality invariant
 function  $W(\phi^r,\,\Gamma)$ of the scalars $\phi^r$ and the
 quantized charges $\Gamma^M$.
 \par
It is straightforward to verify that the generating solution, in the
physical domain where it is well defined (${\bf H}_\ell>0$), is
described by a first order system of the form (\ref{1ord}), with $W$ given by
\begin{align}
W_{gen}&=W_{gen}(\varphi_i,\,\Gamma)=\frac{e^{-\frac{\varphi_1+\varphi_2+\varphi_3}{2}}}{4}\left(k_0+k_1\,e^{\varphi_2+\varphi_3}+k_4\,e^{\varphi_1+\varphi_3}
+k_6\,e^{\varphi_1+\varphi_2}\right)=\nonumber\\&=\frac{e^{-\frac{\varphi_1+\varphi_2+\varphi_3}{2}}}{2\sqrt{2}}
\left(-\epsilon_0\,q_0+\epsilon_1\,p^1\,e^{\varphi_2+\varphi_3}+\epsilon_4\,p^4\,e^{\varphi_1+\varphi_3}
+\epsilon_6\,p^6\,e^{\varphi_1+\varphi_2}\right)\,.\label{Wgen}
\end{align}
Note that, for given charges, it depends on $\epsilon_\ell$, that
is, in light of eq. (\ref{gabela}), only on the $\gamma$-label, the  $\beta$-label being fixed by  the charges
$\epsilon_\ell\,k_\ell$. The value of $W$ on the solution, at radial infinity, is
the ADM mass (\ref{ADMass}) of the solution:
\begin{equation}
\lim_{\tau\rightarrow 0^-}W_{gen}=\lim_{\tau\rightarrow
0^-}e^{-U}\dot{U}=M_{ADM}=\frac{1}{4}\sum_{\ell}
k_\ell\,\label{ADMass2}
\end{equation}
It is useful, at this point, to write the explicit expression of the central and matter charges for the $STU$ model truncation. Using the notation of Sect. \ref{skg4m}, the definitions (\ref{ZZI}) and  the identification (\ref{stupara}), we find for $p^0=q_\ell=0$, $\ell=1,4,6$:
\begin{align}
Z&=e^{\frac{\mathcal{K}}{2}}\,(q_0-p^1\,tu-p^4\,su-p^6\,st)\,,\nonumber\\
Z_1&=e^{\frac{\mathcal{K}}{2}}\,(q_0-p^1\,tu-p^4\,\bar{s}u-p^6\,\bar{s}t)\,,\nonumber\\
Z_4&=e^{\frac{\mathcal{K}}{2}}\,(q_0-p^1\,\bar{t}u-p^4\,{s}u-p^6\,{s}\bar{t})\,,\nonumber\\
Z_6&=e^{\frac{\mathcal{K}}{2}}\,(q_0-p^1\,{t}\bar{u}-p^4\,{s}\bar{u}-p^6\,{s}{t})\,.\nonumber
\end{align}
On the dilatonic generating solutions the above charges read:
\begin{align}
Z&=\frac{1}{2\sqrt{2}}\,e^{-\frac{\varphi_1+\varphi_2+\varphi_3}{2}}\,(q_0+p^1\,e^{\varphi_2+\varphi_3}+
p^4\,e^{\varphi_1+\varphi_3}+p^6\,e^{\varphi_1+\varphi_2})\,,\nonumber\\
Z_1&=\frac{1}{2\sqrt{2}}\,e^{-\frac{\varphi_1+\varphi_2+\varphi_3}{2}}\,(q_0+p^1\,e^{\varphi_2+\varphi_3}-
p^4\,e^{\varphi_1+\varphi_3}-p^6\,e^{\varphi_1+\varphi_2})\,,\nonumber\\
Z_4&=\frac{1}{2\sqrt{2}}\,e^{-\frac{\varphi_1+\varphi_2+\varphi_3}{2}}\,(q_0-p^1\,e^{\varphi_2+\varphi_3}+
p^4\,e^{\varphi_1+\varphi_3}-p^6\,e^{\varphi_1+\varphi_2})\,,\nonumber\\
Z_6&=\frac{1}{2\sqrt{2}}\,e^{-\frac{\varphi_1+\varphi_2+\varphi_3}{2}}\,(q_0-p^1\,e^{\varphi_2+\varphi_3}-
p^4\,e^{\varphi_1+\varphi_3}+p^6\,e^{\varphi_1+\varphi_2})\,.\nonumber
\end{align}
It is a known result that regular BPS black holes are described by a fake superpotential which is the modulus of the central charge $W=W_{BPS}=|Z|$. On the other hand, regular non-BPS solutions with $I_4>0$ are described by a $W$ which coincides with the modulus of one of the matter charges: $W=W_{nBPS,I_4>0}=|Z_\ell|,\,\ell=1,4,6$. More subtle is the first order  description of regular non-BPS solutions with $I_4<0$, for which an explicit duality invariant expression for $W$ is not known\footnote{In \cite{Bossard:2009we} $W^2$ is characterized as a root of an degree six polynomial while in \cite{Andrianopoli:2009je,Andrianopoli:2010bj} an implicit integral form of $W$ is given.}. A non-duality-invariant form $W_{nBPS,I_4<0}$ of $W$ is given in  \cite{Ceresole:2007wx}, which describes the \emph{seed} (or generating) solution of this class in $D=4$ \cite{Gimon:2007mh}. We can consider the $q_0<0,\,p^\ell>0$ representative of the corresponding orbit for which $W_{nBPS,I_4<0}$ reads
\begin{equation}
W_{nBPS,I_4<0}=e^{\frac{\mathcal{K}}{2}}\,\left(-q_0+\frac{p^1}{2}\,(\bar{t}u+u\bar{t})+
\frac{p^4}{2}\,(\bar{s}u+u\bar{s})+\frac{p^6}{2}\,(\bar{t}s+s\bar{t})\right)\,.
\end{equation}
The above expression  on the dilatonic generating solution becomes
\begin{equation}
W_{nBPS,I_4<0}=\frac{1}{2\sqrt{2}}\,e^{-\frac{\varphi_1+\varphi_2+\varphi_3}{2}}\,(-q_0+p^1\,e^{\varphi_2+\varphi_3}+
p^4\,e^{\varphi_1+\varphi_3}+p^6\,e^{\varphi_1+\varphi_2})\,.\label{Wnbpsm0}
\end{equation}
Consider first the regular solutions $k_\ell>0$ for which, as we have seen, the $\gamma$ and $\beta$-labels coincide. We see that the $BPS$ orbit $O_{6H^*}$ with $\gamma$-label $(0,0,0,4)$ has $\epsilon_0=-1$ and $\epsilon_1=\epsilon_4=\epsilon_6=1$ and, from (\ref{Wgen}), we find the known result:
\begin{equation}
W_{gen}=|W_{gen}|=|Z|\,,
\end{equation}
namely the fake superpotential for the regular BPS black holes is the modulus of the central charge $Z$.
As far as the  non-BPS orbit $\bar{O}'_{7H^*}$ is concerned, its representative is obtained by  inverting the signs of two of the $\epsilon_\ell$ with respect to the BPS case. Again from (\ref{Wgen}) we see that:
\begin{equation}
W_{gen}=|W_{gen}|=|Z_{\ell'}|\,,
\end{equation}
where  $|Z_{\ell'}|$ is the largest among all the $|Z_\ell|$ and $|Z|$ for the given set of charges.
Consider now the non-BPS orbit $\bar{O}''_{8H^*}$ with $\gamma$-label $(0,0,2,2)$ and the regular  representative
with $\epsilon_\ell=+1$, $\ell=0,1,4,6$, so that $q_0<0,\,p^1,p^4,p^6>0$. Comparing (\ref{Wgen}) to (\ref{Wnbpsm0}) we indeed find that:
\begin{equation}
W_{gen}=|W_{gen}|=W_{nBPS,I_4<0}\,.
\end{equation}
For other signs of $q_0,\,p^1,\,p^4,\,p^6$ within the same $I_4<0$ orbit of $G_4$, we can use the corresponding $W_{gen}$ as a definition of $W_{nBPS,I_4<0}$ on the generating solution, i.e. the expression (\ref{Wgen}) in which $\epsilon_\ell$ are chosen so that $\epsilon=-\epsilon_0\epsilon_1\epsilon_4\epsilon_6=-1$.\par
For all the regular representatives the ADM mass reads:
\begin{equation}
M_{ADM}=\lim_{\tau\rightarrow 0^-}W_{gen}=\frac{1}{2\sqrt{2}}\,(|q_0|+\sum_{\ell=1,4,6} |p^\ell|)=W_{max}\,,\label{Wmax}
\end{equation}
and it is clearly the largest among the fake superpotentials $|Z|,\,|Z_\ell|,\,W_{nBPS,I_4<0}$ computed on the same charges at infinity. Take for instance the BPS regular solution with $q_0,\,p^1,\,p^4,\,p^6>0$:
\begin{equation}
M_{ADM}=\lim_{\tau\rightarrow 0^-}|Z|=\frac{1}{2\sqrt{2}}\,(q_0+\sum_{\ell=1,4,6} p^\ell)=W_{max}>\lim_{\tau\rightarrow 0^-}|Z_\ell|,\,\lim_{\tau\rightarrow 0^-} W_{nBPS,I_4<0}\,,
\end{equation}
being
\begin{align}
\lim_{\tau\rightarrow 0^-}|Z_1|&=\frac{1}{2\sqrt{2}}\,|(q_0+p^1-p^4-p^6)|\,\,;\,\,\,\lim_{\tau\rightarrow 0^-}|Z_4|=\frac{1}{2\sqrt{2}}\,|(q_0-p^1+p^4-p^6)|\,,\nonumber\\
\lim_{\tau\rightarrow 0^-}|Z_6|&=\frac{1}{2\sqrt{2}}\,|(q_0-p^1-p^4+p^6)|\,,\nonumber\\
\lim_{\tau\rightarrow 0^-}W_{nBPS,I_4<0}&=\frac{1}{2\sqrt{2}}\,\vert(-\epsilon_0\,q_0+\epsilon_1\,p^1+\epsilon_4\,p^4+\epsilon_6\,p^6)
\vert_{\epsilon_0\epsilon_1\epsilon_4\epsilon_6=+1}\,.
\end{align}
This suggests a characterization of regularity in terms of the black hole asymptotics \cite{Bossard:2009we}: Regular BPS and non-BPS  solutions should satisfy a \emph{generalized BPS bound}, i.e. their ADM mass should be larger than any of the fake superpotentials $|Z|,\,|Z_\ell|,\,W_{nBPS,I_4<0}$, computed on the same charges at infinity. For extremal solutions the bound is saturated and $M_{ADM}$ should coincide with the largest of these values, which is $W_{max}$ in (\ref{Wmax}). This condition is not satisfied by representatives for which some of the $k_\ell$, $\ell=0,1,4,6$, are negative. These include the orbits for which the $\gamma$ and $\beta$-labels are different, but also the two orbits $\hat{O}'_{7H^*}$ and $\hat{O}''_{8H^*}$. In these cases:
\begin{equation}
M_{ADM}=\frac{1}{4}\sum_\ell k_\ell<\frac{1}{4}\sum_\ell |k_\ell|=W_{max}\,,
\end{equation}
and the generalized BPS bound is not satisfied. These are the solutions which exhibit a singularity at finite $\tau$ and, by acting on them with the ${\rm SO}(1,1)^4$ isotropy group, the $k_\ell$ can be rescaled so as to obtain a representative of the same orbit with negative ADM mass, which is clearly unphysical.

\subsection{Non-Extremal Solutions}
As pointed out earlier, a diagonalizable $L_0$ can always be
$H^*$-rotated into the Cartan subalgebra in the coset
$\mathcal{M}_N$, i.e. in the space
$\prod_{\ell}[\mathfrak{sl}(2)\ominus \mathfrak{so}(1,1)]_\ell$. In
order for the solution non to have a true space-time singularity at
some finite value of the radial parameter, $L_0$ must have real
eigenvalues only, and thus be expressed as a combination of the
non-compact Cartan generators in the coset space:
\begin{equation}
L_0=k_0\,T_0+\sum_{i=1}^3 k_i\,{\bf h}_i\,,
\end{equation}
where we have chosen as a basis for the non-compact Cartan
generators in the coset $\{T_0,\,{\bf h}_i\}$.\par Upon imposing the
regularity condition (\ref{regcond}) we still find 3 orbits.
\paragraph{The Schwarzschild Orbit}
It corresponds to choosing $k_0=1$ and $k_i\equiv 0$. In this case
$c^2=1/4$ and
\begin{equation}
U=c\,\tau\,\,,\,\,\,\phi^r=\mathcal{Z}^M=a\equiv 0 \,.
\end{equation}
With reference to the conventions defined in Sect. \ref{sbhg}, we
can calculate the horizon area to be:
\begin{equation}
A_H=4\pi\,\lim_{\tau\rightarrow -\infty}\frac{c^2\,e^{-2U}}{{\rm
sinh}^2(c\tau)}=4\pi\,(r^+)^2=4\pi\,(2c)^2\,,
\end{equation}
from which we deduce that $r^+=r_0+c=2c$, $r^-=r_0-c=0$. From the
general relation between $\tau$ and $r$ we find:
\begin{equation}
e^U=e^{c\tau}=\sqrt{1-\frac{2c}{r}}\,,
\end{equation} so that:
\begin{equation}
ds^2=-\left(1-\frac{2c}{r}\right)\,dt^2+\frac{dr^2}{\left(1-\frac{2c}{r}\right)}+r^2\,(d\theta^2+\sin^2(\theta)\,d\varphi^2)\,,
\end{equation}
and we retrieve the familiar Schwarzschild metric for $c=
GM/\tilde{c}^2$ ($\tilde{c}$ being the speed of light).
 The signatures of
the relevant tensor classifiers are:
\begin{align}
{\rm Sign}(\mathcal{T}_{xy})&=(1_+,\,12_-)\,,\nonumber\\
{\rm Sign}(T_{(21)}^{aA,bB})&=(4_+,\,24_-)\,,\nonumber\\
{\rm Sign}(T_{(84)}^{aA,bB})&=(27_+,\,1_-)\,,\nonumber\\
{\rm Sign}(\mathfrak{T}_{xy})&=(0_+,\,1_-)\,,\nonumber\\
{\rm Sign}(\mathbb{T}_{\alpha \beta})&=(0_+,\,1_-)\,,\nonumber\\
{\rm Sign}(\mathfrak{T}_{(84)}^{(AB),(CD)})&=(0_+,\,1_-)\,,\nonumber\\
{\rm Sign}(\mathbb{T}^{(3,21)}_{aA,bB})&=(24_+,\,4_-)\,,\nonumber\\
{\rm Sign}(\mathbb{T}^{(3,84)}_{aA,bB})&=(27_+,\,1_-)\,.\nonumber\\
\end{align}
\paragraph{Second (singular) orbit.}
It corresponds to taking $k_0=k_2=k_3=0$ and $k_1=1$. In this case
we still have $c^2=1/4$ and the only non vanishing field is
$\varphi_1=\tau$. We can compute on the space-time metric
\begin{equation}
R_{\mu\nu\rho\sigma}R^{\mu\nu\rho\sigma}=\frac{12}{c^2}\,{\rm
sinh}^8(c\tau)\,,
\end{equation}
which explodes at $\tau\rightarrow -\infty$, signalling a naked
singularity, with no horizon to cover it: $A_H=0$.\par The
signatures of the relevant tensor classifiers are:
\begin{align}
{\rm Sign}(\mathcal{T}_{xy})&=(5_+,\,8_-)\,,\nonumber\\
{\rm Sign}(T_{(21)}^{aA,bB})&=(12_+,\,16_-)\,,\nonumber\\
{\rm Sign}(T_{(84)}^{aA,bB})&=(19_+,\,9_-)\,,\nonumber\\
{\rm Sign}(\mathfrak{T}_{xy})&=(0_+,\,1_-)\,,\nonumber\\
{\rm Sign}(\mathbb{T}_{\alpha \beta})&=(0_+,\,1_-)\,,\nonumber\\
{\rm Sign}(\mathfrak{T}_{(84)}^{(AB),(CD)})&=(0_+,\,1_-)\,,\nonumber\\
{\rm Sign}(\mathbb{T}^{(3,21)}_{aA,bB})&=(16_+,\,12_-)\,,\nonumber\\
{\rm Sign}(\mathbb{T}^{(3,84)}_{aA,bB})&=(19_+,\,9_-)\,.\nonumber\\
\end{align}
\paragraph{Third (singular) orbit.}
It corresponds to choosing $k_0=k_i=1$. In this case $c^2=1$ and the
only non vanishing fields are:
\begin{equation}
\varphi_i=\tau\,\,,\,\,\,U=\frac{\tau}{2}\,.
\end{equation}
The horizon area is still zero and
\begin{equation}
R_{\mu\nu\rho\sigma}R^{\mu\nu\rho\sigma}=\frac{3\,e^{-6\tau}}{1024}
\,(1-e^{2\tau})^6\,(5-10\,e^{2\tau}+69\,e^{4\tau})\,,
\end{equation}
which diverges as $\tau\rightarrow -\infty$, signalling a true
space-time naked singularity.\par The signatures of the relevant
tensor classifiers are:
\begin{align}
{\rm Sign}(\mathcal{T}_{xy})&=(6_+,\,7_-)\,,\nonumber\\
{\rm Sign}(T_{(21)}^{aA,bB})&=(14_+,\,14_-)\,,\nonumber\\
{\rm Sign}(T_{(84)}^{aA,bB})&=(15_+,\,13_-)\,,\nonumber\\
{\rm Sign}(\mathfrak{T}_{xy})&=(1_+,\,0_-)\,,\nonumber\\
{\rm Sign}(\mathbb{T}_{\alpha \beta})&=(1_+,\,0_-)\,,\nonumber\\
{\rm Sign}(\mathfrak{T}_{(84)}^{(AB),(CD)})&=(1_+,\,0_-)\,,\nonumber\\
{\rm Sign}(\mathbb{T}^{(3,21)}_{aA,bB})&=(14_+,\,14_-)\,,\nonumber\\
{\rm Sign}(\mathbb{T}^{(3,84)}_{aA,bB})&=(13_+,\,15_-)\,.\nonumber\\
\end{align}
\section{Orbits with Higher Degree of Nilpotency}\label{owhdn}
In this section we briefly discuss  single center solutions whose
Lax matrix belong to some of the  $G^{\mathbb{C}}$-orbits with
degree of nilpotency higher than 3, identified with the
$\alpha^{(5)},\dots \alpha^{(15)}$ labels. The corresponding  $H^*$
orbits are described in Tables \ref{alpha5}-\ref{alpha15} and in
Table \ref{orbitstc}.  \par The regularity condition given earlier
rules these orbits out. We shall give some examples of single center
solutions which are indeed lifted to singular space-times. In light
of the analysis in \cite{Bossard:2011kz}, these solutions can be
viewed as singular limits of multicenter ones in which two or more
centers coincide. The Noether charge matrix will then be the sum of
the charges associated with each center. In this resect it is then
useful to express representatives of these higher-degree orbits as
sums of representatives of lower-degree ones discussed in the
previous section. In a forthcoming work we shall analyze the all
these higher degree $H^*$-orbits in terms of multicenter
representatives.
\par
Let us choose as non-compact Cartan subalgebra $\mathcal{C}$ of $\mathfrak{f}_4$  the one in $\mathfrak{H}^*\bigcap\mathfrak{K}$ generated by the $\mathcal{J}_\ell $ generators defined in Sect. \ref{generatingsol}. The generators $\bar{H}_i$, $i=1,2,3,4$,  corresponding to  the orthonormal basis $(\epsilon_i)$ of the $\mathfrak{f}_{4(4)}$-root space are:
\begin{equation}
\bar{H}_1=\mathcal{J}_0+\mathcal{J}_1\,\,;\,\,\,\bar{H}_2=-\mathcal{J}_0+\mathcal{J}_1\,\,;\,\,\,\bar{H}_3=\mathcal{J}_4+\mathcal{J}_6\,\,;\,\,\,\bar{H}_4=-\mathcal{J}_4+\mathcal{J}_6\,.
\end{equation}
Diagonalizing the adjoint action of this  basis on $\mathfrak{K}^*$
we can build a basis of the space  consisting of the shift
generators $\hat{E}_k$ and $\hat{F}_k=\hat{E}^T_k$ corresponding to
the 24 $\mathfrak{f}_{4(4)}$-roots listed in Table \ref{f44posr}.
These generators are listed in Table \ref{Tablehat} of Appendix
\ref{apptech}. As examples we shall work out in detail the
(singular) single center solutions corresponding to the
$\alpha^{(5)}$ $H^*$-orbit and the diagonal (i.e. having equal
$\gamma$- and $\beta$ -labels) $H^*$-orbits within the
$\alpha^{(7)}$ $G^{\mathbb{C}}$-orbit.
 \paragraph{The orbit $\alpha^{(5)}$.}
 The degree of nilpotency of this orbit is $5$.
 We consider a representative of this orbit of the form:
 \begin{equation}
 L_0=L^{(1)}_0+L^{(2)}_0\,\,;\,\,\,L^{(1)}_0=-2\,\hat{F}_{20}\,\,;\,\,\,L^{(2)}_0=2\, \hat{E}_{19}\,.
 \end{equation}
The components $L^{(1)}_0$ and $L^{(2)}_0$ both belong to the  orbit $O_{3H^*}$ and thus generate small, critical BPS black holes. They do not commute. \par
The solution reads:
\begin{align}
e^{-2U}&= 1-2 \tau^2\,\,;\,\,\,e^{-\varphi_1}=e^{-\varphi_2}=e^{-\varphi_3}=1+2 \tau^2\,\,;\,\,\,\alpha^2=\alpha^3=-\frac{\sqrt{2}\,\tau}{1+2\,\tau^2}=\frac{\lambda^5}{\sqrt{2}\tau}\,,\nonumber\\
\mathcal{Z}^2&=\frac{\sqrt{2} \tau}{2 \tau^2-1}=-\mathcal{Z}^3=-\frac{1}{\tau}\,\mathcal{Z}_4=\frac{1}{\tau}\,\mathcal{Z}_6\,,
\end{align}
all other fields being zero.
We see that $e^{-U}$ vanishes at finite $\tau$ and the four dimensional space-time has a true singularity.
It is tempting to interpret this solution as the singular limit of a two-center one, each center being a small BPS black hole described by $L^{(1)}_0$ and $L^{(2)}_0$, respectively. In this case the electric-magnetic charge vectors $\Gamma^{(1)}$ and $\Gamma^{(2)}$ are:
\begin{align}
\Gamma^{(1)}&=(p^\Lambda,q_\Sigma)=(0,0,0,0,0,0,0,0,0,-\sqrt{2},0,0,0,0)\,,\nonumber\\
\Gamma^{(2)}&=(p^\Lambda,q_\Sigma)=(0,0,0,0,0,0,0,0,0,0,\sqrt{2},0,0,0)\,.
\end{align}
The two charges are mutually local: $\Gamma^{(1)\,T}\mathbb{C}\Gamma^{(2)}=0$.

 \paragraph{The orbit $\alpha^{(7)}$.}
 The degree of nilpotency of this orbit is $5$.
 We consider first a representative of the orbit $O_{11H^*}$, identified by the $\gamma$- and $\beta$-labels
being both $(1,0,2,4)$, in the form:
   \begin{equation}
 L_0=L^{(1)}_0+L^{(2)}_0\,\,;\,\,\,L^{(1)}_0=-2\,\hat{F}_{16}\,\,;\,\,\,L^{(2)}_0=\sqrt{3}\,\hat{E}_{2}-\sqrt{3}\,\hat{E}_{12}\,.
 \end{equation}
One can verify that $L^{(1)}_0$ and $L^{(1)}_1$, which are non-commuting,  lie in the orbits $O_{1H^*}$ and $O'_{2H^*}$, respectively. The corresponding electric-magnetic charges $\Gamma^{(1)},\,\Gamma^{(2)}$ are mutually local: $\Gamma^{(1)\,T}\mathbb{C}\Gamma^{(2)}=0$.\par
The solution reads:
{\small \begin{align}
e^{-4U}&= -(\tau+1)^4 \left(3 \tau^2-1\right)\,\,;\,\,\,e^{-2 \varphi_1}=-\frac{\{3 \tau [\tau (4 \tau (\tau+2)+7)+4]+4\}^2}{16 (\tau+1)^4 \left(3 \tau^2-1\right)}\,,\nonumber\\ e^{-2\varphi_2}&=e^{2\varphi_3}=\frac{(\tau+1)^2}{1-3 \tau^2}\,\,;\,\,\,\alpha^1=-\frac{\sqrt{3} \tau \left(4 \tau^3-7 \tau-4\right)}{12 \tau^4+24 \tau^3+21 \tau^2+12 \tau+4}\,,\,\,\,\nonumber\\
\mathcal{Z}^4&=\frac{\tau \left(4 \tau^2+6 \tau+3\right)}{2 \sqrt{2} (\tau+1)^3}\,,\,\,\mathcal{Z}^6=-\frac{\tau (6 \tau+5)}{2 \sqrt{2} (\tau+1) \left(3 \tau^2-1\right)}\,,\,\,\mathcal{Z}_4=\frac{\sqrt{\frac{3}{2}} \tau \left(4 \tau^2+2 \tau-1\right)}{-6 \tau^3-6 \tau^2+2 \tau+2}\,,\nonumber\\ \mathcal{Z}_6&=\frac{\sqrt{\frac{3}{2}} \tau (2 \tau+1)}{2 (\tau+1)^3}\,\,,\,\,\,a=-\frac{\sqrt{3} \tau^4}{(\tau+1)^2 \left(3 \tau^2-1\right)}\,,
\end{align}}
 Notice that, even if $a(\tau)\neq 0$, the NUT charge, which is proportional to $n=-e^{-4U}\,(\dot{a}+\mathcal{Z}^T\mathbb{C}\dot{\mathcal{Z}})$, vanishes and  the $D=4$ metric is diagonal.
 We see that the $D=4$ space-time is singular since $e^{-U}$ vanishes a finite $\tau$.\par
 Next we consider the other orbit in the diagonal of Table \ref{alpha7}: $O'_{12H^*}$ with $\gamma$- and $\beta$-labels
 both equal to $(0,1,2,2)$.
 The representative we choose has the form:
   \begin{equation}
 L_0=L^{(1)}_0+L^{(2)}_0\,\,;\,\,\,L^{(1)}_0=2\,\hat{E}_{13}\,\,;\,\,\,L^{(2)}_0=\sqrt{3}\,\hat{F}_{15}-
 \sqrt{3}\,\hat{E}_{2}\,.
 \end{equation}
 One can verify that $L^{(1)}_0$ and $L^{(1)}_1$, which are non-commuting,  lie in the orbits $O_{1H^*}$ and $O_{3H^*}$, respectively. The corresponding electric-magnetic charges $\Gamma^{(1)},\,\Gamma^{(2)}$ are mutually local: $\Gamma^{(1)\,T}\mathbb{C}\Gamma^{(2)}=0$.\par
 The solution reads:
{\small \begin{align}
e^{-4U}&= (\tau-1)^2 \left(1-4 \tau^2\right)=e^{-2 \varphi_1}\,\,;\,\,\, e^{-2\varphi_2}=\frac{1-2 \tau}{(\tau-1)^2 (2 \tau+1)}\,,\,\,e^{-2\varphi_3}=\frac{\left(1-4 \tau^3\right)^2}{(\tau-1)^2 \left(1-4 \tau^2\right)}\nonumber\\ \alpha^4&=\sqrt{3}\,\tau\,,\,\,\alpha^6=\frac{\sqrt{3} \tau (2 \tau-1)}{4 \tau^3-1}\,,\,\,\,\nonumber\\
\mathcal{Z}^1&=\frac{\sqrt{\frac{3}{2}} \tau}{(\tau-1)^2 (2 \tau+1)}\,,\,\,\mathcal{Z}_0=-\frac{\sqrt{\frac{3}{2}} \tau}{(\tau-1)^2 (2 \tau+1)}\,,\,\,\mathcal{Z}_4=\frac{(3-2 \tau) \tau^2}{\sqrt{2} (\tau-1)^2 (2 \tau+1)}\,,\nonumber\\ \mathcal{Z}_6&=\frac{\tau \left(4 \tau^3-3 \tau+2\right)}{\sqrt{2} (\tau-1)^2 \left(4 \tau^2-1\right)}\,,
\end{align}}
 We see that the $D=4$ space-time is singular since $e^{-U}$ vanishes a finite $\tau$.

\section{ Concluding Remarks}
In this work we have considered the description of  static black holes in a specific $\mathcal{N}=2$ model in terms of geodesics on a pseudo-quaternionic K\"ahler symmetric manifold. We have posed the general question: Given a geodesic on this manifold, can it be related to one describing a black hole solution in $D=4$?
We showed that answering this question requires a classification of the initial velocity vectors of  geodesics with respect to the action of the isotropy group $H^*$, which is what we have accomplished.\par
By referring to the general arguments in \cite{Fre:2011ns} we expect this answer, given in terms of classification of $H^*$-orbits of vectors on the tangent space at the origin, to apply to all the $\mathcal{N}=2$ models lying in the same Tits-Satake universality class as the one considered here. The $c^*$-map chain of embedding for these models read:
\begin{align}
\mathcal{M}_{D=5}&=\frac{{\rm SL}(3,\mathbb{C})}{{\rm
SU}(3)}\stackrel{r-map}{\longrightarrow}\mathcal{M}_{SK}=\frac{{\rm
SU}(3,3)}{{\rm
U}(3)\times{\rm SU}(3)}\stackrel{c*-map}{\longrightarrow}\frac{{\rm E}_{6(2)}}{{\rm
SL}(2,\mathbb{R})\times {\rm
SU}(3,3)}\,,\nonumber\\
\mathcal{M}_{D=5}&=\frac{{\rm SU}^*(6)}{{\rm
Sp}(6)}\stackrel{r-map}{\longrightarrow}\mathcal{M}_{SK}=\frac{{\rm
SO}^*(12)}{{\rm
U}(6)}\stackrel{c*-map}{\longrightarrow}\frac{{\rm E}_{7(-5)}}{{\rm
SL}(2,\mathbb{R})\times {\rm
SO}^*(12)}\,,\nonumber\\
\mathcal{M}_{D=5}&=\frac{{\rm E}_{6(-26)}}{{\rm F}_{4(-52)}}\stackrel{r-map}{\longrightarrow}\mathcal{M}_{SK}=\frac{{\rm E}_{7(-25)}}{{\rm E}_{6(-78)}\times {\rm U}(1)}\stackrel{c*-map}{\longrightarrow}\frac{{\rm E}_{8(-24)}}{{\rm
SL}(2,\mathbb{R})\times {\rm E}_{7(-25)}}\,,\nonumber\\
\end{align}
As anticipated in the introduction, we leave a formal proof of this property to a future investigation.\par
It would be interesting to understand the observed $\gamma,\,\beta$-label degeneracy of certain orbits, observed here for the first time, in terms of the geometric structure of the isotropy group $H^*$. As the tensor classifiers have proved to be a valuable tool for the orbit-classification, we believe it  worthwhile constructing a complete set of such tensors which would itself be sufficient for a complete classification, with no need of $\alpha,\beta$ and $\gamma$-labels. Such refined analysis would require constructing higher order $H^*$-symmetric, covariant tensors.\par
A next step of our analysis is also to apply this orbit classification to a systematic study of multicenter and/or rotating solutions.

%In this paper, our investigation goes much beyond the  Kostant-Sekiguchi's approach. As we previously pointed out the latter does
%not give a classification of $H^*$-orbits. As we consider $H^*$-orbits of the
%Lax, we realize that  the $\beta$-$\gamma$ classification is not complete.
%This  Kostant-Sekiguchi  classification is not directly relevant to our problem. We will
%classify three-dimensional  geodesics being equivalent to four-dimensional  static black holes through the $H^*$-orbit of the affine \emph{velocity vector}  of
%the geodesic, which is nothing else but the Lax operator. The  Kostant-Sekiguchi classification does not cover them. In particular,
%it does not give the $\gamma$-labels and, therefore, the fine structure....to be continued.
\section{Acknowledgement}
We are grateful to L. Borsten, Pietro Fr\'e,  A. Marrani,  Alexander S. Sorin and Thomas Van Riet for stimulating discussions. The work of W.C. is supported in part by the FWO - Vlaanderen, Project No. G.0651.11, and in part by the Federal Office for Scientific, Technical and Cultural Affairs through the Interuniversity Attraction Poles Programme - Belgian Science Policy P6/11-P, and is supported in part by the Natural Sciences and Engineering Research Council (NSERC) of Canada.
W.C. wish to thank Department of Applied Science and Technology of Politecnico di Torino for its hospitality and financial support.
\appendix
\section{Little group and the signature of the tensor classifiers}\label{AA} In this  Appendix we list the orbits with the
little compact group of $h$ in $H^*$, and the corresponding signatures of the tensor classifiers.

\begin{table}[htbp]
\small\addtolength{\tabcolsep}{-9pt}
\scalefont{.7}
\begin{center}
\begin{equation}
\begin{array}{|c|c|c|c|c|c|}
\hline
\multicolumn{6}{|c|}{\text{Little Group (LG) in $H^*$}}\\  \hline
\text{$H^*$-orbit}& \text{Dim.}& \#\,\text{of semisimple gen.}  &\#\,\text{of non-compact gen.} &\#\,\text{of compact gen.}&\textrm{compact part of   (LG)}\nn\\ \hline

\mathcal{O}_{1H^*} & 10 & 10 & 7 & 3&\textrm{SO}(3)\nn\\ \hline

\mathcal{O}_{2H^*}& 12 & 12 & 8 & 4& \textrm{SO}(3)\times \textrm{U}(1)\nn\\ \hline
\mathcal{O}_{3H^*}& 12 & 12 & 8 & 4& \textrm{SO}(3)\times \textrm{U}(1)\nn\\ \hline

\mathcal{O}'_{2H^*}& 10 & 10 & 7 & 3&\textrm{SO}(3)\nn\\ \hline
\mathcal{O}'_{3H^*}& 10 & 10 & 7 & 3&\textrm{SO}(3)\nn\\ \hline

\mathcal{O}_{4H^*}&  10 & 10 & 7 & 3&\textrm{SO}(3)\nn\\ \hline
\mathcal{O}_{5H^*}& 10 & 10 & 7 & 3&\textrm{SO}(3)\nn\\ \hline

\mathcal{O}'_{4H^*} & 6 & 6 & 5 & 1&\textrm{U}(1)\nn\\ \hline
\bar{\mathcal{O}}'_{5H^*} & 6 & 6 & 5 & 1&\textrm{U}(1)\nn\\ \hline
\hat{\mathcal{O}}'_{5H^*}& 6 & 6 & 5 & 1&\textrm{U}(1)\nn\\ \hline

\mathcal{O}_{6H^*}& 22 & 22 &13 & 9&\textrm{U}(3)\nn\\  \hline

\mathcal{O}_{7H^*} & 22 & 22 &13 & 9&\textrm{U}(3)\nn\\  \hline

\mathcal{O}_{8H^*} & 22 & 22 &13 & 9&\textrm{U}(3)\nn\\  \hline

\mathcal{O}'_{6H^*} & 14 & 14 &  9 & 5&\textrm{SO}(3)\times\textrm{U}(1)^2\nn\\ \hline

\bar{\mathcal{O}}'_{7H^*}&14 & 14 &  9 & 5&\textrm{SO}(3)\times\textrm{U}(1)^2\nn\\ \hline

\hat{\mathcal{O}}'_{7H^*}& 14 & 14 &  9 & 5&\textrm{SO}(3)\times\textrm{U}(1)^2\nn\\ \hline

\mathcal{O}'_{8H^*}&14 & 14 &  9 & 5&\textrm{SO}(3)\times\textrm{U}(1)^2\nn\\ \hline

\mathcal{O}''_{6H^*}& 10 & 10 & 7 & 3&\textrm{SO}(3)\nn\\ \hline

\mathcal{O}''_{7H^*}& 10 & 10 & 7 & 3&\textrm{SO}(3)\nn\\ \hline

\bar{\mathcal{O}}''_{8H^*}&0 & 10 & 7 & 3&\textrm{SO}(3)\nn\\ \hline

\hat{\mathcal{O}}''_{8H^*}& 0 & 10 & 7 & 3&\textrm{SO}(3)\nn\\ \hline

\mathcal{O}_{9H^*}& 0 & 10 & 7 & 3&\textrm{SO}(3)\nn\\ \hline

\mathcal{O}_{10H^*}&6 & 6 & 5 & 1&\textrm{U}(1)\nn\\ \hline

\mathcal{O}_{11H^*}& 6 & 6 & 5 & 1&\textrm{U}(1)\nn\\ \hline

\mathcal{O}_{12H^*}& 6 & 6 & 5 & 1&\textrm{U}(1)\nn\\ \hline

\mathcal{O}'_{11H^*}& 6 & 6 & 5 & 1&\textrm{U}(1)\nn\\ \hline

\mathcal{O}'_{12H^*}& 6 & 6 & 5 & 1&\textrm{U}(1)\nn\\ \hline

\mathcal{O}_{13H^*}& 4 & 4 & 4 & 0&\{{\bf 1}\}\nn\\  \hline

\mathcal{O}_{14H^*}& 6 & 6 & 5 & 1&\textrm{U}(1)\nn\\ \hline

\mathcal{O}_{15H^*}&6 & 6 & 5 & 1&\textrm{U}(1)\nn\\ \hline

\mathcal{O}'_{14H^*}& 4 & 4 & 4 & 0&\{{\bf 1}\}\nn\\  \hline

\mathcal{O}'_{15H^*}& 4 & 4 & 4 & 0&\{{\bf 1}\}\nn\\  \hline

\mathcal{O}_{16H^*}& 12 & 12 & 8 & 4& \textrm{SO}(3)\times \textrm{U}(1)\nn\\ \hline

\mathcal{O}_{17H^*}&12 & 12 & 8 & 4& \textrm{SO}(3)\times \textrm{U}(1)\nn\\ \hline

\mathcal{O}_{18H^*}& 12 & 12 & 8 & 4& \textrm{SO}(3)\times \textrm{U}(1)\nn\\ \hline

\mathcal{O}'_{16H^*}& 8 & 8 &6 & 2&\textrm{U}(1)^2\nn\\ \hline

\bar{\mathcal{O}}'_{17H^*}& 8 & 8 &6 & 2&\textrm{U}(1)^2\nn\\ \hline

\hat{\mathcal{O}}'_{17H^*}&8 & 8 &6 & 2&\textrm{U}(1)^2\nn\\ \hline

\mathcal{O}'_{18H^*}& 8 & 8 &6 & 2&\textrm{U}(1)^2\nn\\ \hline

\mathcal{O}''_{16H^*}& 6 & 6 & 5 & 1&\textrm{U}(1)\nn\\ \hline

\mathcal{O}''_{17H^*}&6 & 6 & 5 & 1&\textrm{U}(1)\nn\\ \hline

\bar{\mathcal{O}}''_{18H^*}& 6 & 6 & 5 & 1&\textrm{U}(1)\nn\\ \hline

\hat{\mathcal{O}}''_{18H^*}& 6 & 6 & 5 & 1&\textrm{U}(1)\nn\\ \hline

\mathcal{O}_{19H^*} & 10 & 10 & 7 & 3&\textrm{SO}(3)\nn\\ \hline

\mathcal{O}_{20 H^*} & 10 & 10 & 7 & 3&\textrm{SO}(3)\nn\\ \hline

\mathcal{O}'_{19 H^*}& 6 & 6 & 5 & 1&\textrm{U}(1)\nn\\ \hline

\bar{\mathcal{O}}'_{20 H^*}& 6 & 6 & 5 & 1&\textrm{U}(1)\nn\\ \hline

\hat{\mathcal{O}}'_{20 H^*}& 6 & 6 & 5 & 1&\textrm{U}(1)\nn\\ \hline

\mathcal{O}_{21 H^*}& 4 & 4 & 4 & 0&\{{\bf 1}\}\nn\\  \hline

\mathcal{O}_{22 H^*}& 8 & 8 &6 & 2&\textrm{U}(1)^2\nn\\ \hline

\mathcal{O}_{23 H^*}& 8 & 8 &6 & 2&\textrm{U}(1)^2\nn\\ \hline

\mathcal{O}'_{22 H^*}& 4 & 4 & 4 & 0&\{{\bf 1}\}\nn\\  \hline

\mathcal{O}'_{23 H^*}& 4 & 4 & 4 & 0&\{{\bf 1}\}\nn\\  \hline

\mathcal{O}_{24 H^*}& 4 & 4 & 4 & 0&\{{\bf 1}\}\nn\\  \hline

\mathcal{O}_{25 H^*}& 4 & 4 & 4 & 0&\{{\bf 1}\}\nn\\  \hline

\mathcal{O}'_{24 H^*}& 6 & 6 & 5 & 1&\textrm{U}(1)\nn\\ \hline

\mathcal{O}'_{25 H^*}& 6 & 6 & 5 & 1&\textrm{U}(1)\nn\\ \hline

\mathcal{O}_{26 H^*}& 4 & 4 & 4 & 0&\{{\bf 1}\}\nn\\  \hline

\end{array}
\end{equation}
\end{center}
\caption{Part of the little groups of $h$, connected to the
identity} \label{Hlittle0}
\end{table}

\begin{table}
\begin{center}
\small\addtolength{\tabcolsep}{-1pt}
\scalefont{.75}
\begin{tabular}{|c|r|r|r|r|r|r|r|r|r|r|}
\hline \multirow{1}{*}{$H^*$-Orbits} &
\multicolumn{8}{|c|}{$(n_{+},m_{-})$-signature  of quadratic and
quartic tensors}&\multicolumn{1}{|c|}{Solution} \cr \hline &
$\mathcal{T}_{xy}$ & $T_{(21)}^{a A,b B}$ &
$T_{(84)}^{a A,b
B}$&$\mathfrak{T}_{xy}$&$\mathbb{T}_{\alpha\beta}$&$\mathfrak{T}_{(84)}^{(AB),(CD)}$&$\mathbb{T}^{(3,21)}_{aA,\,bB}$&
$\mathbb{T}^{(3,84)}_{aA,\,bB}$ & \cr \hline $\mathcal{O}_{1H^*}$& $
(0_+,0_{-})$&$(0_+,0_{-})$ & $
(1_+,0_{-})$&$(0_+,0_{-})$&$(0_+,0_{-})$&$(0_+,0_{-})$&&&BPS \cr
\hline

$\mathcal{O}_{2H^*}$& $(0_+,0_{-})$& $(2_+,3_{-}) $&  $(4_+,1_{-})$  &$(0_+,0_{-})$&$(0_+,0_{-})$&$(0_+,0_{-})$&&&BPS \cr \hline

$\mathcal{O}_{3H^*}$& $ (0_+,0_{-})$&$ (3_+,2_{-}) $& $(3_+,2_{-})$&$(0_+,0_{-})$&$(0_+,0_{-})$&$(0_+,0_{-})$&&&BPS \cr \hline

$\mathcal{O}'_{2H^*}$& $ (2_+,1_{-})$&$ (0_+,0_{-}) $& $(3_+,1_{-})$&$(0_+,0_{-})$&$(0_+,0_{-})$&$(2_+,1_{-})$&$(0_+,0_{-})$&$(0_+,0_{-})$&non-BPS \cr \hline

$\mathcal{O}'_{3H^*}$& $ (1_+,2_{-})$&$ (0_+,0_{-}) $& $(3_+,1_{-})$ &$(0_+,0_{-})$&$(0_+,0_{-})$&$(2_+,1_{-})$&$(0_+,0_{-})$&$(0_+,0_{-})$&non-BPS \cr \hline

$\mathcal{O}_{4H^*}$& $ (0_+,0_{-})$&$ (2_+,6_{-}) $& $(7_+,1_{-})$ &$(0_+,0_{-})$&$(0_+,0_{-})$&$(0_+,0_{-})$&&&BPS \cr \hline

$\mathcal{O}_{5H^*}$& $ (0_+,0_{-})$&$ (4_+,4_{-}) $& $(5_+,3_{-})$&$(0_+,0_{-})$&$(0_+,0_{-})$&$(0_+,0_{-})$&&&BPS \cr \hline

$\mathcal{O}'_{4H^*}$& $ (3_+,1_{-})$&$ (2_+,3_{-}) $& $(5_+,1_{-})$&$(0_+,0_{-})$&$(0_+,0_{-})$&$(2_+,1_{-})$&$(0_+,0_{-})$&$(0_+,0_{-})$&non-BPS \cr \hline

$\bar{\mathcal{O}}'_{5H^*}$ & $ (1_+,3_{-})$&$ (2_+,3_{-}) $& $(5_+,1_{-})$ &$(0_+,0_{-})$&$(0_+,0_{-})$&$(2_+,1_{-})$&$(0_+,0_{-})$&$(0_+,0_{-})$&non-BPS \cr \hline

%9

$\hat{\mathcal{O}}'_{5H^*}$& $ (2_+,2_{-})$&$ (3_+,2_{-}) $& $(4_+,2_{-})$ &$(0_+,0_{-})$&$(0_+,0_{-})$&$(2_+,1_{-})$&$(0_+,0_{-})$&$(0_+,0_{-})$&non-BPS \cr \hline
%10

$\mathcal{O}_{6H^*}$& $ (0_+,0_{-})$&$ (2_+,12_{-}) $& $(13_+,1_{-})$&$(0_+,0_{-})$&$(0_+,1_{-})$&$(0_+,0_{-})$&&&BPS \cr \hline

$\mathcal{O}_{7H^*}$& $ (0_+,0_{-})$&$ (6_+,8_{-}) $& $(9_+,5_{-})$&$(0_+,0_{-})$&$(0_+,1_{-})$&$(0_+,0_{-})$&&&BPS \cr \hline
%12

$\mathcal{O}_{8H^*}$& $ (0_+,0_{-})$&$ (7_+,7_{-}) $& $(8_+,6_{-})$ &$(0_+,0_{-})$&$(1_+,0_{-})$&$(0_+,0_{-})$&&&BPS \cr \hline
%13

$\mathcal{O}'_{6H^*}$& $ (5_+,1_{-})$&$ (4_+,6_{-}) $& $(9_+,1_{-})$ &$(0_+,1_{-})$&$(0_+,0_{-})$&$(2_+,1_{-})$&$(0_+,0_{-})$&$(0_+,0_{-})$&non-BPS \cr \hline

$\bar{\mathcal{O}}'_{7H^*}$& $ (1_+,5_{-})$&$ (4_+,6_{-}) $& $(9_+,1_{-})$ &$(0_+,1_{-})$&$(0_+,0_{-})$&$(2_+,1_{-})$&$(0_+,0_{-})$&$(0_+,0_{-})$&non-BPS \cr \hline

$\hat{\mathcal{O}}'_{7H^*}$& $ (3_+,3_{-})$&$ (6_+,4_{-}) $& $(7_+,3_{-})$ &$(0_+,1_{-})$&$(0_+,0_{-})$&$(2_+,1_{-})$&$(0_+,0_{-})$&$(0_+,0_{-})$&non-BPS \cr \hline
%16

$\mathcal{O}'_{8H^*}$& $ (3_+,3_{-})$&$ (5_+,5_{-}) $& $(6_+,4_{-})$ &$(1_+,0_{-})$&$(0_+,0_{-})$&$(2_+,1_{-})$&$(0_+,0_{-})$&$(0_+,0_{-})$&non-BPS \cr \hline
%17

$\mathcal{O}''_{6H^*}$& $ (5_+,1_{-})$&$ (2_+,6_{-}) $& $(8_+,1_{-})$ &$(0_+,0_{-})$&$(0_+,0_{-})$&$(2_+,1_{-})$&$(0_+,0_{-})$&$(0_+,1_{-})$&non-BPS \cr \hline

$\mathcal{O}''_{7H^*}$& $ (3_+,3_{-})$&$ (4_+,4_{-}) $& $(6_+,3_{-})$&$(0_+,0_{-})$&$(0_+,0_{-})$&$(2_+,1_{-})$&$(0_+,0_{-})$&$(0_+,1_{-})$&non-BPS \cr \hline

$\bar{\mathcal{O}}''_{8H^*}$& $ (1_+,5_{-})$&$ (2_+,6_{-}) $& $(8_+,1_{-})$ &$(0_+,0_{-})$&$(0_+,0_{-})$&$(2_+,1_{-})$&$(0_+,0_{-})$&$(1_+,0_{-})$&non-BPS \cr \hline

$\hat{\mathcal{O}}''_{8H^*}$& $ (3_+,3_{-})$&$ (4_+,4_{-}) $& $(6_+,3_{-})$ &$(0_+,0_{-})$&$(0_+,0_{-})$&$(2_+,1_{-})$&$(0_+,0_{-})$&$(1_+,0_{-})$&non-BPS \cr \hline
%21

 $\mathcal{O}_{9H^*}$& $ (5_+,5_{-})$&$ (7_+,5_{-}) $&
$(7_+,5_{-})$
&$(2_+,1_{-})$&$(0_+,0_{-})$&$(2_+,1_{-})$&$(0_+,0_{-})$&$(0_+,0_{-})$&non-BPS
\cr \hline

$\mathcal{O}_{10H^*}$& $ (3_+,3_{-})$&$ (6_+,5_{-}) $& $(7_+,4_{-})$&$(1_+,0_{-})$&$(0_+,0_{-})$&$(2_+,1_{-})$&$(0_+,0_{-})$&$(1_+,1_{-})$&non-BPS \cr \hline
%23

$\mathcal{O}_{11H^*}$& $ (5_+,4_{-})$&$ (7_+,7_{-}) $& $(9_+,6_{-})$ &$(1_+,0_{-})$&$(1_+,0_{-})$&$(2_+,1_{-})$&$(3_+,2_{-})$&$(3_+,2_{-})$&non-BPS \cr \hline
%24

$\mathcal{O}_{12H^*}$& $ (4_+,5_{-})$&$ (7_+,7_{-}) $& $(9_+,6_{-})$ &$(1_+,0_{-})$&$(1_+,0_{-})$&$(2_+,1_{-})$&$(2_+,3_{-})$&$(2_+,3_{-})$&non-BPS \cr \hline
%25

$\mathcal{O}'_{11H^*}$& $ (5_+,5_{-})$&$ (6_+,8_{-}) $& $(9_+,5_{-})$ & $(1_+,0_{-})$&$(0_+,0_{-})$&$(2_+,1_{-})$&$(0_+,0_{-})$&$(2_+,2_{-})$&non-BPS \cr \hline

$\mathcal{O}'_{12H^*}$& $ (5_+,5_{-})$&$ (7_+,7_{-}) $& $(8_+,6_{-})$ &$(1_+,0_{-})$&$(0_+,0_{-})$&$(2_+,1_{-})$&$(0_+,0_{-})$&$(2_+,2_{-})$&non-BPS \cr \hline

$\mathcal{O}_{13H^*}$& $ (5_+,5_{-})$&$ (7_+,6_{-}) $& $(8_+,5_{-})$ &$(2_+,1_{-})$&$(0_+,0_{-})$&$(2_+,1_{-})$&$(0_+,0_{-})$&$(1_+,1_{-})$&non-BPS \cr \hline

$\mathcal{O}_{14H^*}$& $ (5_+,5_{-})$&$ (8_+,7_{-}) $& $(9_+,6_{-})$ &$(2_+,1_{-})$&$(0_+,0_{-})$&$(2_+,1_{-})$&$(0_+,0_{-})$&$(2_+,2_{-})$&non-BPS \cr \hline

$\mathcal{O}_{15H^*}$& $ (5_+,5_{-})$&$ (8_+,7_{-}) $& $(9_+,6_{-})$ &$(2_+,1_{-})$&$(0_+,0_{-})$&$(2_+,1_{-})$&$(0_+,0_{-})$&$(2_+,2_{-})$&non-BPS \cr \hline

%30

$\mathcal{O}'_{14H^*}$& $ (5_+,5_{-})$&$ (7_+,7_{-}) $& $(9_+,6_{-})$&$(1_+,0_{-})$&$(0_+,0_{-})$&$(2_+,1_{-})$&$(2_+,3_{-})$&$(2_+,3_{-})$&non-BPS \cr \hline

$\mathcal{O}'_{15H^*}$& $ (5_+,5_{-})$&$ (7_+,7_{-}) $& $(9_+,6_{-})$&$(1_+,0_{-})$&$(0_+,0_{-})$&$(2_+,1_{-})$&$(3_+,2_{-})$&$(3_+,2_{-})$&non-BPS \cr \hline

$\mathcal{O}_{16H^*}$& $ (5_+,5_{-})$&$ (8_+,8_{-}) $& $(10_+,6_{-})$ &$(2_+,1_{-})$&$(0_+,0_{-})$&$(2_+,1_{-})$&$(0_+,0_{-})$&$(2_+,2_{-})$&non-BPS \cr \hline

$\mathcal{O}_{17H^*}$& $ (5_+,5_{-})$&$ (8_+,8_{-}) $& $(10_+,6_{-})$ &$(2_+,1_{-})$&$(0_+,0_{-})$&$(2_+,1_{-})$&$(0_+,0_{-})$&$(2_+,2_{-})$&non-BPS \cr \hline

$\mathcal{O}_{18H^*}$& $ (5_+,5_{-})$&$ (8_+,8_{-}) $& $(10_+,6_{-})$  &$(2_+,1_{-})$&$(0_+,0_{-})$&$(2_+,1_{-})$&$(0_+,0_{-})$&$(2_+,2_{-})$&non-BPS \cr \hline

%35

$\mathcal{O}'_{16H^*}$& $ (5_+,5_{-})$&$ (8_+,8_{-}) $& $(10_+,6_{-})$ &$(1_+,0_{-})$&$(1_+,0_{-})$&$(2_+,1_{-})$&$(2_+,4_{-})$&$(2_+,4_{-})$&non-BPS \cr \hline

$\bar{\mathcal{O}}'_{17H^*}$& $ (5_+,5_{-})$&$ (8_+,8_{-}) $& $(10_+,6_{-})$ &$(1_+,0_{-})$&$(1_+,0_{-})$&$(2_+,1_{-})$&$(3_+,3_{-})$&$(3_+,3_{-})$&non-BPS \cr \hline

$\hat{\mathcal{O}}'_{17H^*}$& $ (5_+,5_{-})$&$ (8_+,8_{-}) $& $(10_+,6_{-})$&$(1_+,0_{-})$&$(1_+,0_{-})$&$(2_+,1_{-})$&$(4_+,2_{-})$&$(4_+,2_{-})$&non-BPS \cr \hline
%38

$\mathcal{O}'_{18H^*}$& $ (5_+,5_{-})$&$ (8_+,8_{-}) $& $(10_+,6_{-})$&$(1_+,0_{-})$&$(0_+,1_{-})$&$(2_+,1_{-})$&$(3_+,3_{-})$&$(3_+,3_{-})$&non-BPS \cr \hline
%39

$\mathcal{O}''_{16H^*}$& $ (5_+,5_{-})$&$ (8_+,8_{-}) $& $(10_+,6_{-})$ &$(2_+,1_{-})$&$(0_+,0_{-})$&$(2_+,1_{-})$&$(2_+,3_{-})$&$(3_+,3_{-})$&non-BPS \cr \hline

$\mathcal{O}''_{17H^*}$& $ (5_+,5_{-})$&$ (8_+,8_{-}) $& $(10_+,6_{-})$&$(2_+,1_{-})$&$(0_+,0_{-})$&$(2_+,1_{-})$&$(3_+,2_{-})$&$(4_+,2_{-})$&non-BPS \cr \hline

$\bar{\mathcal{O}}''_{18H^*}$& $ (5_+,5_{-})$&$ (8_+,8_{-}) $& $(10_+,6_{-})$ &$(2_+,1_{-})$&$(0_+,0_{-})$&$(2_+,1_{-})$&$(2_+,3_{-})$&$(2_+,4_{-})$&non-BPS \cr \hline

%42

$\hat{\mathcal{O}}''_{18H^*}$& $ (5_+,5_{-})$&$ (8_+,8_{-}) $& $(10_+,6_{-})$ &$(2_+,1_{-})$&$(0_+,0_{-})$&$(2_+,1_{-})$&$(3_+,2_{-})$&$(3_+,3_{-})$&non-BPS \cr \hline
%43

$\mathcal{O}_{19H^*}$& $ (6_+,6_{-})$&$ (8_+,13_{-}) $& $(14_+,7_{-})$&$(1_+,0_{-})$&$(1_+,1_{-})$&$(2_+,1_{-})$&$(7_+,7_{-})$&$(7_+,7_{-})$&non-BPS \cr \hline

$\mathcal{O}_{20 H^*}$& $ (6_+,6_{-})$&$ (10_+,11_{-}) $& $(12_+,9_{-})$ &$(1_+,0_{-})$&$(1_+,1_{-})$&$(2_+,1_{-})$&$(7_+,7_{-})$&$(7_+,7_{-})$&non-BPS \cr \hline
%45

$\mathcal{O}'_{19 H^*}$& $ (8_+,6_{-})$&$ (9_+,10_{-}) $& $(12_+,7_{-})$&$(2_+,1_{-})$&$(1_+,0_{-})$&$(2_+,1_{-})$&$(7_+,5_{-})$&$(7_+,5_{-})$&non-BPS \cr \hline

$\bar{\mathcal{O}}'_{20 H^*}$& $ (7_+,7_{-})$&$ (10_+,9_{-}) $& $(11_+,8_{-})$ &$(2_+,1_{-})$&$(1_+,0_{-})$&$(2_+,1_{-})$&$(6_+,6_{-})$&$(6_+,6_{-})$&non-BPS \cr \hline

$\hat{\mathcal{O}}'_{20 H^*}$& $ (6_+,8_{-})$&$ (9_+,10_{-}) $& $(12_+,7_{-})$ &$(2_+,1_{-})$&$(1_+,0_{-})$&$(2_+,1_{-})$&$(5_+,7_{-})$&$(5_+,7_{-})$&non-BPS \cr \hline
%48

$\mathcal{O}_{21 H^*}$& $ (7_+,7_{-})$&$ (10_+,9_{-}) $& $(11_+,8_{-})$&$(2_+,1_{-})$&$(0_+,0_{-})$&$(2_+,1_{-})$&$(6_+,6_{-})$&$(6_+,6_{-})$&non-BPS \cr \hline

$\mathcal{O}_{22 H^*}$& $ (7_+,7_{-})$&$ (10_+,10_{-}) $& $(12_+,8_{-})$ &$(2_+,1_{-})$&$(1_+,0_{-})$&$(2_+,1_{-})$&$(6_+,6_{-})$&$(6_+,6_{-})$&non-BPS \cr \hline

$\mathcal{O}_{23 H^*}$& $ (7_+,7_{-})$&$ (10_+,10_{-}) $& $(12_+,8_{-})$ &$(2_+,1_{-})$&$(0_+,1_{-})$&$(2_+,1_{-})$&$(6_+,6_{-})$&$(6_+,6_{-})$&non-BPS \cr \hline
%51

$\mathcal{O}'_{22 H^*}$& $ (7_+,8_{-})$&$ (10_+,10_{-}) $& $(12_+,8_{-})$&$(2_+,1_{-})$&$(0_+,0_{-})$&$(2_+,1_{-})$&$(7_+,6_{-})$&$(7_+,6_{-})$&non-BPS \cr \hline

$\mathcal{O}'_{23 H^*}$& $ (8_+,7_{-})$&$ (10_+,10_{-}) $& $(12_+,8_{-})$ &$(2_+,1_{-})$&$(0_+,0_{-})$&$(2_+,1_{-})$&$(6_+,7_{-})$&$(6_+,7_{-})$&non-BPS \cr \hline
%53

$\mathcal{O}_{24 H^*}$& $ (8_+,8_{-})$&$ (11_+,11_{-}) $& $(13_+,9_{-})$ &$(2_+,1_{-})$&$(1_+,0_{-})$&$(2_+,1_{-})$&$(7_+,9_{-})$&$(7_+,9_{-})$&non-BPS \cr \hline

$\mathcal{O}_{25 H^*}$& $ (8_+8_{-})$&$ (11_+,11_{-}) $& $(13_+,9_{-})$ &$(2_+,1_{-})$&$(1_+,0_{-})$&$(2_+,1_{-})$&$(9_+,7_{-})$&$(9_+,7_{-})$&non-BPS \cr \hline
%55

$\mathcal{O}'_{24 H^*}$& $ (8_+,8_{-})$&$ (11_+,11_{-}) $& $(13_+,9_{-})$ &$(2_+,1_{-})$&$(1_+,1_{-})$&$(2_+,1_{-})$&$(8_+,8_{-})$&$(8_+,8_{-})$&non-BPS \cr \hline

$\mathcal{O}'_{25 H^*}$& $ (8_+,8_{-})$&$ (11_+,11_{-}) $& $(13_+,9_{-})$ &$(2_+,1_{-})$&$(1_+,1_{-})$&$(2_+,1_{-})$&$(8_+,8_{-})$&$(8_+,8_{-})$&non-BPS \cr \hline

$\mathcal{O}_{26 H^*}$& $ (9_+,9_{-})$&$ (12_+,12_{-}) $& $(14_+,10_{-})$ &$(2_+,1_{-})$&$(1_+,1_{-})$&$(2_+,1_{-})$&$(10_+,10_{-})$&$(10_+,10_{-})$&non-BPS \cr \hline

\end{tabular}
\caption{Signature of the relevant tensor classifiers for the various $H^*$-orbits.}    \label{orbitstc}
\end{center}
\end{table}

\section{Generators of $\mathfrak{f}_{4(4)}$ in the ${\bf 26}$}\label{apptech}
Here we present the generators of  $\mathfrak{f}_{4(4)}$ in terms of matrices $e_{i,j}$ whose only non vanishing entry is a $1$ on the $i^{th}$ row and $j^{th}$ column.
\begin{eqnarray}H_{1}&=&\frac{1}{2}(2 e_{1,1}+e_{2,2}+e_{3,3}+e_{4,4}+e_{5,5}+e_{6,6}+e_{7,7}+e_{9,9}+e_{11,11}-e_{16,16}-e_{18,18} \nonumber\\&&-e_{20,20}-e_{21,21}-e_{22,22}-e_{23,23}-e_{24,24}-e_{25,25}-2 e_{26,26})\nonumber\\
H_{2}&=&\frac{1}{2}( e_{2,2}+e_{3,3}+e_{4,4}-e_{5,5}+e_{6,6}-e_{7,7}+2e_{8,8}-e_{9,9}-e_{11,11}+e_{16,16} +e_{18,18}\nonumber\\&&-2e_{19,19}+e_{20,20}-e_{21,21}+e_{22,22}-e_{23,23}-e_{24,24}-e_{25,25})\nonumber\\
H_{3}&=&\frac{1}{2}( e_{2,2}+e_{3,3}-e_{4,4}+e_{5,5}-e_{6,6}+e_{7,7}-e_{9,9}-e_{10,10}-e_{11,11}+e_{16,16}-2 e_{17,17} \nonumber\\&&+e_{18,18}-e_{20,20}+e_{21,21}-e_{22,22}+e_{23,23}-e_{24,24}-e_{25,25})\nonumber\\
H_{4}&=&\frac{1}{2}( e_{2,2}-e_{3,3}+e_{4,4}+e_{5,5}-e_{6,6}-e_{7,7}+e_{9,9}-e_{11,11}++2e_{12,12}-2e_{15,15}+ e_{16,16} \nonumber\\&&-e_{18,18}+e_{20,20}+e_{21,21}-e_{22,22}-e_{23,23}+e_{24,24}-e_{25,25})\\
E_{1}&=&e_{1, 8} + e_{5, 16} - e_{7, 18} + e_{9, 20} - e_{11, 22} -
  e_{19, 26}\nonumber\\
E_{2}&=& e_{1, 10} + e_{4, 16} - e_{6, 18} - e_{9, 21} + e_{11, 23} -
  e_{17, 26}
\nonumber\\
E_{3}&=&-e_{4, 5} - e_{6, 7} + e_{8, 10} - e_{17, 19} - e_{20, 21} - e_{22, 23}\nonumber\\
E_{4}&=&e_{1, 12} + e_{3, 16} + e_{6, 20} + e_{7, 21} + e_{11, 24} - e_{15, 26}\nonumber\\
E_{5}&=&-e_{3, 5} - e_{6, 9} + e_{8, 12} - e_{15, 19} - e_{18, 21} -
  e_{22, 24}\nonumber\\
E_{6}&=&-e_{3, 4} + e_{7, 9} + e_{10, 12} - e_{15, 17} - e_{18, 20} + e_{23, 24}\nonumber\\
E_{7}&=&e_{1,
      13} + \frac{e_{2, 16}}{\sqrt{2}} + \frac{e_{3,
          18}}{\sqrt{2}} + \frac{e_{4, 20}}{\sqrt{2}} + \frac{e_{5,
          21}}{\sqrt{2}} + \frac{e_{6, 22}}{\sqrt{2}} + \frac{e_{7,
          23}}{\sqrt{2}} + \frac{e_{9, 24}}{\sqrt{2}} + \frac{e_{11,
          25}}{\sqrt{2}} - e_{13, 26}\nonumber\\
E_{8}&=&-\frac{e_{2, 5}}{\sqrt{2}} + \frac{e_{3,
          7}}{\sqrt{2}} - \frac{e_{4, 9}}{\sqrt{2}} + \frac{e_{6,
          11}}{\sqrt{2}} + e_{8, 13} -
  e_{13, 19} + \frac{e_{16, 21}}{\sqrt{2}} - \frac{e_{18,
          23}}{\sqrt{2}} + \frac{e_{20, 24}}{\sqrt{2}} - \frac{e_{22,
          25}}{\sqrt{2}}\nonumber\\
E_{9}&=&-\frac{e_{2, 4}}{\sqrt{2}} + \frac{e_{3,
          6}}{\sqrt{2}} + \frac{e_{5, 9}}{\sqrt{2}} - \frac{e_{7,
          11}}{\sqrt{2}} + e_{10, 13} -
  e_{13, 17} + \frac{e_{16, 20}}{\sqrt{2}} - \frac{e_{18,
          22}}{\sqrt{2}} - \frac{e_{21, 24}}{\sqrt{2}} + \frac{e_{23,
          25}}{\sqrt{2}}\nonumber\\
E_{10}&=&-\frac{e_{2, 3}}{\sqrt{2}} - \frac{e_{4, 6}}{\sqrt{2}} - \frac{e_{5,
          7}}{\sqrt{2}} - \frac{e_{9, 11}}{\sqrt{2}} + e_{12, 13} -
  e_{13, 15} + \frac{e_{16, 18}}{\sqrt{2}} + \frac{e_{20,
          22}}{\sqrt{2}} + \frac{e_{21, 23}}{\sqrt{2}} + \frac{e_{24,
          25}}{\sqrt{2}}\nonumber\\
E_{11}&=&e_{1, 15} - e_{2, 18} - e_{4, 22} - e_{5, 23} - e_{9, 25} -
  e_{12, 26}\nonumber\\
E_{12}&=&-e_{2, 7} - e_{4, 11} + e_{8, 15} - e_{12, 19} - e_{16, 23} -
   e_{20, 25}\nonumber\\
 E_{13}&=&-e_{2, 6} + e_{5, 11} + e_{10, 15} - e_{12, 17} -
  e_{16, 22} + e_{21, 25}
\end{eqnarray}

\begin{eqnarray}E_{14}&=&e_{2, 9} + e_{3, 11} + e_{8, 17} - e_{10, 19} + e_{16, 24} +
  e_{18, 25}\nonumber\\
E_{15}&=&e_{1, 17} - e_{2, 20} + e_{3, 22} + e_{5, 24} - e_{7, 25} -
  e_{10, 26}\nonumber\\
E_{16}&=&(e_{(1,19)}-e_{(2,21)}+e_{(3,23)}-e_{(4,24)}+e_{(6,25)}-e_{(8,26)})\nonumber\\
E_{17}&=& -\frac{e_{1, 6}}{\sqrt{2}} + \frac{e_{2,
            8}}{\sqrt{2}} - \frac{e_{5, 13}}{2} + \frac{1}{2} \sqrt{3} e_{5,
        14} - \frac{e_{7, 15}}{\sqrt{2}} + \frac{e_{9,
            17}}{\sqrt{2}} + \frac{e_{10, 18}}{\sqrt{2}} - \frac{e_{12,
            20}}{\sqrt{2}} - \frac{e_{13,
            22}}{2} + \nonumber\\&&+\frac{1}{2} \sqrt{3} e_{14,
        22} - \frac{e_{19, 25}}{\sqrt{2}} + \frac{e_{21, 26}}{\sqrt{2}}\nonumber\\
E_{18}&=&\frac{e_{1, 4}}{\sqrt{2}} + \frac{e_{3, 8}}{\sqrt{2}} - \frac{e_{5,
          12}}{\sqrt{2}} + \frac{e_{7, 13}}{2} + \frac{1}{2} \sqrt{3} e_{7,
      14} + \frac{e_{10, 16}}{\sqrt{2}} + \frac{e_{11,
          17}}{\sqrt{2}} + \frac{e_{13, 20}}{2} + \frac{1}{2} \sqrt{3} e_{14,
      20} -\nonumber\\&&- \frac{e_{15, 22}}{\sqrt{2}} - \frac{e_{19,
          24}}{\sqrt{2}} - \frac{e_{23, 26}}{\sqrt{2}}\nonumber\\
E_{19}&=&\frac{e_{1, 3}}{\sqrt{2}} - \frac{
          e_{4, 8}}{\sqrt{2}} + \frac{e_{5, 10}}{\sqrt{
          2}} + \frac{e_{9, 13}}{2} + \frac{
      1}{2} \sqrt{3} e_{9,
          14} + \frac{e_{11, 15}}{\sqrt{2}} + \frac{e_{12, 16}}{\sqrt{
          2}} + \frac{e_{13,
   18}}{2} + \frac{1}{2} \sqrt{3} e_{14, 18} +\nonumber\\&&+ \frac{e_{17, 22}}{\sqrt{
      2}} + \frac{e_{19, 23}}{\sqrt{2}} - \frac{e_{24, 26}}{\sqrt{2}}\nonumber\\
E_{20}&=&-\frac{e_{1, 2}}{\sqrt{2}} - \frac{e_{6, 8}}{\sqrt{2}} + \frac{e_{7,
          10}}{\sqrt{2}} + \frac{e_{9, 12}}{\sqrt{2}} - \frac{e_{11,
          13}}{2} + \frac{1}{2} \sqrt{3} e_{11,
      14} - \frac{e_{13, 16}}{2} + \frac{1}{2} \sqrt{3} e_{14,
      16} + \nonumber\\&&+\frac{e_{15, 18}}{\sqrt{2}} + \frac{e_{17,
          20}}{\sqrt{2}} + \frac{e_{19, 21}}{\sqrt{2}} + \frac{e_{25,
          26}}{\sqrt{2}}\nonumber\\
E_{21}&=&-\frac{e_{1, 7}}{\sqrt{2}} - \frac{e_{2, 10}}{\sqrt{2}} + \frac{e_{4,
          13}}{2} - \frac{1}{2} \sqrt{3} e_{4,
      14} + \frac{e_{6, 15}}{\sqrt{2}} + \frac{e_{8,
          18}}{\sqrt{2}} + \frac{e_{9, 19}}{\sqrt{2}} - \frac{e_{12,
          21}}{\sqrt{2}} - \frac{e_{13, 23}}{2} + \nonumber\\&&+ \frac{1}{2} \sqrt{3} e_{14,
      23} + \frac{e_{17, 25}}{\sqrt{2}} - \frac{e_{20, 26}}{\sqrt{2}}\nonumber\\
E_{22}&=&\frac{e_{1, 5}}{\sqrt{2}} - \frac{e_{3, 10}}{\sqrt{2}} + \frac{e_{4,
          12}}{\sqrt{2}} - \frac{e_{6, 13}}{2} - \frac{1}{2} \sqrt{3} e_{6,
      14} + \frac{e_{8, 16}}{\sqrt{2}} + \frac{e_{11,
          19}}{\sqrt{2}} + \frac{e_{13, 21}}{2} + \frac{1}{2} \sqrt{3} e_{14,
      21} - \nonumber\\&&-\frac{e_{15, 23}}{\sqrt{2}} + \frac{e_{17,
          24}}{\sqrt{2}} + \frac{e_{22, 26}}{\sqrt{2}}\nonumber\\
E_{23}&=&\frac{e_{1, 11}}{\sqrt{2}} - \frac{e_{2, 13}}{2} - \frac{1}{2} \sqrt{3} e_{2,
      14} - \frac{e_{3, 15}}{\sqrt{2}} - \frac{e_{4,
          17}}{\sqrt{2}} - \frac{e_{5, 19}}{\sqrt{2}} - \frac{e_{8,
          22}}{\sqrt{2}} + \frac{e_{10, 23}}{\sqrt{2}} + \frac{e_{12,
          24}}{\sqrt{2}} +\nonumber\\&&+ \frac{e_{13, 25}}{2} + \frac{1}{2} \sqrt{3} e_{14,
      25} + \frac{e_{16, 26}}{\sqrt{2}}\nonumber\\
E_{24}&=&-\frac{e_{1, 9}}{\sqrt{2}} - \frac{e_{2, 12}}{\sqrt{2}} + \frac{e_{3,
          13}}{2} - \frac{1}{2} \sqrt{3} e_{3,
      14} - \frac{e_{6, 17}}{\sqrt{2}} - \frac{e_{7,
          19}}{\sqrt{2}} - \frac{e_{8, 20}}{\sqrt{2}} + \frac{e_{10,
          21}}{\sqrt{2}} - \frac{e_{13, 24}}{2} +\nonumber\\&&+ \frac{1}{2} \sqrt{3} e_{14,
      24} + \frac{e_{15, 25}}{\sqrt{2}} - \frac{e_{18, 26}}{\sqrt{2}}\\
\end{eqnarray}
In the chosen basis for the fundamental representation the shift generators corresponding to negative roots $F_\alpha=E_{-\alpha}$ are $F_\alpha=(E_\alpha)^T$.\par
Let us now give the $\mathfrak{H}^*=\mathfrak{sl}(2)\oplus \mathfrak{sp}'(6)$ generators in terms of the shift generators $\hat{E}_{\beta'}$ corresponding to the positive roots $\beta'$, relative to the basis (\ref{CHstarb}) of the Cartan subalgebra $\mathcal{C}_{H^*}$. As usual $\beta'_i$ represent the $\mathfrak{sp}'(6)$-simple roots and $\beta'_4$ the  $\mathfrak{sl}(2)$ simple  roots:
\begin{align}
\hat{E}_{\beta'_1}&=\sqrt{2} \left(J_7+J_{10}\right)\,,\nonumber\\
\hat{E}_{\beta'_2}&=\sqrt{2} \left(J_{17}+J_{21}\right)\,,\nonumber\\
\hat{E}_{\beta'_3}&=-\frac{J_1+J_2-J_5+J_6+J_{12}-J_{13}-J_{15}+J_{16}}{\sqrt{2}}\,,\nonumber\\
\hat{E}_{\beta'_1+\beta'_2}&=\sqrt{2} \left(J_{18}+J_{22}\right)\,,\nonumber\\
\hat{E}_{\beta'_2+\beta'_3}&=\sqrt{2} \left(J_{20}-J_{24}\right)\,,\nonumber\\
\hat{E}_{\beta'_1+\beta'_2+\beta'_3}&=\sqrt{2} \left(J_{19}-J_{23}\right)\,,\nonumber\\
\hat{E}_{2\beta'_2+\beta'_3}&=\frac{-J_1+J_2+J_5+J_6+J_{12}+J_{13}+J_{15}+J_{16}}{\sqrt{2}}\,,\nonumber\\
\hat{E}_{\beta'_1+2\beta'_2+\beta'_3}&=\sqrt{2} \left(J_8+J_9\right)\,,\nonumber\\
\hat{E}_{2\beta'_1+2\beta'_2+\beta'_3}&=\frac{J_1-J_2+J_5+J_6+J_{12}+J_{13}-J_{15}-J_{16}}{\sqrt{2}}\,,\nonumber\\
\hat{E}_{\beta'_4}&=\frac{1}{2} \left(J_1+J_2+J_5-J_6-J_{12}+J_{13}-J_{15}+J_{16}\right)\,,
\end{align}
where the $J_\alpha$, $\alpha=1,\dots, 24$, are defined as $J_\alpha\equiv \frac{1}{2}\,(E_\alpha-\eta E_\alpha^T\eta)$. The corresponding negative-root generators are obtained through transposition: $\hat{E}_{-\beta'}=\hat{E}_{\beta'}^T$. \par
Finally in Table \ref{Tablehat} we list the shift generators $\hat{E}_\alpha$ (the positive roots being represented by the corresponding number
in Table \ref{f44posr}). These roots are referred to the Cartan subalgebra $\mathcal{C}$ defined in Sect. \ref{owhdn}.
\begin{table}
\begin{center}
\small\addtolength{\tabcolsep}{-1pt}
\scalefont{.8}
\begin{tabular}{|c|r|r|r|r|r|r|r|r|r|r|}
  \hline
  $ \hat{E}_{1}=N_{1}^{+}=\frac{1}{2}(H_{2}-H_{3}-2 K_{3})$& $\hat{F}_{1}=N_{1}^{-}=\frac{1}{2}(H_{2}-H_{3}+2 K_{3})$ \\ \hline
 $ \hat{E}_{2}=\frac{1}{2}( K_{16}+ K_{12}- K_{5}+ K_{2}+ K_{15}- K_{1}- K_{6}+ K_{13})$& $\hat{F}_{2}=\frac{1}{2}( K_{16}- K_{12}+ K_{5}- K_{2}- K_{15}- K_{1}- K_{6}+ K_{13})$ \\ \hline
$ \hat{E}_{5}=\frac{1}{2}( K_{16}- K_{12}- K_{5}+ K_{2}- K_{15}+ K_{1}+ K_{6}+ K_{13})$& $\hat{F}_{5}=\frac{1}{2}( K_{16}+ K_{12}+ K_{5}- K_{2}+ K_{15}+ K_{1}+ K_{6}+ K_{13})$ \\ \hline
$ \hat{E}_{6}=N_{3}^{+}=\frac{1}{2}(H_{1}-H_{4}-2 K_{4})$& $\hat{F}_{6}=N_{3}^{-}=\frac{1}{2}(H_{1}-H_{4}+2 K_{4})$ \\ \hline
$ \hat{E}_{8}= K_{8}- K_{9}$& $\hat{F}_{8}= K_{8}+ K_{9}$ \\ \hline
$ \hat{E}_{9}= K_{7}+ K_{10}$& $\hat{F}_{9}= K_{7}- K_{10}$ \\ \hline
$ \hat{E}_{12}=\frac{1}{2}( K_{16}+ K_{12}+ K_{5}+ K_{2}- K_{15}+ K_{1}- K_{6}- K_{13})$& $\hat{F}_{12}=\frac{1}{2}( K_{16}- K_{12}- K_{5}- K_{2}+ K_{15}+ K_{1}- K_{6}- K_{13})$ \\ \hline
 $ \hat{E}_{13}=N_{4}^{+}=\frac{1}{2}(H_{1}+H_{4}+2 K_{11})$& $\hat{F}_{13}=N_{4}^{-}=\frac{1}{2}(H_{1}+H_{4}-2 K_{11})$ \\ \hline
 $ \hat{E}_{15}=\frac{1}{2}( K_{16}- K_{12}+ K_{5}+ K_{2}+ K_{15}- K_{1}+ K_{6}- K_{13})$& $\hat{F}_{15}=\frac{1}{2}( K_{16}+ K_{12}- K_{5}- K_{2}- K_{15}- K_{1}+ K_{6}- K_{13})$ \\ \hline
 $ \hat{E}_{16}=N_{2}^{+}=\frac{1}{2}(H_{2}+H_{3}-2 K_{14})$& $\hat{F}_{16}=N_{2}^{-}=\frac{1}{2}(H_{2}+H_{3}+2 K_{14})$ \\ \hline
 $ \hat{E}_{19}=- K_{22}- K_{18}$& $\hat{F}_{19}=- K_{22}+ K_{18}$ \\ \hline
 $ \hat{E}_{20}= K_{21}+ K_{17}$& $\hat{F}_{20}= K_{21}- K_{17}$ \\ \hline
 $ \hat{E}_{23}= K_{23}- K_{19}$& $\hat{F}_{23}= K_{23}+ K_{19}$ \\ \hline
 $ \hat{E}_{24}= K_{24}- K_{20}$& $\hat{F}_{24}= K_{24}+ K_{20}$ \\ \hline

\end{tabular}
\caption{The shift generators $\hat{E}_k$ and $\hat{F}_k=\hat{E}^T_k$ .}    \label{Tablehat}
\end{center}
\end{table}

\bibliography{F4}
\bibliographystyle{utphysmodb}
\end{document}